\documentclass[12pt,a4paper]{article}
\usepackage{ifthen} 
\newboolean{pdflatex}
\setboolean{pdflatex}{true} 
\newboolean{articletitles}
\setboolean{articletitles}{true}
\newboolean{uprightparticles}
\setboolean{uprightparticles}{false}
\def\paperauthors{LHCb Collaboration}
\def\paperasciititle{Multiplicity Dependence of ratio of production of psi(2S) over J/psi in pp collisions at center of mass energy 13 TeV} 
\def\papertitle{Multiplicity dependence of $\sigma_{\psitwos}/\sigma_{\jpsi}$ in $pp$ collisions at $\sqrt{s}=13$ TeV} 
\def\paperkeywords{{High Energy Physics}, {LHCb}} 
\def\papercopyright{\the\year\ CERN for the benefit of the LHCb collaboration} 
\def\paperlicence{CC BY 4.0 licence}
\def\paperlicenceurl{https://creativecommons.org/licenses/by/4.0/}

\usepackage[top=1in, bottom=1.25in, left=1in, right=1in]{geometry}

\usepackage{microtype}
\usepackage{lineno}  
\usepackage{xspace} 
\usepackage{caption} 

\usepackage{graphicx}  
\usepackage{color}
\usepackage{colortbl}
\graphicspath{{./figs/}} 

\usepackage{amsmath} 
\usepackage{amssymb}
\usepackage{amsfonts}
\usepackage{upgreek} 

\newcommand*\patchAmsMathEnvironmentForLineno[1]{%
\expandafter\let\csname old#1\expandafter\endcsname\csname #1\endcsname
\expandafter\let\csname oldend#1\expandafter\endcsname\csname
end#1\endcsname
 \renewenvironment{#1}%
   {\linenomath\csname old#1\endcsname}%
   {\csname oldend#1\endcsname\endlinenomath}%
}
\newcommand*\patchBothAmsMathEnvironmentsForLineno[1]{%
  \patchAmsMathEnvironmentForLineno{#1}%
  \patchAmsMathEnvironmentForLineno{#1*}%
}
\AtBeginDocument{%
\patchBothAmsMathEnvironmentsForLineno{equation}%
\patchBothAmsMathEnvironmentsForLineno{align}%
\patchBothAmsMathEnvironmentsForLineno{flalign}%
\patchBothAmsMathEnvironmentsForLineno{alignat}%
\patchBothAmsMathEnvironmentsForLineno{gather}%
\patchBothAmsMathEnvironmentsForLineno{multline}%
\patchBothAmsMathEnvironmentsForLineno{eqnarray}%
}

\usepackage{hyperxmp}

\usepackage[pdftex,
            pdfauthor={\paperauthors},
            pdftitle={\paperasciititle},
            pdfkeywords={\paperkeywords},
            pdfcopyright={Copyright (C) \papercopyright},
            pdflicenseurl={\paperlicenceurl}]{hyperref}

\usepackage[colorinlistoftodos,textsize=scriptsize]{todonotes}

\usepackage[bottom,flushmargin,hang,multiple]{footmisc}

\usepackage[all]{hypcap} 

\usepackage{xspace} 
\usepackage{upgreek}

\def\lhcb   {\mbox{LHCb}\xspace}

\def\MagUp {\mbox{\em Mag\kern -0.05em Up}\xspace}

\ifthenelse{\boolean{uprightparticles}}%
{

 \def\Ppsi        {\ensuremath{\uppsi}\xspace}

 \def\PDelta      {\ensuremath{\Delta}\xspace}                 
 \def\PXi         {\ensuremath{\Xi}\xspace}                 
 \def\PLambda     {\ensuremath{\Lambda}\xspace}                 
 \def\PSigma      {\ensuremath{\Sigma}\xspace}                 
 \def\POmega      {\ensuremath{\Omega}\xspace}                 
 \def\PUpsilon    {\ensuremath{\Upsilon}\xspace}
 \let\oldPi\Pi
 \def\PPi         {\ensuremath{\oldPi}\xspace}

 \def\PB      {\ensuremath{\mathrm{B}}\xspace}                 
                  
 \def\PD      {\ensuremath{\mathrm{D}}\xspace}

 \def\PJ      {\ensuremath{\mathrm{J}}\xspace}                 
 \def\PK      {\ensuremath{\mathrm{K}}\xspace}

 \def\Pi      {\ensuremath{\mathrm{i}}\xspace}

 \def\Ps      {\ensuremath{\mathrm{s}}\xspace}

 \def\thebaroffset{0.0em}
}
{

 \def\Ppsi        {\ensuremath{\psi}\xspace}                 
                  
 \mathchardef\PDelta="7101
 \mathchardef\PXi="7104
 \mathchardef\PLambda="7103
 \mathchardef\PSigma="7106
 \mathchardef\POmega="710A
 \mathchardef\PUpsilon="7107
 \mathchardef\PPi="7105
                  
 \def\PB      {\ensuremath{B}\xspace}                 
                  
 \def\PD      {\ensuremath{D}\xspace}

 \def\PJ      {\ensuremath{J}\xspace}                 
 \def\PK      {\ensuremath{K}\xspace}

 \def\Pi      {\ensuremath{i}\xspace}

 \def\Ps      {\ensuremath{s}\xspace}

 \def\thebaroffset{0.18em}
}
\newcommand{\offsetoverline}[2][\thebaroffset]{\kern #1\overline{\kern -#1 #2}}%

\makeatletter
\ifcase \@ptsize \relax
  \newcommand{\miniscule}{\@setfontsize\miniscule{4}{5}}
\or
  \newcommand{\miniscule}{\@setfontsize\miniscule{5}{6}}
\or
  \newcommand{\miniscule}{\@setfontsize\miniscule{5}{6}}
\fi
\makeatother

\DeclareRobustCommand{\optbar}[1]{\shortstack{{\miniscule (\rule[.5ex]{1.25em}{.18mm})}
  \\ [-.7ex] $#1$}}

\def\squark    {{\ensuremath{\Ps}}\xspace}

\def\KorKbar {\kern \thebaroffset\optbar{\kern -\thebaroffset \PK}{}\xspace}

\def\D       {{\ensuremath{\PD}}\xspace}

\def\DorDbar {\kern \thebaroffset\optbar{\kern -\thebaroffset \PD}\xspace}

\def\Dp      {{\ensuremath{\D^+}}\xspace}
\def\Dm      {{\ensuremath{\D^-}}\xspace}

\def\DpDm    {\ensuremath{\Dp {\kern -0.16em \Dm}}\xspace}

\def\B       {{\ensuremath{\PB}}\xspace}

\def\BorBbar {\kern \thebaroffset\optbar{\kern -\thebaroffset \PB}\xspace}

\def\Bd      {{\ensuremath{\B^0}}\xspace}

\def\BdorBdbar {\kern \thebaroffset\optbar{\kern -\thebaroffset \Bd}\xspace}

\def\Bs      {{\ensuremath{\B^0_\squark}}\xspace}

\def\BsorBsbar {\kern \thebaroffset\optbar{\kern -\thebaroffset \Bs}\xspace}

\def\jpsi     {{\ensuremath{{\PJ\mskip -3mu/\mskip -2mu\Ppsi}}}\xspace}
\def\psitwos  {{\ensuremath{\Ppsi{(2S)}}}\xspace}

\def\Y#1S{\ensuremath{\PUpsilon{(#1S)}}\xspace}
\def\OneS  {{\Y1S}\xspace}
\def\TwoS  {{\Y2S}\xspace}
\def\ThreeS{{\Y3S}\xspace}

\def\LorLbar     {\kern \thebaroffset\optbar{\kern -\thebaroffset \PLambda}\xspace}

\def\BF         {{\ensuremath{\mathcal{B}}}\xspace}
\def\BR         {\BF}

\def\AT#1     {\ensuremath{A_{\mathrm{T}}^{#1}}\xspace}           

\def\C#1      {\ensuremath{\mathcal{C}_{#1}}\xspace}                       
\def\Cp#1     {\ensuremath{\mathcal{C}_{#1}^{'}}\xspace}                    
\def\Ceff#1   {\ensuremath{\mathcal{C}_{#1}^{\mathrm{(eff)}}}\xspace}        
\def\Cpeff#1  {\ensuremath{\mathcal{C}_{#1}^{'\mathrm{(eff)}}}\xspace}       
\def\Ope#1    {\ensuremath{\mathcal{O}_{#1}}\xspace}                       
\def\Opep#1   {\ensuremath{\mathcal{O}_{#1}^{'}}\xspace}                    

\newcommand{\nospaceunit}[1]{\ensuremath{\text{#1}}}       
\newcommand{\aunit}[1]{\ensuremath{\text{\,#1}}}       

\newcommand{\tev}{\aunit{Te\kern -0.1em V}\xspace}
\newcommand{\gev}{\aunit{Ge\kern -0.1em V}\xspace}
\newcommand{\mev}{\aunit{Me\kern -0.1em V}\xspace}
\newcommand{\kev}{\aunit{ke\kern -0.1em V}\xspace}
\newcommand{\ev}{\aunit{e\kern -0.1em V}\xspace}
 
\newcommand{\mevc}{\ensuremath{\aunit{Me\kern -0.1em V\!/}c}\xspace}
\newcommand{\gevc}{\ensuremath{\aunit{Ge\kern -0.1em V\!/}c}\xspace}
\newcommand{\mevcc}{\ensuremath{\aunit{Me\kern -0.1em V\!/}c^2}\xspace}
\newcommand{\gevcc}{\ensuremath{\aunit{Ge\kern -0.1em V\!/}c^2}\xspace}

\def\mum  {\ensuremath{\,\upmu\nospaceunit{m}}\xspace}

\def\pb {\aunit{pb}\xspace}
\def\invpb {\ensuremath{\pb^{-1}}\xspace}

\def\deriv {\ensuremath{\mathrm{d}}}

\def\gsim{{~\raise.15em\hbox{$>$}\kern-.85em
          \lower.35em\hbox{$\sim$}~}\xspace}
\def\lsim{{~\raise.15em\hbox{$<$}\kern-.85em
          \lower.35em\hbox{$\sim$}~}\xspace}

\def\pt         {\ensuremath{p_{\mathrm{T}}}\xspace}

\def\evtgen     {\mbox{\textsc{EvtGen}}\xspace}

\def\geant      {\mbox{\textsc{Geant4}}\xspace}

\def\photos     {\mbox{\textsc{Photos}}\xspace}

\def\pythia     {\mbox{\textsc{Pythia}}\xspace}

\def\tell1  {TELL1\xspace}
\def\ukl1   {UKL1\xspace}

\usepackage{cite} 
\usepackage{mciteplus}

\usepackage{longtable} 
\usepackage{multirow}
\usepackage{float}
\usepackage{url}
\usepackage{amsmath}
\begin{document}
\newcommand{\lhcborcid}[1]{\href{https://orcid.org/#1}{\hspace*{0.1em}\raisebox{-0.45ex}{\includegraphics[width=1em]{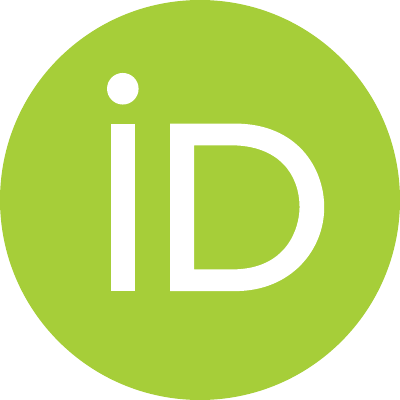}}}}
\renewcommand{\thefootnote}{\fnsymbol{footnote}}
\setcounter{footnote}{1}

\begin{titlepage}
\pagenumbering{roman}
\vspace*{-1.5cm}
\centerline{\large EUROPEAN ORGANIZATION FOR NUCLEAR RESEARCH (CERN)}
\vspace*{1.5cm}
\noindent
\begin{tabular*}{\linewidth}{lc@{\extracolsep{\fill}}r@{\extracolsep{0pt}}}
\ifthenelse{\boolean{pdflatex}}
{\vspace*{-1.5cm}\mbox{\!\!\!\includegraphics[width=.14\textwidth]{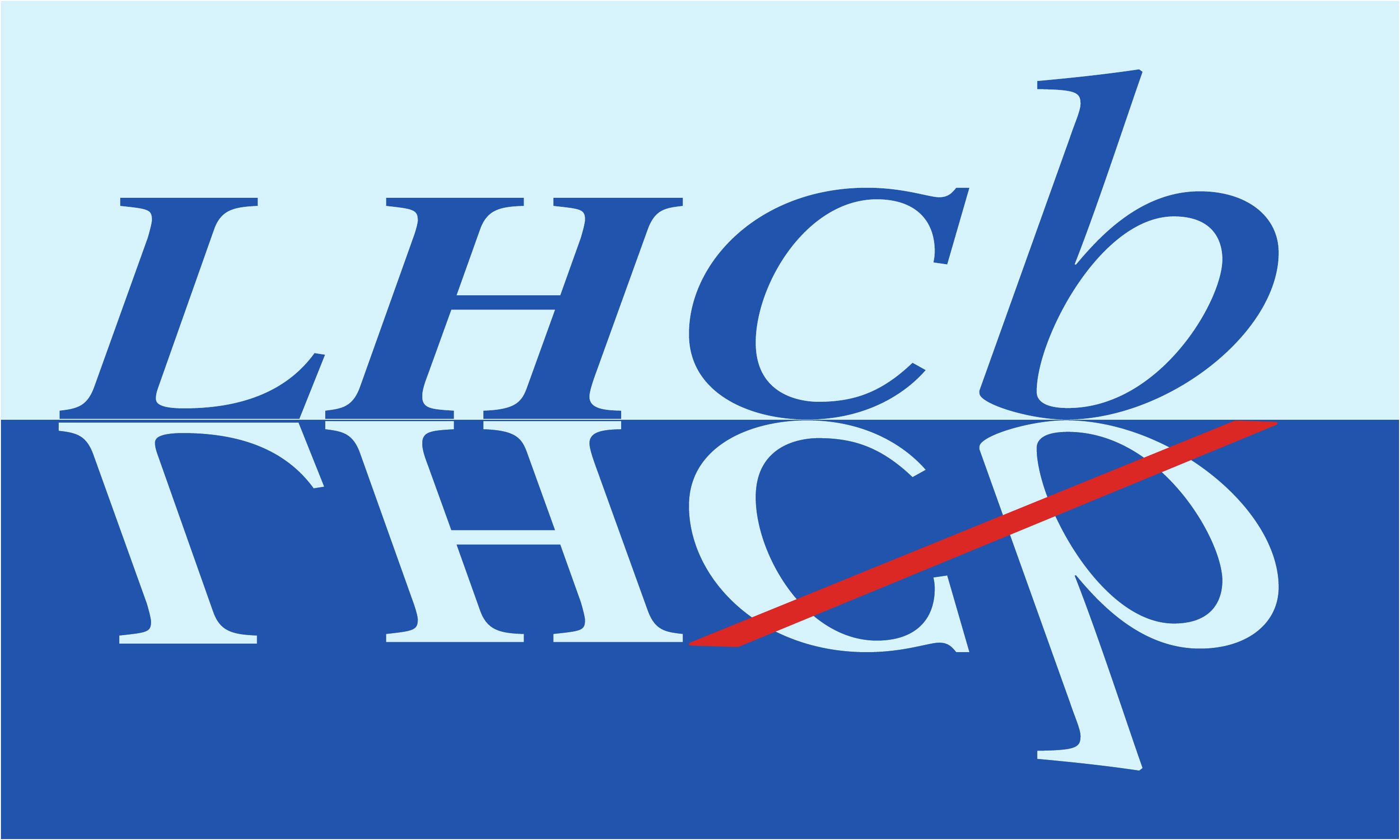}} & &}%
{\vspace*{-1.2cm}\mbox{\!\!\!\includegraphics[width=.12\textwidth]{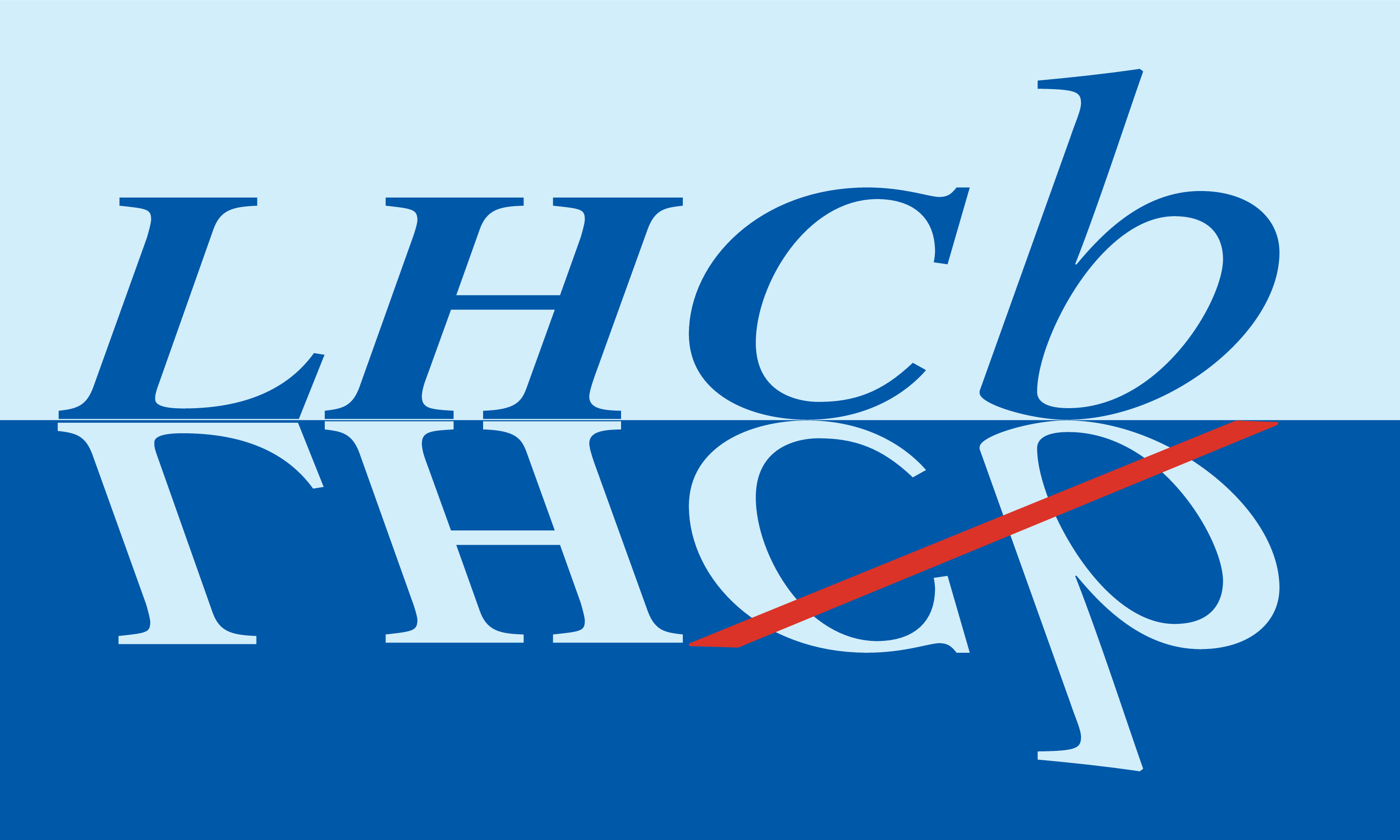}} & &}%
\\
 & & CERN-EP-2023-277 \\ 
 & & LHCb-PAPER-2023-035 \\  
 & & May 22, 2024 \\ 
 & & \\
\end{tabular*}
\vspace*{2.0cm}
{\normalfont\bfseries\boldmath\huge
\begin{center}
  \papertitle 
\end{center}
}
\vspace*{2.0cm}
\begin{center}
\paperauthors\footnote{Authors are listed at the end of this paper.}
\end{center}
\vspace{\fill}

\begin{abstract}
  \noindent
  The ratio of production cross-sections of $\psi(2S)$ over $J/\psi$ mesons as a function of charged-particle multiplicity in proton-proton collisions at a centre-of-mass energy \mbox{$\sqrt{s}=13 \tev$} is measured with a data sample collected by the LHCb detector, corresponding to an integrated luminosity of 658 pb$^{-1}$. The ratio is measured for both prompt and non-prompt $\psi(2S)$ and $J/\psi$ mesons. When there is an overlap between the rapidity ranges over which multiplicity and charmonia production are measured, a multiplicity-dependent modification of the ratio is observed for prompt mesons. No significant multiplicity dependence is found when the ranges do not overlap. For non-prompt production, the \psitwos-to-\jpsi production ratio is roughly independent of multiplicity, irrespective of the rapidity range over which the multiplicity is measured.
  The results are compared to predictions of the co-mover model and agree well except in the low multiplicity region. The ratio of production cross-sections of $\psi(2S)$ over $J/\psi$ mesons are cross-checked with other measurements in di-lepton channels and found to be compatible.
\end{abstract}
\vspace*{2.0cm}
\begin{center}
Published in JHEP 05 (2024) 243
\end{center}
{\footnotesize 
\centerline{\copyright~\papercopyright. \href{\paperlicenceurl}{\paperlicence}.}}
\vspace*{2mm}
\vspace{\fill}
\end{titlepage}
\newpage
\setcounter{page}{2}
\mbox{~}
\renewcommand{\thefootnote}{\arabic{footnote}}
\setcounter{footnote}{0}
\cleardoublepage
\pagestyle{plain}
\setcounter{page}{1}
\pagenumbering{arabic}
\newpage

\section{Introduction}
\label{sec:Introduction}
Heavy quarkonium production has long served as a pivotal tool for probing the intricate dynamics of quantum chromodynamics (QCD). Heavy-ion collisions in particular offer a unique vantage point. Heavy quarks are generated in the initial stages of the collisions, making quarkonium production a powerful probe for the evolution of heavy-ion interactions. Quarkonium production in heavy-ion collisions are influenced by several mechanisms. These include phenomena such as Debye screening, which suppresses the production in quark-gluon plasma (QGP)\cite{MATSUI1986416} and recombination processes~\cite{Zhao:2020jqu}, which lead to an enhancement at higher energies\cite{Thews:2000rj}. To reliably interpret the results from heavy-ion measurements, reference measurements without these nuclear matter effects, such as in proton-proton ($pp$) collisions, are required.

In heavy-ion collisions, quarkonium production is affected by initial-state effects like modification of parton distribution function (PDF)~\cite{AtashbarTehrani:2017mzi} and energy losses through multi-particle scattering~\cite{Arleo:2014oha}. While comparing the production rates for excited quarkonium states to those of the ground states can mitigate the initial-state effects listed above, observations highlight a distinct suppression of excited charmonia (\psitwos) and bottomonia (\TwoS, \ThreeS) compared to their ground states $\jpsi(1S)$ and $\OneS$, from collaborations at the SPS~\cite{NA50:2006yzz}, RHIC~\cite{PHENIX:2022nrm} and LHC~\cite{LHCb:2018psc,ALICE:2020vjy, LHCb:2016vqr,CMS:2022wfi,ATLAS:2017prf}. This difference points towards the necessity of considering additional effects.

These modifications can arise in the presence of QGP, which, together with the Debye color screening may lead to the deconfinement of quarks and hence, the dissociation of quarkonia. The signatures of QGP are observed in small system collisions including protons, at high charged-particle multiplicities. For instance, the ALICE collaboration reported strangeness enhancement in high-multiplicity $pp$ and $p$Pb collisions at $\sqrt{s}=7$ TeV~\cite{ALICE:2016fzo}, while the CMS collaboration identified significant collective flow in high-multiplicity $pp$ collisions at $\sqrt{s}=13$ TeV~\cite{CMS:2016fnw}. Additionally, long-range correlations between particles emitted in a given rapidity range, resembling anisotropic collective flow, have been detected in high-multiplicity $pp$ and $p$Pb collisions at the LHC~\cite{ALICE:2012eyl,ATLAS:2012cix,CMS:2012qk,CMS:2010ifv}. With an increasing number of QGP-like signatures being observed, there is growing interest in investigating quarkonium suppression in high-multiplicity small system collisions as a probe for QGP in such environment.

Another mechanism that affects the quarkonium production is the co-mover effect~\cite{Ferreiro:2012rq}, in which co-moving particles can modify quarkonium properties, resulting in the suppression of these states. Excited states of quarkonium are loosely bound with larger radii, and are more likely to dissociate in the interaction with co-moving particles, which manifests as relative suppression compared to ground states. The suppression by interaction with co-moving particles is correlated to the charged-particle multiplicity since the latter influences the number of co-moving particles. The co-mover model has been successful at describing suppression of charmonia, bottomonia, and $X(3872)$ hadrons in various collision systems ~\cite{Gavin:1996yd,Ferreiro:2018wbd,Esposito:2020ywk,Braaten:2020iqw}. 

This study uses $pp$ collision data at a center-of-mass energy of 13~TeV collected by the LHCb detector in 2016, corresponding to a luminosity of $658\pm13$~pb$^{-1}$, to measure and compare the production cross-sections of prompt and non-prompt (from $b-$hadron decay) \psitwos and \jpsi mesons.  The subsequent determination of cross-section ratios across different multiplicity regions help to understand the relation between charmonium production, QGP and the co-mover effect. 

To distinguish between the influence of the co-mover effect on the production and that of QGP, several types of multiplicity variables are needed. In this analysis, the \psitwos-to-\jpsi cross-section ratio is measured as a function of local multiplicity, which is the multiplicity close to PV, as well as forward (backward) multiplicity, which is measured in same (opposite) rapidity range over which the charmonia production is measured. The usage of backward multiplicity allows the co-mover effect to be removed so that the existence of QGP can be tested.

\section{Detector and data samples}
\label{Data and Monte Carlo samples}
The LHCb detector~\cite{LHCb-DP-2008-001,LHCb-DP-2014-002} is a single-arm forward spectrometer covering the pseudorapidity range $2 < \eta < 5$, designed primarily for the study of particles containing $b$ or $c$ quarks. The detector includes a high-precision tracking system consisting of a silicon-strip vertex detector surrounding the $pp$ interaction region~\cite{LHCb-DP-2014-001}, a large-area silicon-strip detector located upstream of a dipole magnet with a bending power of about 4 Tm, and three stations of silicon-strip detectors and straw drift tubes~\cite{LHCb-DP-2013-003} placed downstream of the magnet. The tracking system provides a measurement of momentum, $p$, of charged particles with a relative uncertainty that varies from $0.5\%$ at low momentum to $1.0\%$ at 200~\gevc. The minimum distance of a track to a primary vertex (PV), the impact parameter, is measured with a resolution of $(15 + 29/\pt)\mum$, where \pt is expressed in \gevc. Different types of charged hadrons are distinguished using information from two ring-imaging Cherenkov detectors~\cite{LHCb-DP-2012-003}. Photons, electrons and hadrons are identified by a calorimeter system consisting of scintillating-pad (SPD) and preshower detectors, an electromagnetic calorimeter and a hadronic calorimeter. Muons are identified by a system composed of alternating layers of iron and multiwire proportional chambers~\cite{LHCb-DP-2012-002}.

Simulated signal samples are used to study the efficiencies. In the simulation, $pp$ collisions are generated using \pythia~\cite{Sjostrand:2006za} with a specific LHCb configuration~\cite{LHCb:2011dpk}. The prompt-charmonium production is simulated in \pythia with contributions from both the leading order color-singlet and color-octet mechanisms~\cite{LHCb:2011dpk,Bargiotti:2007zz}. Decays of hadronic particles are described by \evtgen~\cite{Lange:2001uf}, in which final state radiation is generated using \photos~\cite{Golonka:2005pn}. The interaction of the generated particles with the detector and its response are implemented using the \geant toolkit~\cite{GEANT4:2002zbu,Clemencic:2011zza}.  The charmonia are generated unpolarised since all the LHC measurements so far indicate that the polarisation is small~\cite{LHCb-PAPER-2013-008,LHCb-PAPER-2013-067}.

\section{Selection of \boldmath{\jpsi} and \boldmath{\psitwos} candidates}
\label{Candidate Reconstruction and selection}

The \jpsi and \psitwos candidates are reconstructed through their decays to the $\mu^+\mu^-$ final state.
The trigger selection consists of a hardware stage and a software stage. The hardware trigger selects events containing two tracks identified as muon tracks. Then in the software stage, the two muons are required to be oppositely charged and have good qualities of the track fit and vertex reconstruction fit. Both muons are required to have a momentum of at least $6 \gevc$ and transverse momentum $\pt > 0.3$~\gevc. The invariant masses of the \jpsi and \psitwos candidates must be within $120 \mevcc $ of the known  masses~\cite{Workman:2022ynf}. 

Additional selections and particle identification (PID) requirements are imposed to suppress the background from random combinations of tracks. Each muon is further required to have $\pt>1.2\gevc$ and $2.0<\eta<4.9$. Events are required to have exactly one PV, with a coordinate along the beam direction, $z_{\mathrm{PV}}$, within a range in which the Vertex Locator (VELO) acceptance is uniform and the vertex reconstruction efficiency is weakly dependent on the track multiplicity. The range varies for the multiplicity variables due to the different rapidity ranges they are measured and is listed in Table~\ref{T1}, which is introduced together with multiplicity variables in Sec~\ref{sec:mul}. 

The pseudo decay time $t_z$ is defined as
\begin{equation}
    \label{PseudoDecayTime}
    t_z = \frac{(z-z_{\mathrm{PV}}) \times{} m}{p_z},
\end{equation}
where $z$ is the decay vertex $z$ coordinate of the charmonium candidate, $p_z$ is the momentum in $z$-direction, and $m$ is the known mass for the corresponding charmonium state~\cite{Workman:2022ynf}. Candidates are required to satisfy $|t_z|<10$~ps and $|\sigma_{t_z}|<0.3$~ps, where $\sigma_{t_z}$ is the uncertainty of $t_z$. The pseudo decay time distribution is used to distinguish the prompt from non-prompt charmonia, since $t_z$ tends to be larger for non-prompt components.

\section{Multiplicity variables}
\label{sec:mul}
The multiplicity variables to be studied are defined as follows. 

The local multiplicity, $N_{\rm tracks}^{\rm PV}$,  is the number of VELO tracks considered for a PV reconstruction, which is a good estimator for prompt charged-particle multiplicity  for events with one PV~\cite{Aaij:2014zzy, Kucharczyk:1756296}. The forward (backward) multiplicity is represented by $N_{\rm fwd}^{\rm PV}$ ($N_{\rm bwd}^{\rm PV}$), corresponding to the number of tracks in the forward (backward) pseudorapidity region of $1.5<\eta<5.2$ ($-5.2<\eta<-1.5$). The forward region roughly coincides with the LHCb acceptance. The two muon tracks are thus included in $N_{\rm tracks}^{\rm PV}$ and $N_{\rm fwd}^{\rm PV}$ but not in $N_{\rm bwd}^{\rm PV}$. The restriction on $z_{\rm PV}$ for different multiplicity variables is listed in Table~\ref{T1}, within which the VELO acceptance is roughly uniform for different multiplicity variables. 
\begin{table}[H]
\caption{Requirements on $z_{\rm PV}$ for different multiplicity variables.}
\begin{center}
\begin{tabular}{ll}
\hline
	Variable & $z_{\rm PV}$ range (mm)\\
\hline
         $N_{\rm tracks}^{\rm PV}$ & [--60, 180]\\
         $N_{\rm fwd}^{\rm PV}$ & [--180, 180] \\
         $N_{\rm bwd}^{\rm PV}$ & [--30, 180] \\
\hline
\end{tabular}
\end{center}
\label{T1}
\end{table}

The multiplicity variables are further nondimensionalised by dividing their corresponding mean values in no-bias data. The no-bias data are recorded with random trigger decisions, which makes its mean values of multiplicity good references. By nondimensionalising the multiplicity variables, the nearly negligible inefficiency can be further canceled. The mean values in no-bias data are summarized in Table~\ref{NoBiasMult}. The forward and backward multiplicity slightly don't sum up to the local multiplicity due to different constraints in the $z_{\rm PV}$ for different multiplicity variables. 
\begin{table}[H]
\caption{Mean values of the three multiplicity variables in no-bias data.}
\begin{center}
\begin{tabular}{ll}
\hline
Variable & mean value\\
\hline
	 $N_{\rm tracks}^{\rm PV}$ & 26.74 \\
	 $N_{\rm fwd}^{\rm PV}$  & 16.53 \\
	 $N_{\rm bwd}^{\rm PV}$  & 11.26 \\
\hline
\end{tabular}
\end{center}
\label{NoBiasMult}
\end{table}

\section{Determination of the cross-section ratio}
\label{Ratio of Cross-Section Determination}
\def\effTot{\ensuremath{\epsilon_{\mathrm{tot}}}\xspace}
\def\effTotJ{\ensuremath{\epsilon_{\mathrm{tot,\jpsi}}}\xspace}
\def\effTotP{\ensuremath{\epsilon_{\mathrm{tot,\psitwos}}}\xspace}
\def\effiTotJ{\ensuremath{\epsilon_{\mathrm{\jpsi},i}}\xspace}
\def\effiTotP{\ensuremath{\epsilon_{\mathrm{\psitwos},i}}\xspace}
The double-differential cross-section for prompt and non-prompt $\jpsi$ and \psitwos production in a given (\pt, $y$) bin and multiplicity range is defined as
\begin{equation}
    \frac{\deriv^2\sigma}{\deriv y\deriv \pt} 
	= \frac{N(\pt,y)}
           {\mathcal{L}\times\effTot(\pt,y)\times \BR \times\Delta y \times \Delta \pt}, 
  \label{CrossSecJ}
\end{equation} 
where $N(\pt,y)$ is the number of signal candidates extracted from a fit to the dimuon invariant mass, $\mathcal{L}$ is the integrated luminosity, \effTot is the total efficiency associated to the selection requirements, and $\Delta\pt$ and $\Delta y$ are the corresponding bin widths. The branching fractions $\BR(\jpsi\rightarrow\mu^+\mu^-)=(5.961\pm0.033)\%$ and $\BR(\psitwos\rightarrow e^+e^-)=(7.93\pm0.17)\times10^{-3}$ are used~\cite{Workman:2022ynf}. For the \psitwos mode, the dielectron decay, which has higher precision, is used under the assumption of lepton universality. The measurement is performed in the kinematic range $2.0<y<4.5$ and $0.3<\pt<20\gevc$, where the lower limit in \pt suppresses contributions from photon-induced production~\cite{Ma:2022rfl}.

The double-differential ratio of prompt or non-prompt production in a certain multiplicity range is further defined by
\begin{equation}
    \frac{\sigma_{\psitwos}(\pt,y)}{\sigma_{\jpsi}(\pt,y)} =
    \frac{N_{\psitwos}(\pt,y)}{N_{\jpsi}(\pt,y)} \times
    \frac{\effTotJ(\pt,y)} {\effTotP(\pt,y)} \times
    \frac{\mathcal{B}_{\jpsi \rightarrow \mu^+ \mu^-}}{\mathcal{B}_{\psitwos\rightarrow e^+e^-}}.
    \label{Rsingle}
\end{equation}
Therefore, the ratio of production over an integrated (\pt, $y$) region can be calculated as $\Sigma_{k}\sigma_{\psitwos,k} / \Sigma_{k}\sigma_{\jpsi,k}$, where $k$ denotes the bin index for kinematic bins $(\pt,y)$.

\begin{figure}[!tbp]
   \begin{center}
     \includegraphics[width=0.49\linewidth]{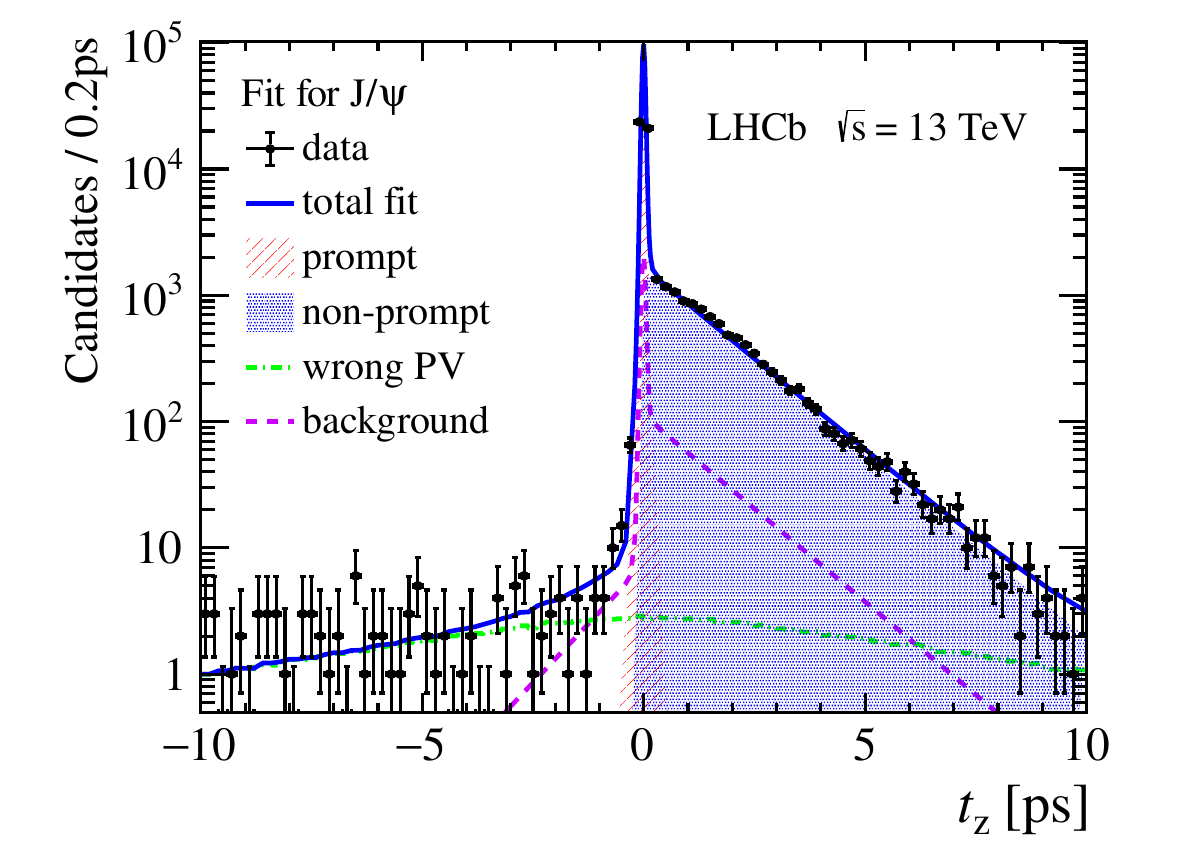}
     \includegraphics[width=0.49\linewidth]{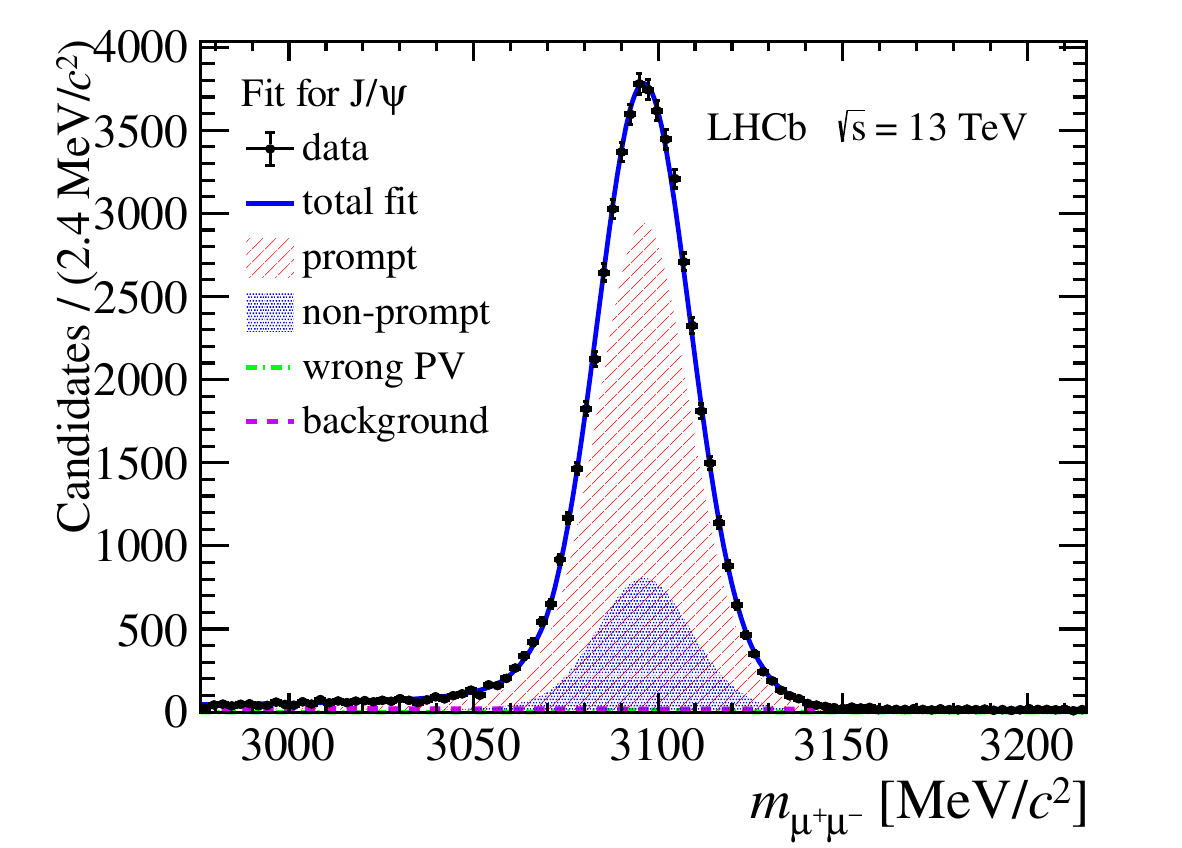}
     \includegraphics[width=0.49\linewidth]{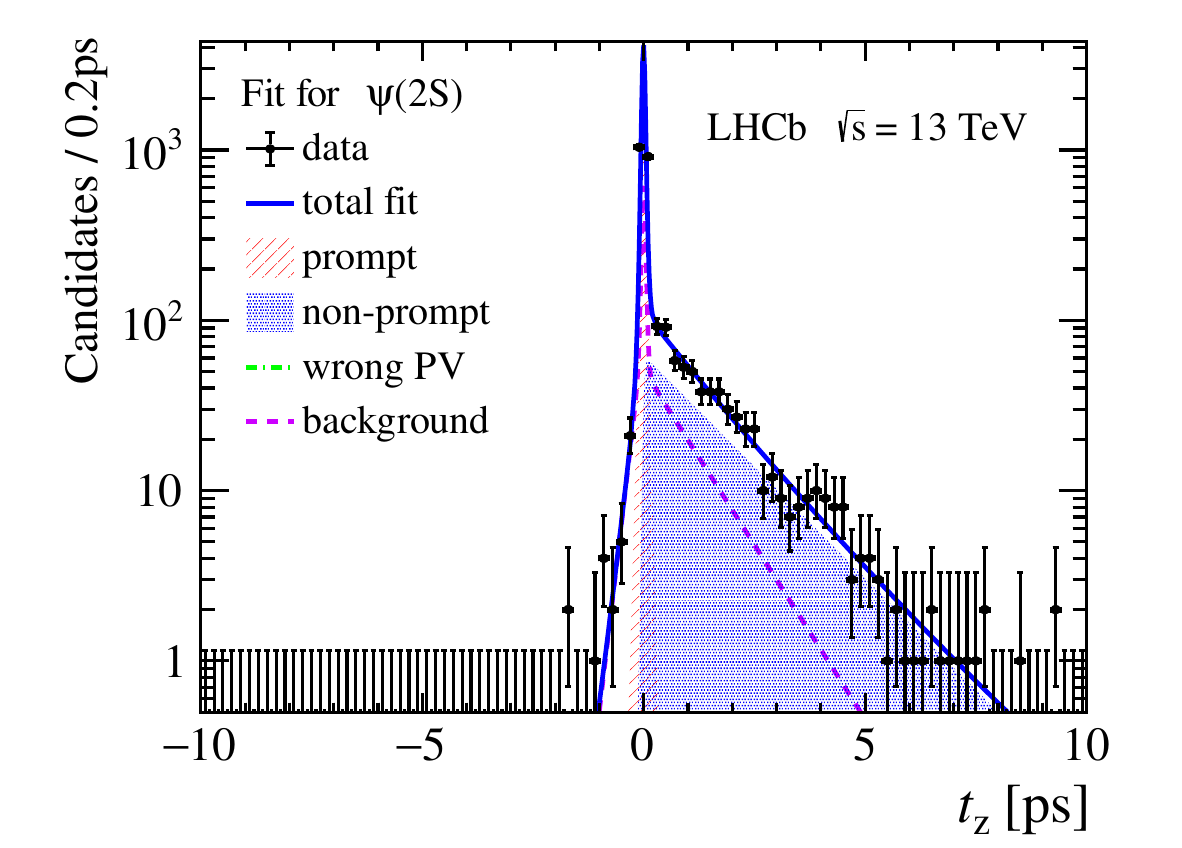}
     \includegraphics[width=0.49\linewidth]{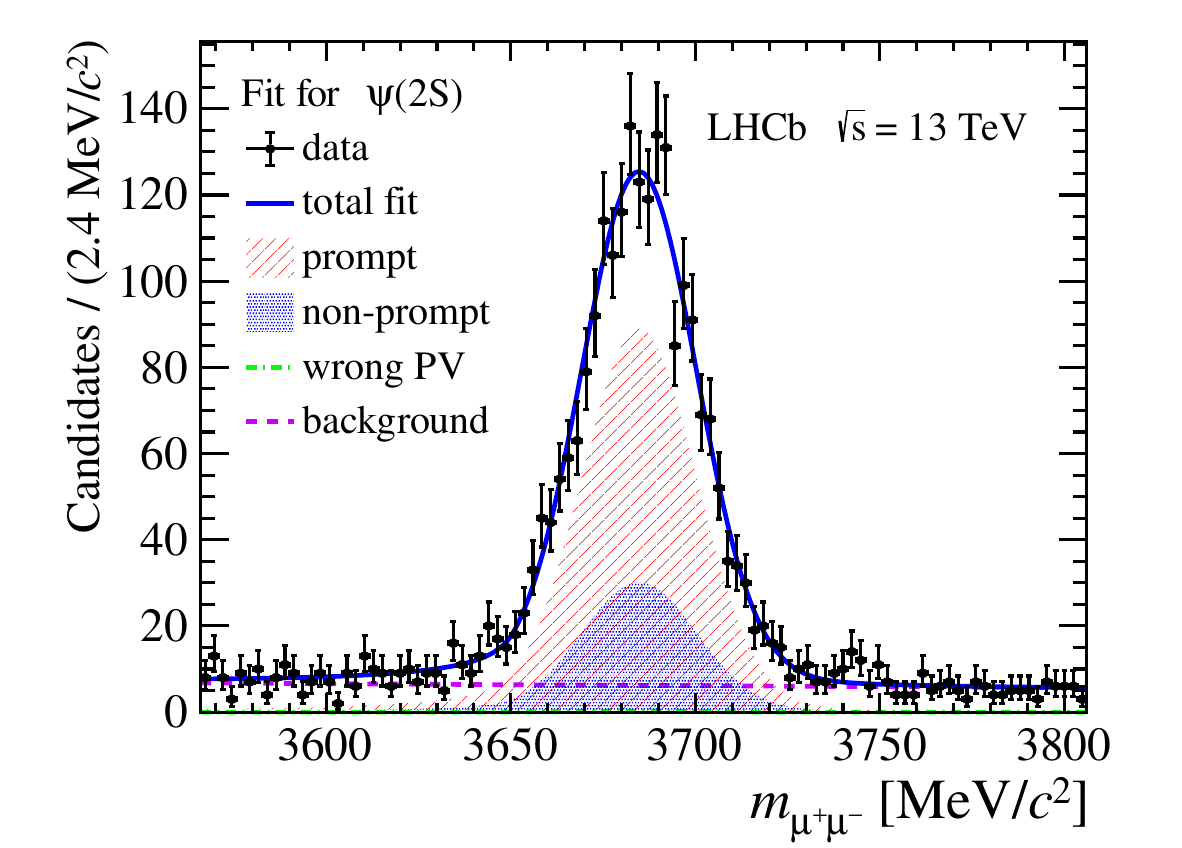}
   \end{center}
   \caption{
   Distributions of the (left) pseudo decay time and (right) mass for (upper) \jpsi and (lower) \psitwos are plotted in the region $4\leq N_{\rm tracks}^{\rm PV} < 20$, $2.8<y<3.5$, and $6<\pt<8\gevc$. The fit results are overlaid.
     }
   \label{fig:2Dtz}
 \end{figure}

The yields of prompt and non-prompt charmonia are extracted by performing a two-dimensional extended unbinned maximum-likelihood fit to the distributions of invariant mass $m_{\mu^+\mu^-}$ and pseudo decay time $t_z$ of the candidates. The mass spectrum of signal is described by the sum of two Crystal Ball functions~\cite{Skwarnicki:1986xj}, which share a common mean but have independent width parameters. The relation between the widths and tail parameters are determined from simulation. The background in the mass spectrum is described by an exponential function whose decay parameter is allowed to vary. The pseudo decay time distribution of prompt and non-prompt charmonia are described by a Dirac delta function and an exponential function, respectively. Both are convolved with a common double-Gaussian function, which accounts for the detector resolution. The background $t_z$ distribution is parameterised with an empirical function consisting of three (two) exponential functions for positive (negative) $t_z$ direction, convolved with a double-Gaussian function. The parameters for the background shapes are determined from fits to data in the mass sideband regions, corresponding to candidates with masses at least $50\mev$ above or below the known charmonia masses, which are mostly populated by background. The background parameters are then fixed in the final two-dimensional fit. A charmonium candidate can also be associated to a wrong PV, resulting in a long tail component in the $t_z$ distribution. Even though the number of PVs is restricted to be exactly one, it is still possible that the true PV is not reconstructed and the candidate is associated to an incorrect PV. Its shape is modelled from data by calculating $t_z$ with respect to the PV of the next event. As an example, the projections in mass and pseudo decay time of two-dimensional fits to the \jpsi and \psitwos distributions are shown in Fig.~\ref{fig:2Dtz}.
The total efficiency \effTot is the product of the geometrical acceptance and the efficiencies of the particle reconstruction, event selection, muon identification and the trigger. The track reconstruction and muon identification efficiencies are determined using data-driven methods~\cite{Anderlini:2202412,LHCb-DP-2013-002}. The trigger efficiency is estimated on simulation and cross-checked using data~\cite{Tolk:2014llp}. The total efficiencies for prompt and non-prompt signals are calculated separately in each (\pt, $y$) and multiplicity bin. They are found to be very similar.
The efficiencies in different multiplicity regions are determined separately, according to the binning scheme in data samples.

\section{Systematic uncertainties}
\def\effTri{\ensuremath{\epsilon_{\mathrm{trigger}}}\xspace}
\def\effTotJ{\ensuremath{\epsilon_{\mathrm{tot,\jpsi}}}\xspace}
\def\effTotP{\ensuremath{\epsilon_{\mathrm{tot,\psitwos}}}\xspace}
\def\pandb{prompt components and components from $b$-hadron decay}
Systematic uncertainties arising from a variety of sources are studied. The systematic uncertainty ranges for the ratios in different $N_{\rm tracks}^{\rm PV}$ bins are summarized in Table~\ref{SysU}. The binning scheme for multiplicity and (\pt, $y$) ensures that in each individual bin, analysis can be performed with enough events.

To study the uncertainties related to the mass fits, the signal invariant mass distribution in each bin is also fitted with a kernel-estimated~\cite{Cranmer:2000du} distribution extracted from simulated samples. In order to account for the difference in resolution between data and simulation, a Gaussian function is used to smear the shape of the signals. The study is performed in each kinematic and multiplicity bin, and the relative difference between the ratio obtained from the default fit model and from the alternative fit model is quoted as a systematic uncertainty for both the prompt and non-prompt ratios.

To estimate the uncertainty due to imperfection of the $t_z$ signal model, it is fitted to the simulated sample, where the true numbers of prompt and non-prompt candidates are known. Then the fit yield is compared to the true numbers and the variation is quoted as a systematic uncertainty. For the $t_z$ background model, the sideband sample which the model is originally fitted to, is replaced by the distribution extracted from the data using the $sPlot$~\cite{Pivk:2004ty} method. The invariant mass and pseudo decay time have a correlation factor of $10^{-3}$, so the invariant mass may be used as a discriminating variable to extract the signal and background distribution of pseudo decay time $t_z$ with the $sPlot$~\cite{Pivk:2004ty} method. The difference between the alternative and default fits is taken as a systematic uncertainty.

The tracking efficiencies are determined from the simulation. Corrections, as a function of $p$ and $\eta$, are determined using \jpsi control samples in data~\cite{DeCian:1402577}. The uncertainty due to the limited size of the control samples are propagated to the results using pseudoexperiments. The multiplicity distribution of simulated samples is weighted to match that of the control samples, and the bias caused by the choice of the multiplicity variable is canceled in the \psitwos-to-\jpsi cross-section production ratio.

The muon identification efficiency is obtained using simulated samples and calibrated with data samples of $\jpsi\rightarrow \mu^+\mu^-$ decays using the tag-and-probe method~\cite{Archilli:2013npa}.  The single-muon identification efficiency is calculated as a function of $p$, $\eta$ and the number of hits in the SPD, $N_{\rm spd}$. The uncertainty due to the limited size of control samples is propagated to the final result. The uncertainty related to the binning scheme is studied by changing the binning scheme of ($p$,$\eta$,$N_{\rm spd}$), and the variation on the ratio of production is quoted as a systematic uncertainty.

The uncertainty due to limited size of the simulated sample used to determine the efficiencies is propagated to the final result. The trigger efficiency is estimated from simulation and is cross-checked by a data-driven method using a fully reconstructed sample. The trigger efficiency in data is determined using a subset of events that would have triggered independently of the \jpsi and \psitwos signal candidates~\cite{LHCb-DP-2012-004}. The relative difference between data and simulation in different multiplicity bins is taken as a systematic uncertainty.
The uncertainties in the charmonium branching ratios and luminosity cancel in the normalised ratios, defined as the ratios in different multiplicity regions divided by the total ratio over all multiplicity regions.
\begin{table}[!tbp]
  \centering
	\caption{Summary of systematic uncertainty ranges for the double-differential \psitwos-to-\jpsi cross-section ratio, in different kinematic bins within corresponding multiplicity intervals in $\%$. Sources marked with $\dagger$ are considered to be correlated beween bins.}
\begin{center}
  \begin{tabular}{l|ccccc}
    \hline
    & \multicolumn{5}{c}{Interval of $N_{\rm tracks}^{\rm PV}$}\\
    Source &  4--20 & 20--45  & 45--70   & 70--95   & 95--200   \\ \hline
    Signal mass shape$\dagger$ & & & & &  \\
    prompt \& non-prompt & 0.1--1.6 & 0.1--1.3 & 0.1--1.1 & 0.1--1.8 & 0.1--3.2 \\ \hline
    Signal $t_z$ shape$\dagger$  & & & & &  \\
    prompt & 0.2--1.8 & 0.2--1.2 & 0.0--1.4 & 0.0--1.8 & 0.1--9.7 \\ 
    non-prompt & 0.0--1.8 & 0.2--2.0 & 0.0--2.3 & 0.1--1.7 & 0.0--4.0 \\ \hline
    Background $t_z$ shape$\dagger$  & & & & &  \\
    prompt & 0.0--1.6 & 0.2--3.1 & 0.0--2.2 & 0.2--6.2 & 0.4--6.4 \\
    non-prompt & 1.2--6.5 & 0.2--5.1 & 0.2--3.4 & 0.6--3.7 & 0.0--9.7 \\ \hline
    Tracking efficiency$\dagger$ & & & & &  \\
    prompt & 0.0--0.8 & 0.0--0.7 & 0.0--0.7 & 0.0--0.7 & 0.0--0.8 \\
    non-prompt & 0.0--1.0 & 0.0--1.1 & 0.0--1.2 & 0.0--1.1 & 0.0--1.1 \\ \hline
    Trigger efficiency$\dagger$ & & & & &  \\
    prompt \& non-prompt & 2.1 & 4.0 & 1.0 & 0.0 & 1.1 \\ \hline
    Muon ID sample size$\dagger$ & & & & &  \\
    prompt & 0.0--0.1 & 0.0--0.1 & 0.0--0.1 & 0.0--0.1 & 0.0--0.1 \\
    non-prompt & 0.0--0.1 & 0.0--0.1 & 0.0--0.1 & 0.0--0.1 & 0.0--0.1 \\
    Muon ID binning$\dagger$ & & & & &  \\
    prompt & 0.1--0.8 & 0.1--0.8 & 0.1--0.9 & 0.1--0.7 & 0.1--0.9 \\
    non-prompt & 0.0--0.9 & 0.0--0.8 & 0.1--0.7 & 0.1--0.8 & 0.1--1.0 \\ \hline
    Simulated sample size & & & & &  \\
    prompt & 1.5--4.4 & 1.1--2.4 & 1.3--2.5 & 2.0--3.4 & 4.0--6.4 \\
    non-prompt & 2.7--5.5 & 1.6--3.2 & 1.6--3.4 & 2.3--4.7 & 4.1--8.9 \\ \hline
    Total & & & & &  \\
    prompt & 3.1--7.7 & 4.3--5.5 & 2.2--3.9 & 2.9--7.7 & 6.1--13 \\
    non-prompt &5.2--10.2 & 5.0--7.6 & 3.3--5.3 & 3.9--8.4 & 7.4--17\\ \hline
\end{tabular}
\end{center}
\label{SysU}
\end{table}

The \psitwos-to-\jpsi cross-section ratios are reported with integration separately over \pt and $y$, and over the full \pt and $y$ range. The systematic uncertainties that are correlated across different bins may cancel each other when adding the production over different kinematic bins, since in some bins the production cross-section may be over-estimated and in other bins it may be under-estimated. The systematic uncertainties for ratios over integrated kinematic bins are obtained in the same way as for each bins. The uncertainties that are not correlated across different bins are propagated to the final result.

\section{Results}
\label{sec:results}

\subsection{\boldmath{\psitwos-to-\jpsi} ratio as a function of multiplicity}
\begin{figure}[H]
    \begin{center}
      \includegraphics[width=0.6\linewidth]{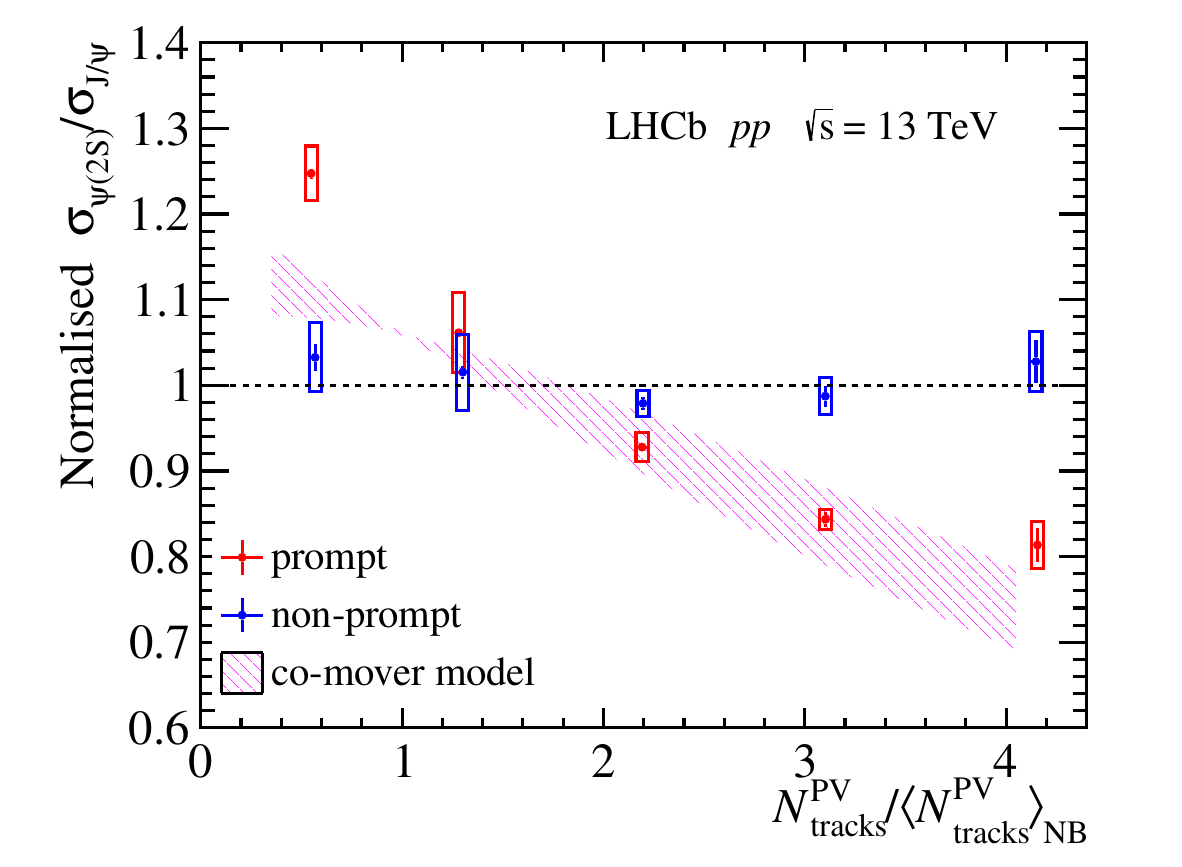}
    \end{center}
	\caption{Normalised production ratio as a function of $N_{\rm tracks}^{\rm PV}/\langle N_{\rm tracks}^{\rm PV}\rangle_{\rm{NB}}$, integrated over the full \pt-$y$ range of $2.0<y<4.5$ and $0.3<\pt<20$~\gevc with $-60 < z_{\rm PV} < 180$~mm. The error bars represent statistical uncertainties and heights of the boxes represent systematical uncertainties. 
      }
    \label{Ratio}
\end{figure}

The measured \psitwos-to-\jpsi production cross-section ratios are presented in various kinematic and multiplicity bins, with the full numerical data details provided in Appendix A~\ref{Sec:Ratio}. 
Subsequently, the normalised ratio as a function of the nondimensionalised multiplicity can be obtained by integrating the double-differential results over the \pt and $y$ bins. The \psitwos-to-\jpsi ratio in each multiplicity bin is normalised to the ratio of total cross-section of \jpsi and \psitwos mesons as
\begin{equation}
\mathrm{Normalised \ } \sigma_{\psitwos,n}/\sigma_{\jpsi,n}=\frac{\sigma_{\psitwos,n}/\sigma_{\jpsi,n}}{\sum_n\sigma_{\psitwos,n}/\sum_n\sigma_{\jpsi,n}},
\label{norm}
\end{equation}
where $n$ is the bin index for multiplicity.
The normalised ratio for the integrated production over \pt and $y$ as a function of the nondimensionalised $N_{\rm tracks}^{\rm PV}$ is illustrated in Fig.~\ref{Ratio}. The multiplicity distribution for \jpsi and \psitwos are quite similar to each other, and the $x$-values are the weighted average of (non-)prompt \jpsi and \psitwos multiplicity in each bin, where the weight is the inverse variation of the multiplicity distribution in corresponding bin. The uncertainty for the $x$-values is nearly negligible. For the sake of visibility, a small width is applied when drawing the systematic uncertainty.

The production ratio from $b$-hadron decays shows little dependence on multiplicity irrespective of the choice of multiplicity variable. For prompt \psitwos and \jpsi,  an evident decrease of ratio with $N_{\rm tracks}^{\rm PV}$ is observed (P-value of slope $<0.0005$), which aligns well with the predictions of the co-mover model, except in the low multiplicity region.

\begin{figure}[H]
  \begin{center}
    \includegraphics[width=0.49\linewidth]{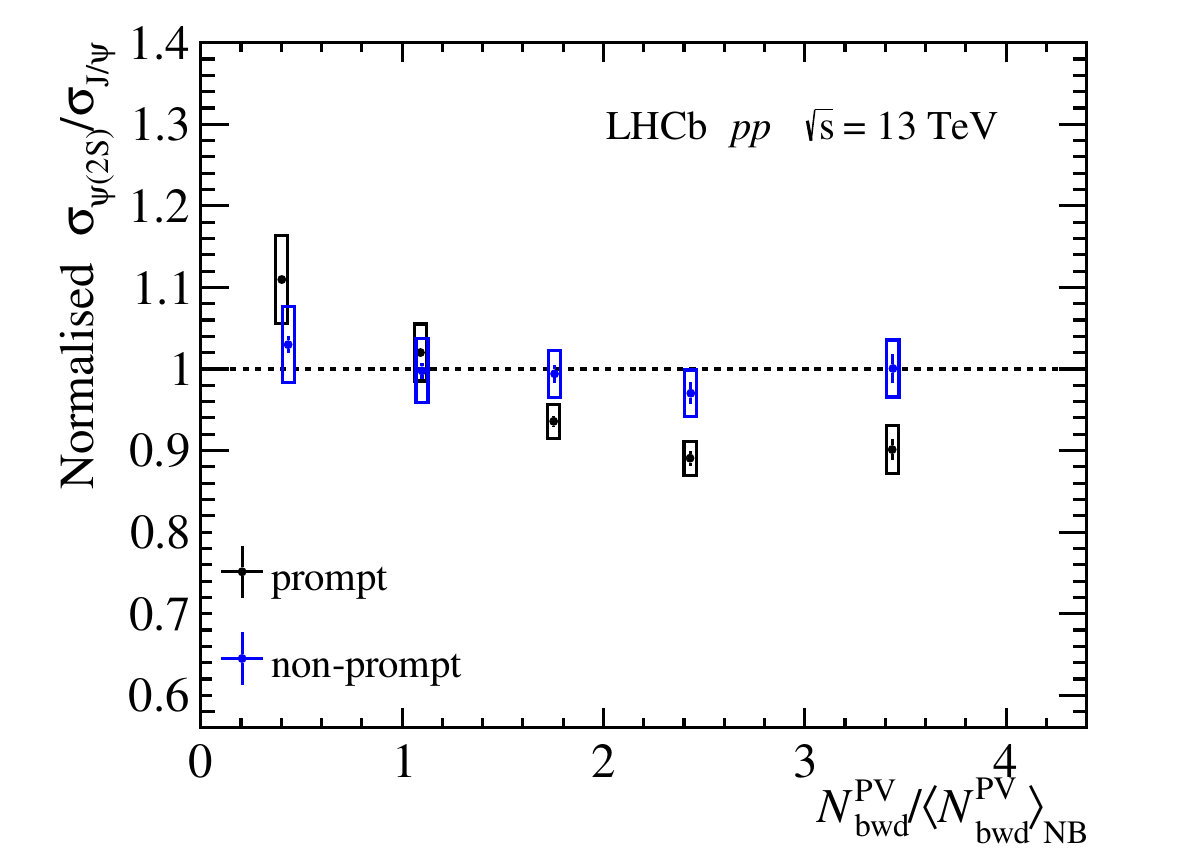}
    \includegraphics[width=0.49\linewidth]{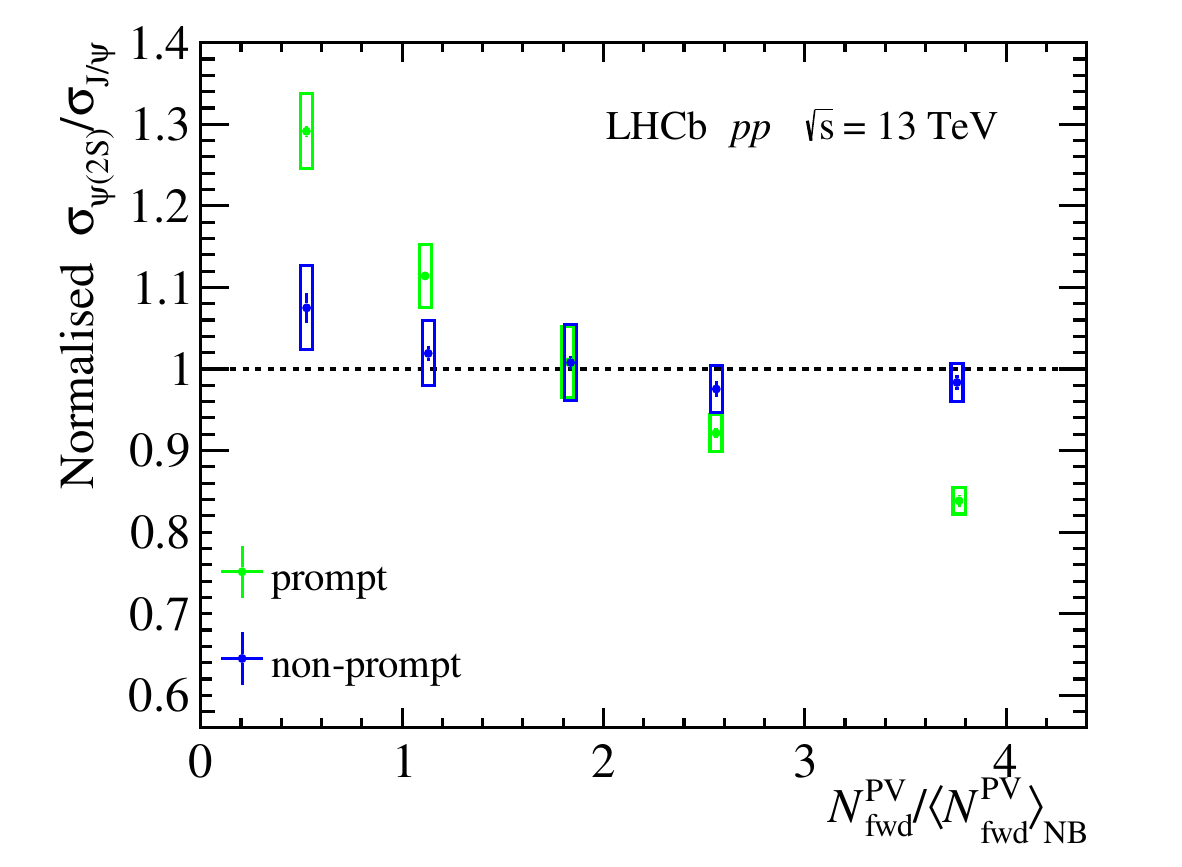}
    \vspace*{-0.5cm}
  \end{center}
	\caption{Normalised production ratio as a function of (left) $N_{\rm bwd}^{\rm PV}/\langle N_{\rm bwd}^{\rm PV}\rangle_{\rm{NB}}$ , with $-30<z_{\rm PV}<180$~mm, and (right) $N_{\rm fwd}^{\rm PV}/\langle N_{\rm fwd}^{\rm PV}\rangle_{\rm{NB}}$, with $-180<z_{\rm PV}<180$~mm, integrated over the full \pt-$y$ range of $2.0<y<4.5$ and $0.3<\pt<20$~\gevc.
    }
  \label{RatioFB}
\end{figure}

\begin{figure}[H] 
    \begin{center}
      \includegraphics[width=0.6\linewidth]{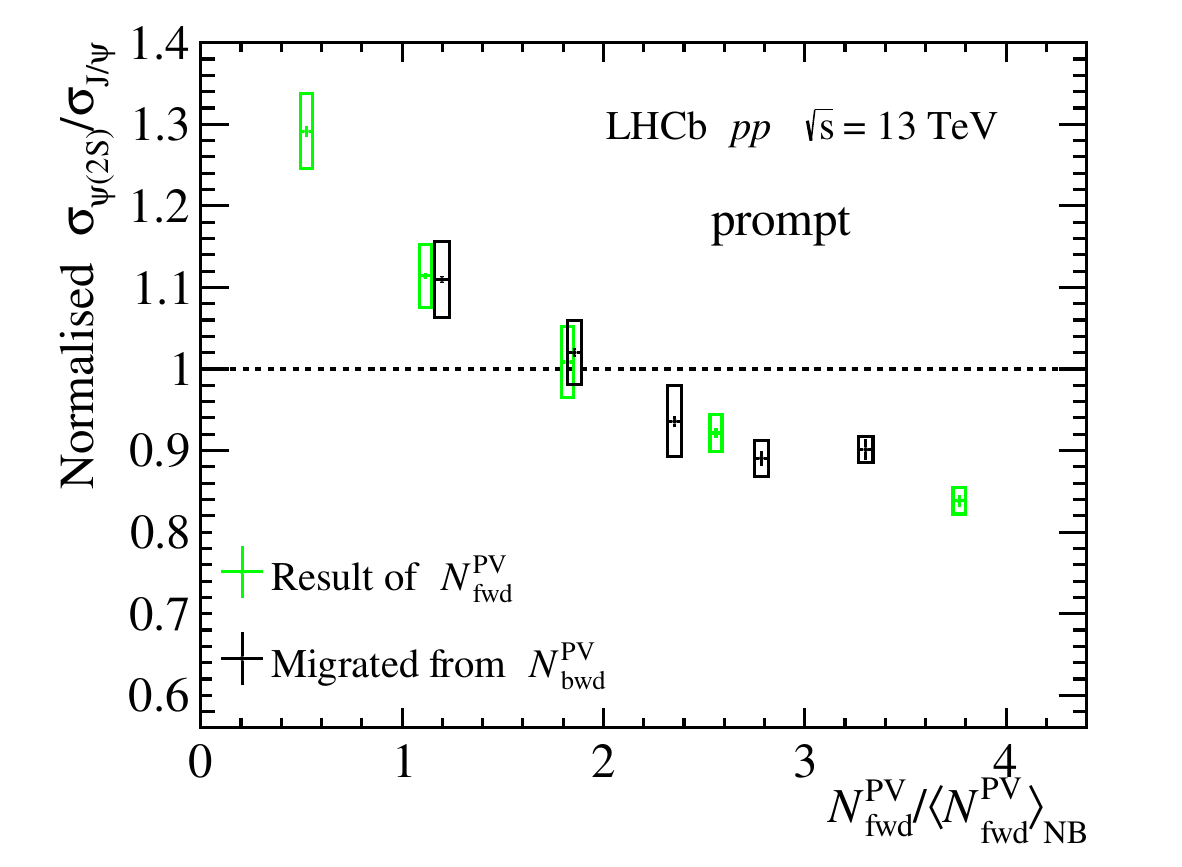}
    \end{center}
	\caption{Normalised production ratio as a function of $N_{\rm fwd}^{\rm PV}/\langle N_{\rm fwd}^{\rm PV}\rangle_{\rm{NB}}$, compared to a translation of the ratio as a function of $N_{\rm bwd}^{\rm PV}/\langle N_{\rm bwd}^{\rm PV}\rangle_{\rm{NB}}$.
      } 
    \label{BtoF}
\end{figure}

Additionally, the ratios are examined as a function of the nondimensionalised $N_{\rm bwd}^{\rm PV}$ and $N_{\rm fwd}^{\rm PV}$ in Fig.~\ref{RatioFB}. 
Notably, the trend is weaker as a function of $N_{\rm bwd}^{\rm PV}$ than $N_{\rm fwd}^{\rm PV}$ and  $N_{\rm tracks}^{\rm PV}$. This suggests a correlation between the relative suppression and local particle multiplicity. According to comover effect~\cite{Ferreiro:2012rq}, the ratio should remain constant across different $N_{\rm bwd}^{\rm PV}$, with $N_{\rm bwd}^{\rm PV}$ being measured in a backward rapidity range and not including the muon tracks from charmonia. The slight decreasing trend of ratio as function of $N_{\rm bwd}^{\rm PV}$ could result from the correlation between $N_{\rm bwd}^{\rm PV}$ and $N_{\rm fwd}^{\rm PV}$, where the correlation factor is 0.54 and 0.51 for prompt \jpsi and \psitwos, respectively. To estimate the impact of this correlation, the mean $N_{\rm fwd}^{\rm PV}$ value within each $N_{\rm bwd}^{\rm PV}$ bin for prompt charmonia is calculated. Then, the normalised ratios across different $N_{\rm bwd}^{\rm PV}$ bins are plotted as a function of $N_{\rm fwd}^{\rm PV}$, as shown in Fig.~\ref{BtoF}. A coherent alignment in the decreasing trend is consistent with the hypothesis that the dependency of the normalised \psitwos-to-\jpsi ratio on $N_{\rm bwd}^{\rm PV}$ arises from $N_{\rm bwd}^{\rm PV}$-$N_{\rm fwd}^{\rm PV}$ correlation. 

\subsection{\boldmath{\psitwos-to-\jpsi} ratio in different \pt and $y$ ranges}
\begin{figure}[H]
  \begin{center}

	  \includegraphics[width=0.48\linewidth]{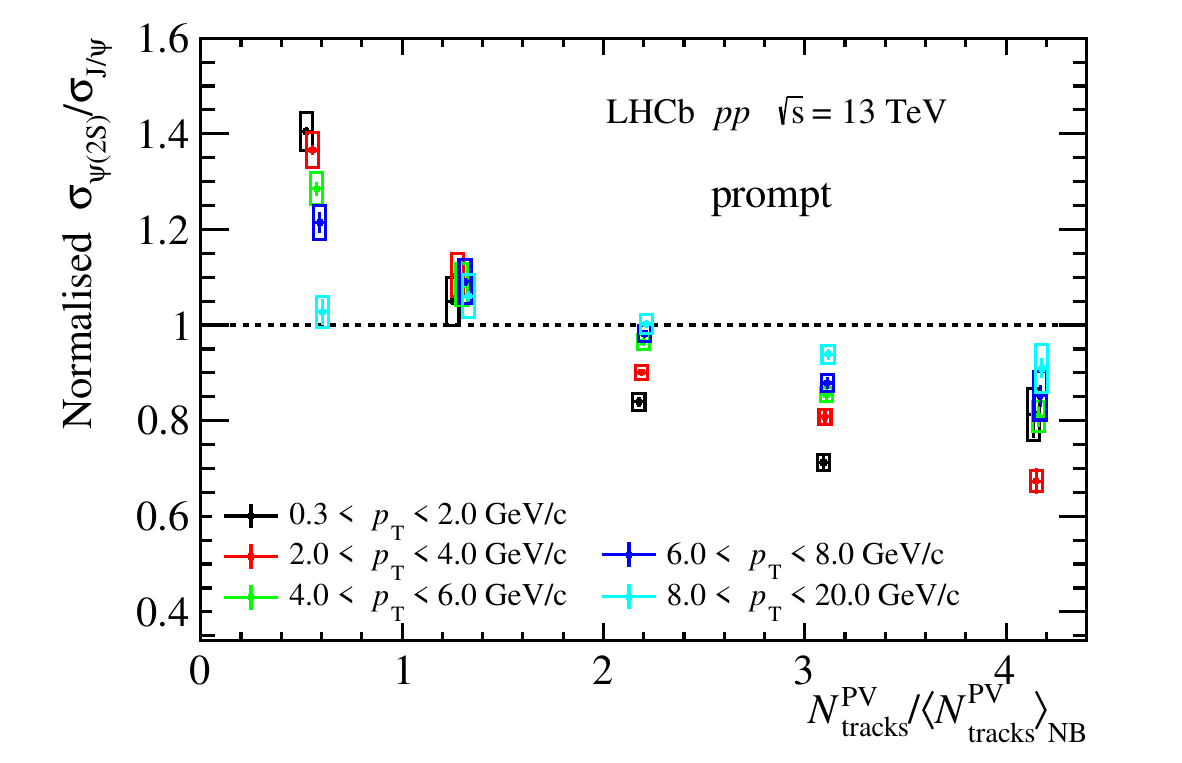}
          \includegraphics[width=0.48\linewidth]{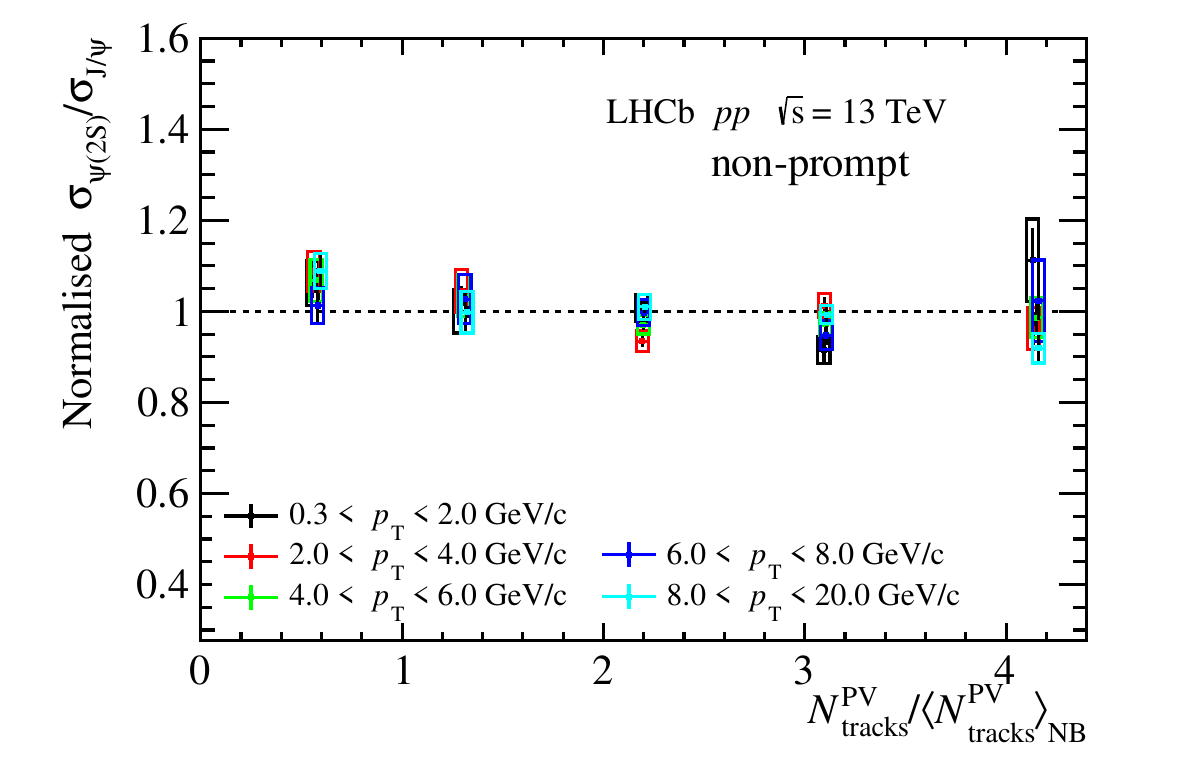}
	  \vspace*{-0.5cm}
  \end{center}
	\caption{Normalised production ratio for (left) prompt and(right) non-prompt charmonia as a function of $N_{\rm tracks}^{\rm PV}/\langle N_{\rm tracks}^{\rm PV}\rangle_{\rm{NB}}$ in different \pt intervals.}
  \label{RatioPT_PVN}
\end{figure}

\begin{figure}[H]
  \begin{center}
	  \vspace*{-0.5cm}
    \includegraphics[width=0.48\linewidth]{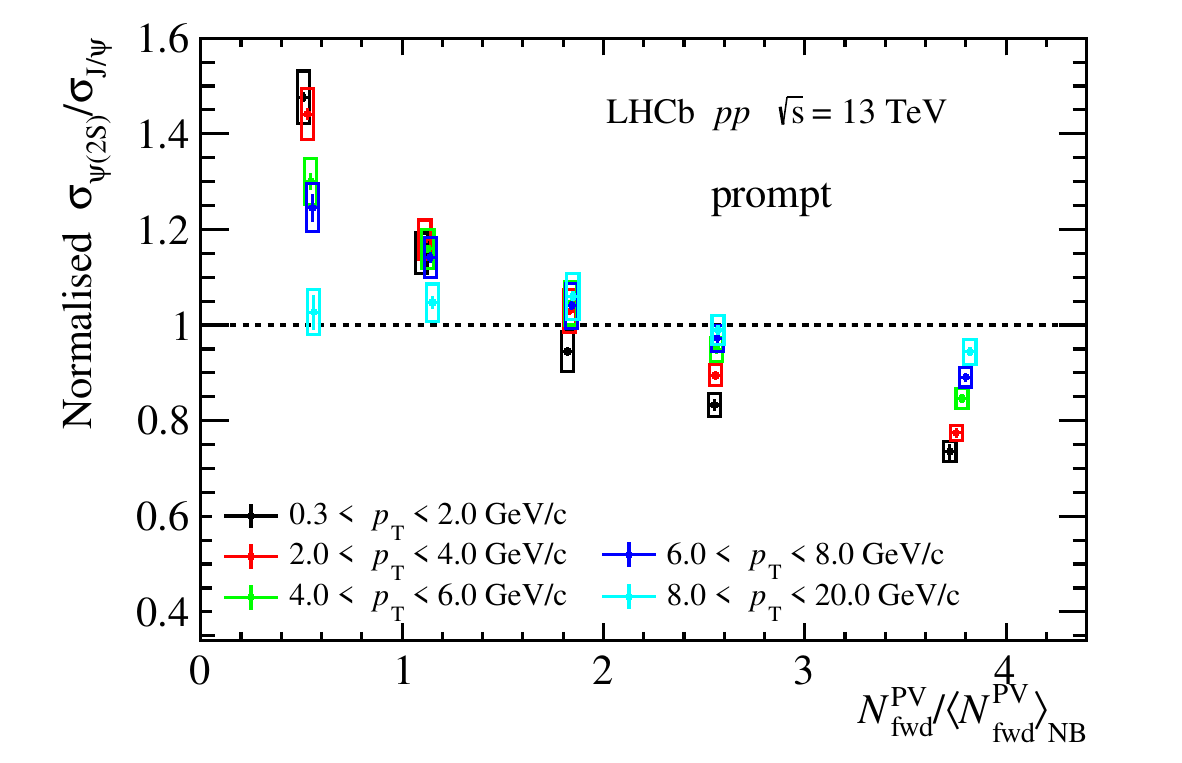}
    \includegraphics[width=0.48\linewidth]{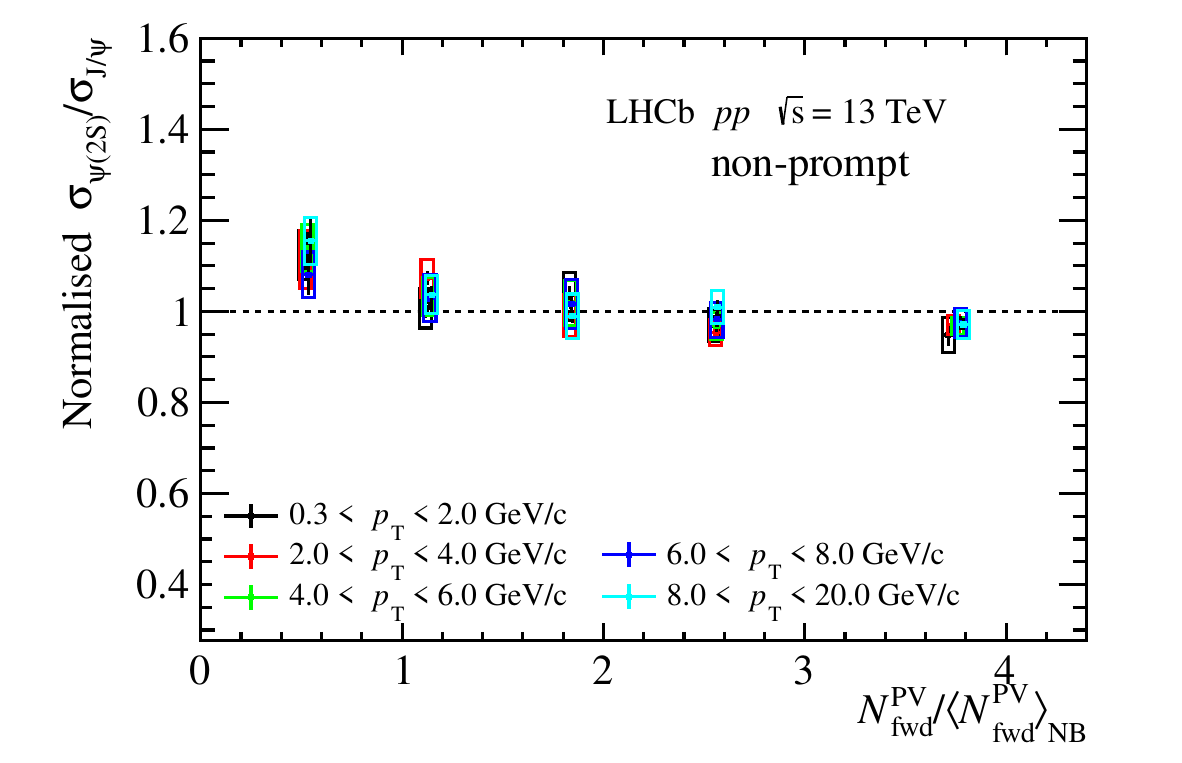}
	  \vspace*{-0.5cm}
  \end{center}
	\caption{Normalised production ratio for (left) prompt and(right) non-prompt charmonia as a function of $N_{\rm fwd}^{\rm PV}/\langle N_{\rm fwd}^{\rm PV}\rangle_{\rm{NB}}$ in different \pt intervals.}
  \label{RatioPT_For}
\end{figure}

\begin{figure}[H]
  \begin{center}
	  \vspace*{-0.5cm}
    \includegraphics[width=0.48\linewidth]{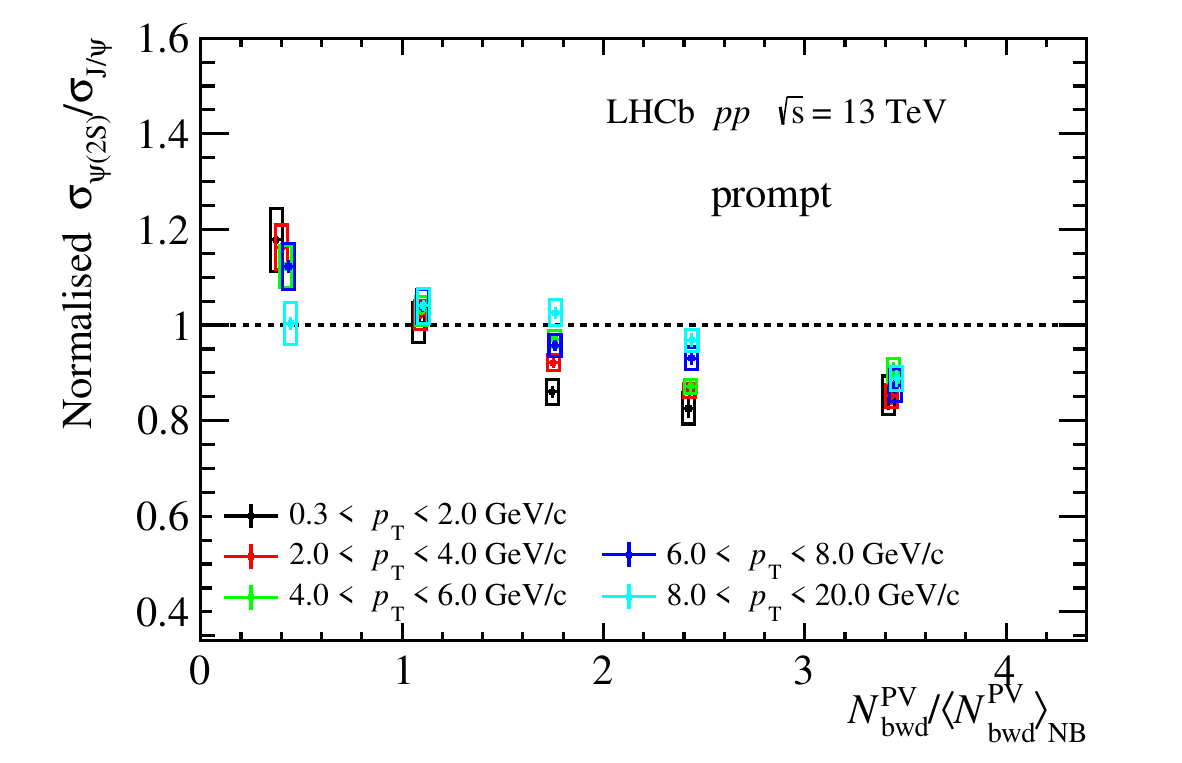}
    \includegraphics[width=0.48\linewidth]{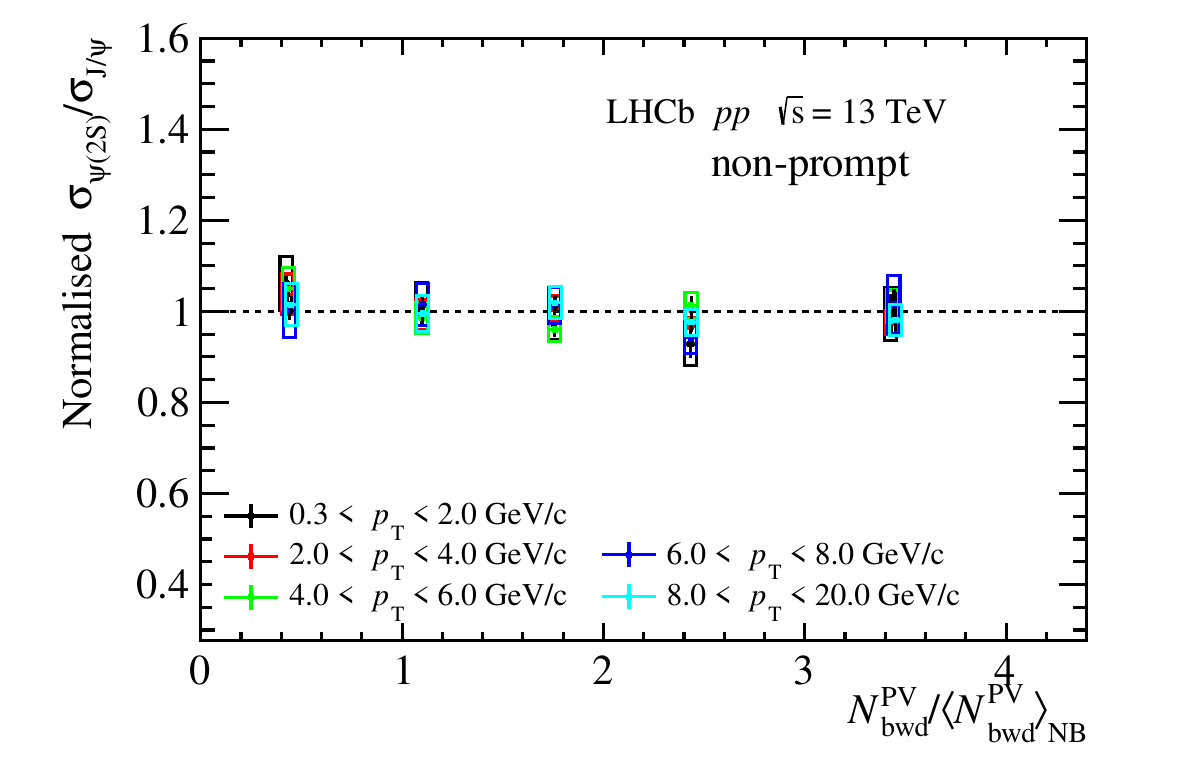}
	  \vspace*{-0.5cm}
  \end{center}
	\caption{Normalised production ratio for (left) prompt and(right) non-prompt charmonia as a function of $N_{\rm bwd}^{\rm PV}/\langle N_{\rm bwd}^{\rm PV}\rangle_{\rm{NB}}$ in different \pt intervals.}
  \label{RatioPT_Back}
\end{figure}

Figures~\ref{RatioPT_PVN},~\ref{RatioPT_For}, and~\ref{RatioPT_Back} present the normalised production ratios, as a function of the multiplicity, in different \pt ranges. The mean values for multiplicity variables in each bin are calculated in different kinematic regions accordingly, where their distributions are slight different. The relative production of \psitwos mesons is suppressed in the lower \pt region. However, in the high-\pt region, the prompt is almost independent of multiplicity as the non-prompt ratio. This is consistent with the measurement on multiplicity dependence of $\it{\Upsilon}(nS)$ by CMS~\cite{CMS:2020fae} and ATLAS~\cite{ATLAS-CONF-2022-023}, where the decreasing trend become slight in the high-\pt region.


\begin{figure}[H]
  \begin{center}

    \includegraphics[width=0.48\linewidth]{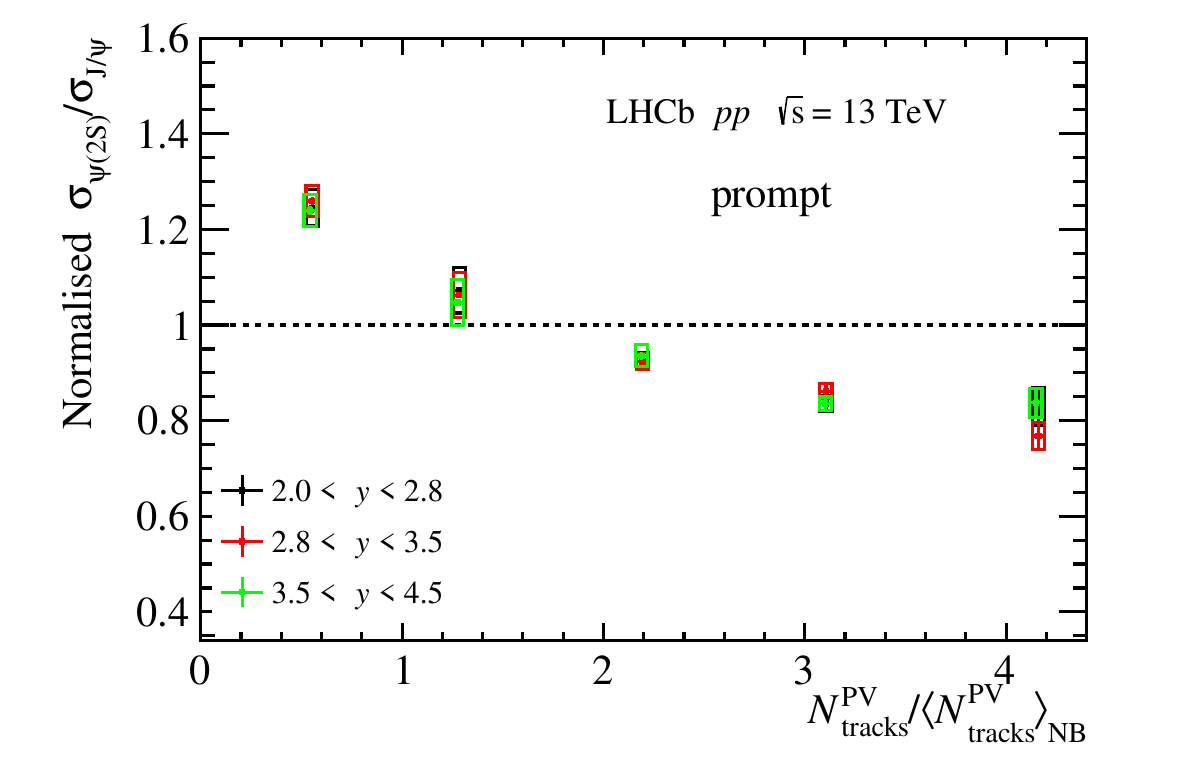}
    \includegraphics[width=0.48\linewidth]{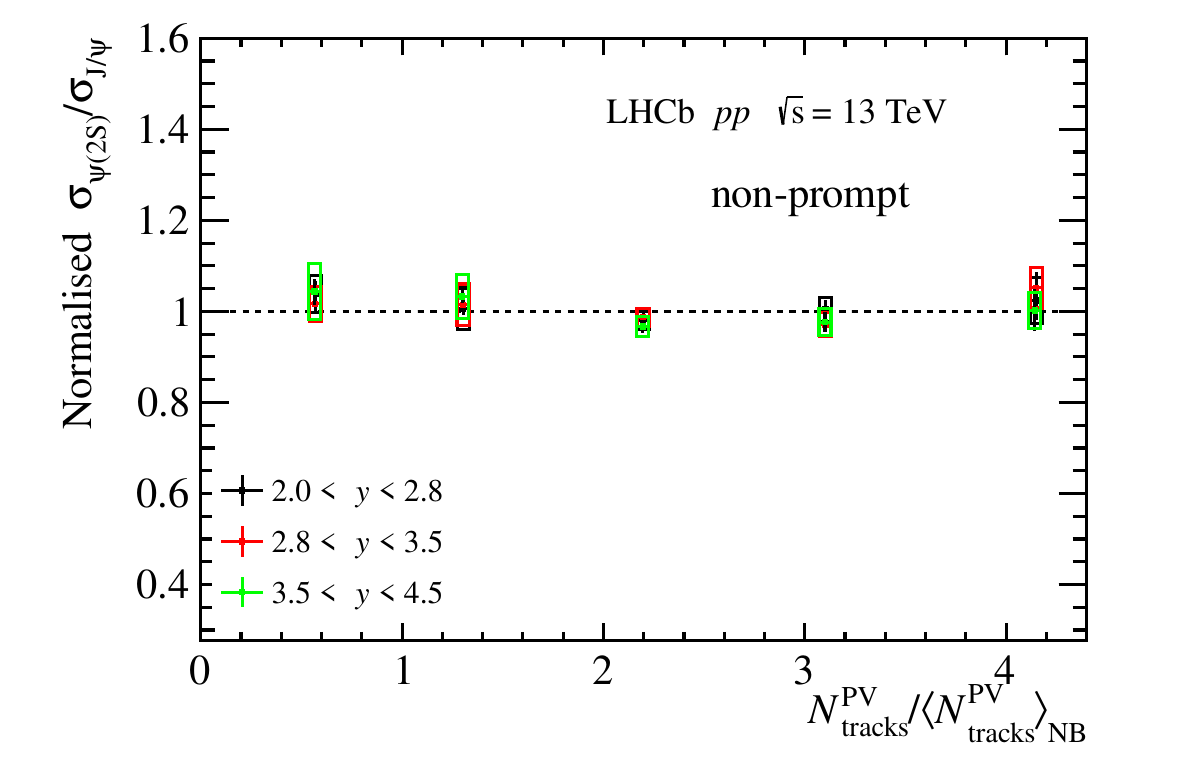}
	  \vspace*{-0.5cm}
  \end{center}
	\caption{Normalised production ratio for (left) prompt and(right) non-prompt charmonia as a function of $N_{\rm tracks}^{\rm PV}/\langle N_{\rm tracks}^{\rm PV}\rangle_{\rm{NB}}$ in different $y$ intervals.}
  \label{RatioY}
\end{figure}

\begin{figure}[H]
  \begin{center}
	  \vspace*{-0.5cm}
    \includegraphics[width=0.48\linewidth]{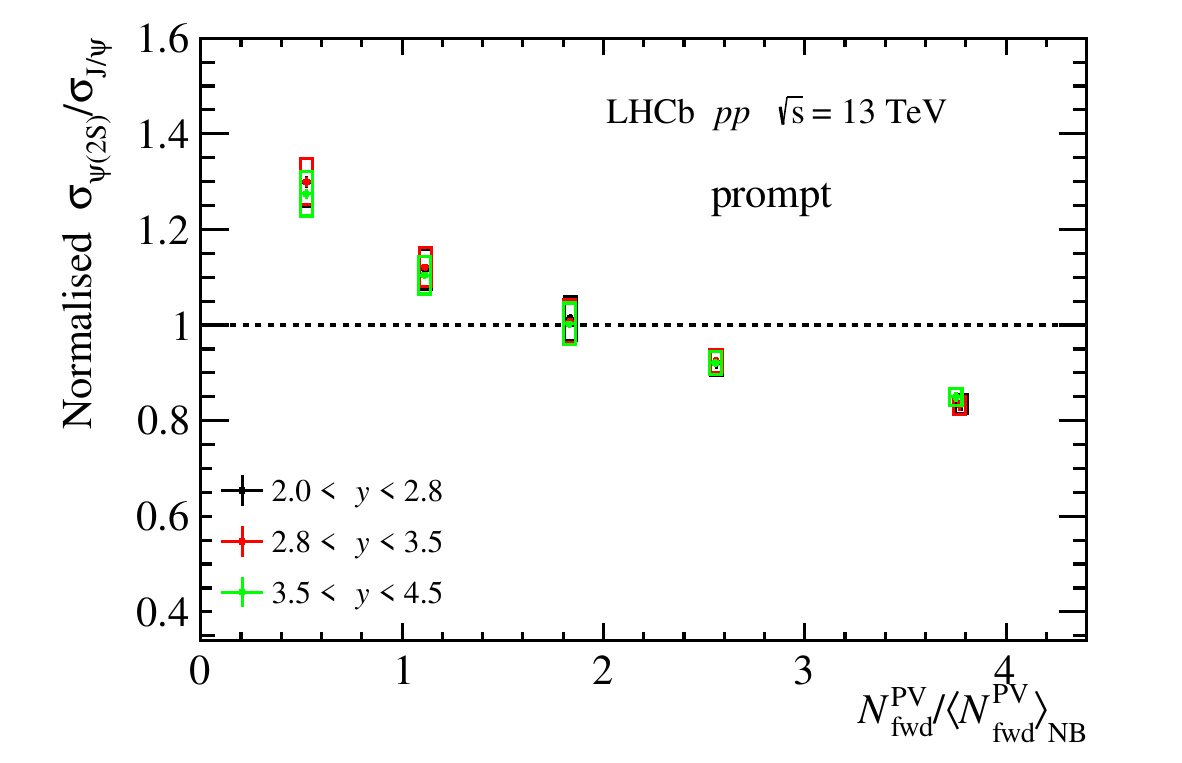}
    \includegraphics[width=0.48\linewidth]{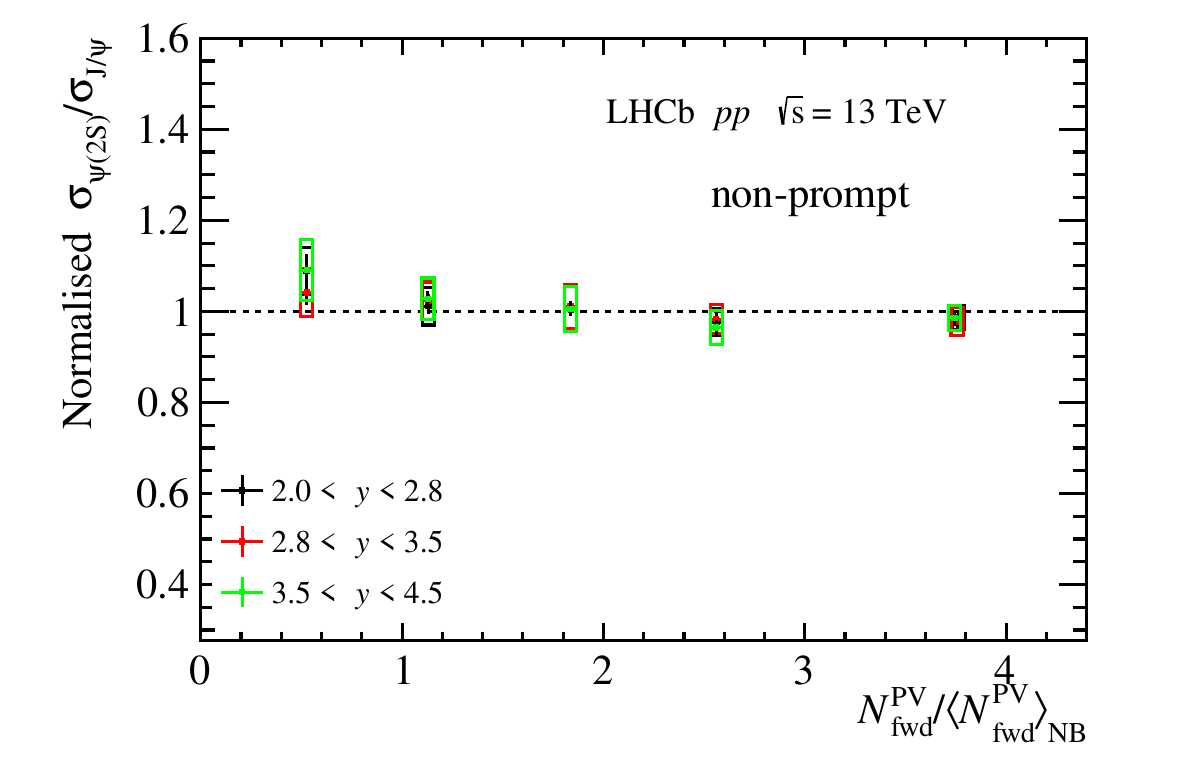}
	  \vspace*{-0.5cm}
  \end{center}
	\caption{Normalised production ratio for (left) prompt and(right) non-prompt charmonia as a function of $N_{\rm fwd}^{\rm PV}/\langle N_{\rm fwd}^{\rm PV}\rangle_{\rm{NB}}$ in different $y$ intervals.}
  \label{RatioY_For}
\end{figure}

\begin{figure}[H]
  \begin{center}
	  \vspace*{-0.5cm}
    \includegraphics[width=0.48\linewidth]{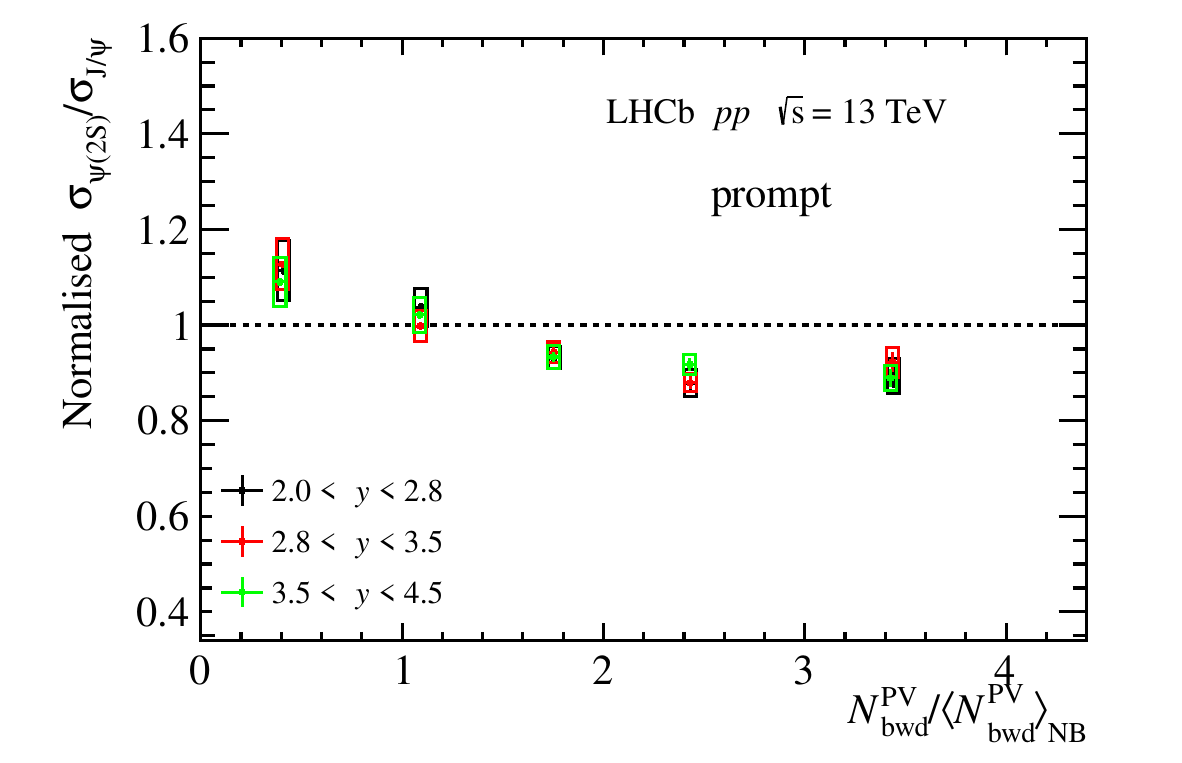}
    \includegraphics[width=0.48\linewidth]{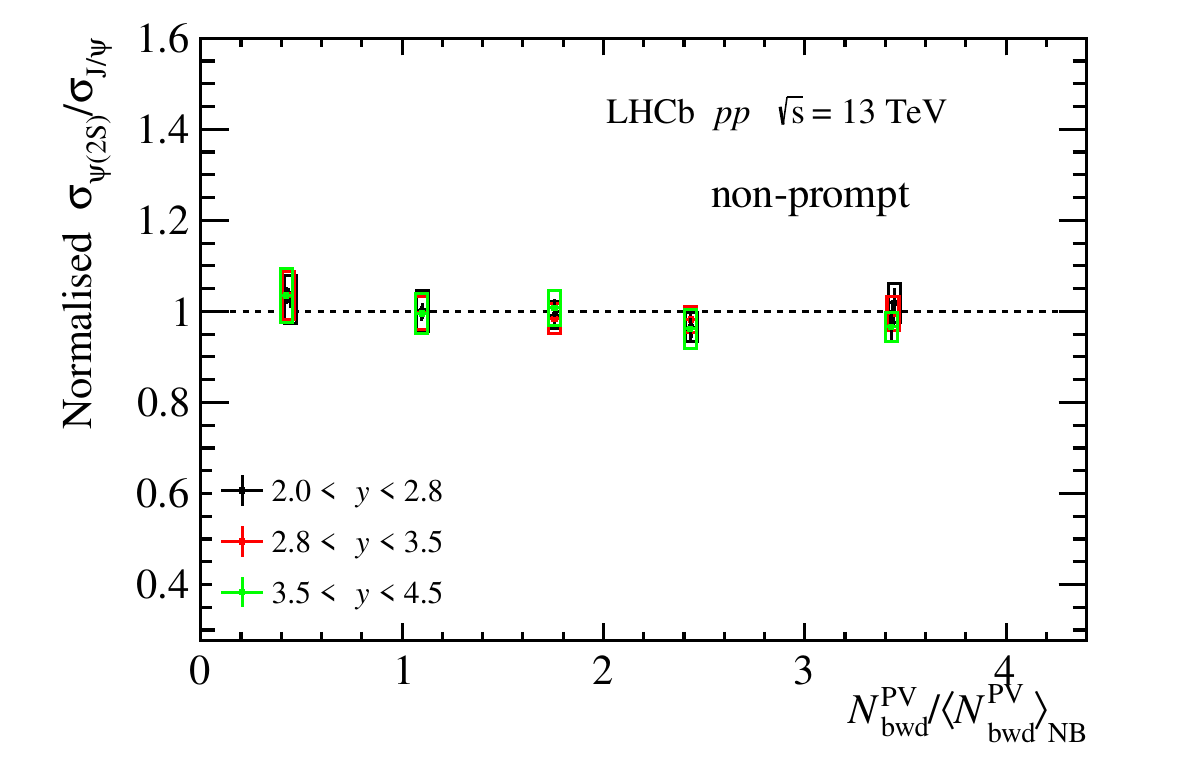}
	  \vspace*{-0.5cm}
  \end{center}
	\caption{Normalised production ratio for (left) prompt and(right) non-prompt charmonia as a function of $N_{\rm bwd}^{\rm PV}/\langle N_{\rm bwd}^{\rm PV}\rangle_{\rm{NB}}$ in different $y$ intervals.}
  \label{RatioY_Back}
\end{figure}

The results for the ratio in different rapidity bins are shown in Figs.~\ref{RatioY},~\ref{RatioY_For} and~\ref{RatioY_Back}. Notably, there is minimal difference across different rapidity regions, for both prompt and non-prompt signals.

\subsection{Comparisons with other measurements}
\label{Compare}
\begin{figure}[H]
  \begin{center}
  \includegraphics[width=0.7\linewidth]{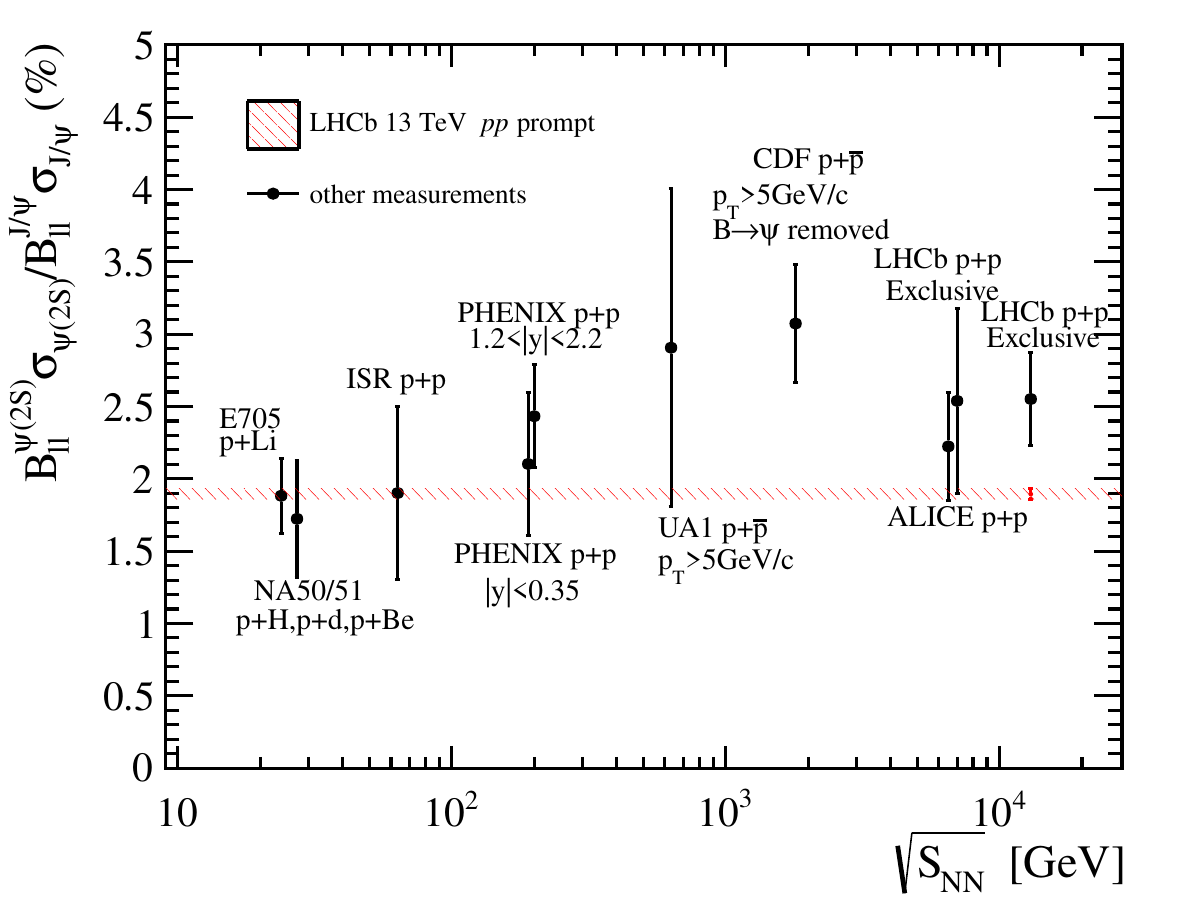 }
  \end{center}
  \caption{Production ratios compared to existing measurements~\cite{PHENIX:2016vmz}.}
\label{compare_total}
\end{figure}

\begin{figure}[H]
  \begin{center}
  \includegraphics[width=0.7\linewidth]{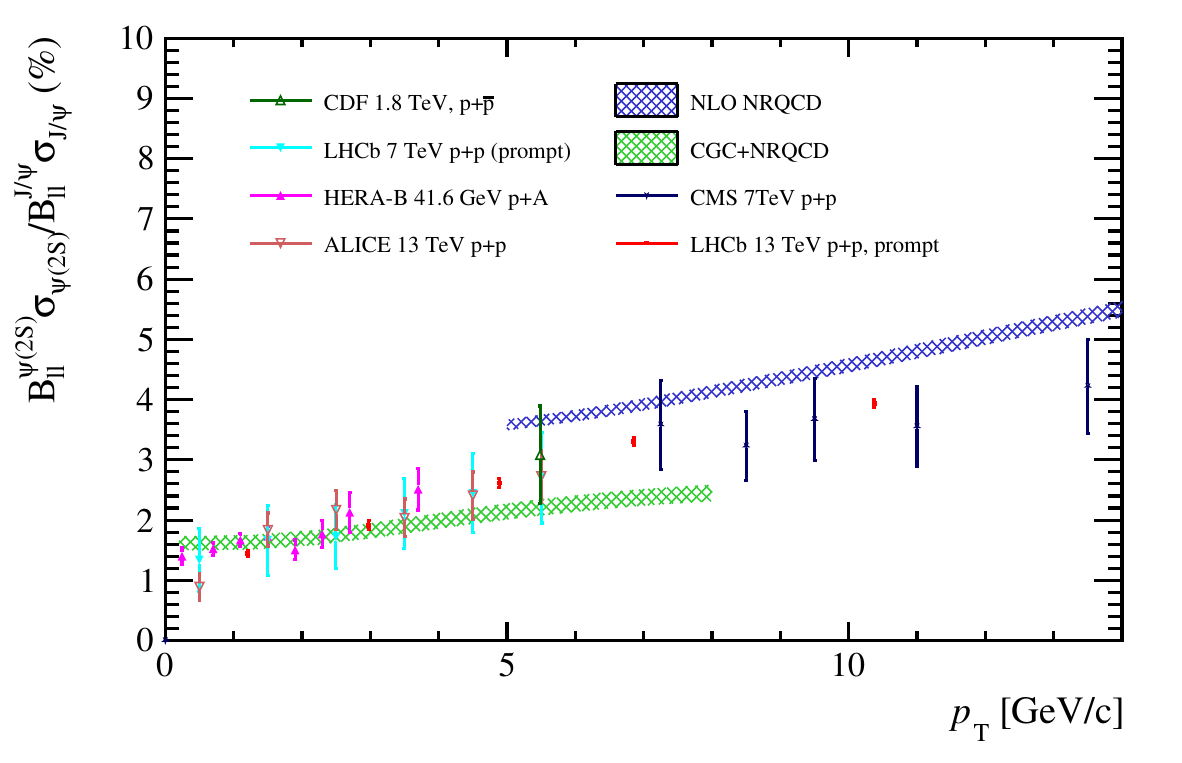}
  \end{center}
  \caption{Production ratios as a function of \pt compared to existing measurements~\cite{PHENIX:2016vmz} and model predictions~\cite{Ma:2014mri}.}
\label{compare_pt}
\end{figure}

Measurements of $\BR_{\psitwos}\times\sigma_{\psitwos}$ with $\BR_{\jpsi}\times\sigma_{\jpsi}$ have been carried out across diverse collision systems and energies by comparing the ratio of $\sigma_{\psitwos}/\sigma_{\jpsi}$ alongside their branching fractions and systematic uncertainties. Remarkably, the ratio is consistent regardless of the collision systems and energies. A comprehensive comparison is shown in Figs.~\ref{compare_total} and~\ref{compare_pt}, revealing strong agreement with other measurements~\cite{PHENIX:2016vmz,NA50:2006rdp,PHENIX:2011gyb,E705:1992vec,NA51:1998uun,Clark:1978mg,UA1:1990eni,CDF:1997ykw,LHCb:2013nqs,CMS:2011rxs,ALICE:2017leg}, which has been summarized in~\cite{PHENIX:2016vmz}. The value of this measurement is
\begin{equation}
		\frac{\BR_{\psitwos}\times\sigma_{\psitwos}}{\BR_{\jpsi}\times\sigma_{\jpsi}} = (1.894\pm0.039)\%.
\end{equation}


\cleardoublepage
\section{Conclusion}
\label{sec:conclusion}

This analysis presents a measurement of the the normalised \psitwos-to-\jpsi cross-section ratio, as a function of different charged-track multiplicity variables, in $pp$ collisions at $\sqrt{s}=13$~TeV. The data sample was collected by the \lhcb detector in 2016 and corresponds to an integrated luminosity of $658\pm 13$\invpb.

Notably, for non-prompt production the cross-section ratio exhibits no dependence on any of these multiplicity variables. In contrast, in prompt production the ratio decreases as a function of $N_{\rm tracks}^{\rm PV}$, which is a trend that aligns well with the predictions of the co-mover model. Furthermore, the ratio for prompt production has a similar trend as a function of $N_{\rm fwd}^{\rm PV}$, which can be attributed to the correlation between the multiplicity and the charmonium production, a phenomenon that can also be ascribed to the co-mover effect. The use of $N_{\rm bwd}^{\rm PV}$ serves to mitigate this correlation, leading to a small yet observable diminishing trend in relation to multiplicity. The small dependence on $N_{\rm bwd}^{\rm PV}$ could arise from the correlation between $N_{\rm bwd}^{\rm PV}$ and $N_{\rm fwd}^{\rm PV}$. The distinct behaviors exhibited by the prompt and non-prompt components underscore the significant influence of interactions with other particles within the collision event on charmonium production, even in $pp$ collisions. Furthermore, the variation of these ratios as a function of different multiplicity variables further solidifies this perspective. 

The multiplicity dependence of the ratios is also measured in distinct \pt and $y$ regions, uncovering an enhanced suppression pattern at low \pt, and an almost multiplicity independence in the high-\pt region for prompt \psitwos-to-\jpsi ratio, regardless of the choice of multiplicity variables. Finally, the results of the ratio $\BR_{\psitwos} \times \sigma_{\psitwos}$ over $\BR_{\jpsi} \times \sigma_{\jpsi}$ are compared with other measurements, and are found to be in agreement with a relative higher precision.

\section*{Acknowledgements}
%
%
We express our gratitute to Elena Gonzalez Ferreiro for the co-mover model predictions and YanQing Ma for the NRQCD model predictions.
We express our gratitude to our colleagues in the CERN
accelerator departments for the excellent performance of the LHC. We
thank the technical and administrative staff at the LHCb
institutes.
We acknowledge support from CERN and from the national agencies:
CAPES, CNPq, FAPERJ and FINEP (Brazil); 
MOST and NSFC (China); 
CNRS/IN2P3 (France); 
BMBF, DFG and MPG (Germany); 
INFN (Italy); 
NWO (Netherlands); 
MNiSW and NCN (Poland); 
MCID/IFA (Romania); 
MICINN (Spain); 
SNSF and SER (Switzerland); 
NASU (Ukraine); 
STFC (United Kingdom); 
DOE NP and NSF (USA).
We acknowledge the computing resources that are provided by CERN, IN2P3
(France), KIT and DESY (Germany), INFN (Italy), SURF (Netherlands),
PIC (Spain), GridPP (United Kingdom), 
CSCS (Switzerland), IFIN-HH (Romania), CBPF (Brazil),
and Polish WLCG (Poland).
We are indebted to the communities behind the multiple open-source
software packages on which we depend.
Individual groups or members have received support from
ARC and ARDC (Australia);
Key Research Program of Frontier Sciences of CAS, CAS PIFI, CAS CCEPP, 
Fundamental Research Funds for the Central Universities, 
and Sci. \& Tech. Program of Guangzhou (China);
Minciencias (Colombia);
EPLANET, Marie Sk\l{}odowska-Curie Actions, ERC and NextGenerationEU (European Union);
A*MIDEX, ANR, IPhU and Labex P2IO, and R\'{e}gion Auvergne-Rh\^{o}ne-Alpes (France);
AvH Foundation (Germany);
ICSC (Italy); 
GVA, XuntaGal, GENCAT, Inditex, InTalent and Prog.~Atracci\'on Talento, CM (Spain);
SRC (Sweden);
the Leverhulme Trust, the Royal Society
 and UKRI (United Kingdom).

\clearpage
\section*{A. Numeric results}
\label{Sec:Ratio}
\begin{table}[H]
\caption{Ratios (\%) of double differential production cross-section for \psitwos to \jpsi in bins of (\pt,$y$) for 4 $\leq$ $N_{\rm tracks}^{\rm PV}$ $<$ 20 (at least 4 tracks required to form a high-quality vertex). The first uncertainties are statistical, the second are systematic.}
\centering
\resizebox{0.7\linewidth}{!}{
\begin{tabular}{|c|ccc|}
\hline
\multicolumn{4}{|c|}{prompt}\\\hline
\pt(\gevc)& $2<y<2.8$& $2.8<y<3.5$& $3.5<y<4.5$ \\
\hline
0-2&$14.95\pm0.19\pm0.54$&$15.18\pm0.15\pm0.44$&$15.54\pm0.15\pm0.48$\\
2-4&$20.45\pm0.30\pm0.72$&$18.97\pm0.22\pm0.54$&$19.49\pm0.24\pm0.56$\\
4-6&$29.18\pm0.60\pm1.02$&$23.22\pm0.39\pm0.68$&$22.75\pm0.45\pm0.75$\\
6-8&$34.60\pm1.08\pm1.45$&$26.74\pm0.76\pm0.97$&$27.89\pm0.93\pm1.16$\\
8-20&$33.05\pm1.39\pm1.52$&$27.67\pm1.19\pm1.15$&$29.83\pm1.65\pm1.70$\\
\hline
\multicolumn{4}{|c|}{non-prompt}\\\hline
\pt(\gevc)& $2<y<2.8$& $2.8<y<3.5$& $3.5<y<4.5$ \\
\hline
0-2&$21.88\pm1.19\pm1.16$&$19.99\pm0.91\pm1.01$&$21.04\pm1.16\pm1.74$\\
2-4&$28.29\pm1.17\pm1.37$&$27.02\pm0.90\pm1.09$&$27.60\pm1.11\pm1.56$\\
4-6&$32.89\pm1.53\pm1.61$&$29.53\pm1.13\pm1.46$&$33.70\pm1.57\pm1.98$\\
6-8&$34.89\pm2.10\pm1.88$&$31.70\pm1.69\pm1.41$&$32.33\pm2.23\pm1.96$\\
8-20&$43.93\pm2.36\pm1.97$&$39.96\pm2.16\pm1.77$&$39.78\pm2.95\pm2.80$\\
\hline
\end{tabular}
}
\label{RatioPVN1}
\end{table}

\begin{table}[H]
\caption{Ratios (\%) of double differential production cross-section for \psitwos to \jpsi in bins of (\pt,$y$) for 20 $\leq$ $N_{\rm tracks}^{\rm PV}$ $<$ 45. The first uncertainties are statistical, the second are systematic.}
\centering
\resizebox{0.7\linewidth}{!}{
\begin{tabular}{|c|ccc|}
\hline
\multicolumn{4}{|c|}{prompt}\\\hline
\pt(\gevc)& $2<y<2.8$& $2.8<y<3.5$& $3.5<y<4.5$ \\
\hline
0-2&$11.50\pm0.14\pm0.54$&$10.92\pm0.11\pm0.53$&$11.62\pm0.11\pm0.63$\\
2-4&$16.48\pm0.17\pm0.74$&$15.33\pm0.13\pm0.65$&$15.81\pm0.14\pm0.67$\\
4-6&$22.41\pm0.25\pm0.98$&$20.68\pm0.18\pm0.86$&$20.62\pm0.21\pm0.88$\\
6-8&$29.18\pm0.40\pm1.41$&$26.20\pm0.32\pm1.14$&$25.11\pm0.38\pm1.16$\\
8-20&$35.01\pm0.53\pm1.65$&$29.31\pm0.43\pm1.35$&$28.46\pm0.54\pm1.39$\\
\hline
\multicolumn{4}{|c|}{non-prompt}\\\hline
\pt(\gevc)& $2<y<2.8$& $2.8<y<3.5$& $3.5<y<4.5$ \\
\hline
0-2&$19.03\pm0.57\pm1.04$&$20.24\pm0.49\pm1.00$&$20.72\pm0.56\pm1.14$\\
2-4&$26.92\pm0.55\pm1.30$&$25.65\pm0.43\pm1.24$&$27.27\pm0.56\pm1.37$\\
4-6&$34.28\pm0.68\pm1.58$&$28.18\pm0.49\pm1.33$&$29.32\pm0.67\pm1.55$\\
6-8&$34.61\pm0.83\pm1.89$&$32.17\pm0.71\pm1.57$&$34.12\pm1.02\pm2.38$\\
8-20&$41.47\pm0.84\pm1.99$&$36.17\pm0.78\pm1.78$&$34.52\pm1.04\pm1.80$\\
\hline
\end{tabular}
}
\label{RatioPVN2}
\end{table}

\begin{table}[H]
\caption{Ratios (\%) of double differential production cross-section for \psitwos to \jpsi in bins of (\pt,$y$) for 45 $\leq$ $N_{\rm tracks}^{\rm PV}$ $<$ 70. The first uncertainties are statistical, the second are systematic.}
\centering
\resizebox{0.7\linewidth}{!}{
\begin{tabular}{|c|ccc|}
\hline
\multicolumn{4}{|c|}{prompt}\\\hline
\pt(\gevc)& $2<y<2.8$& $2.8<y<3.5$& $3.5<y<4.5$ \\
\hline
0-2&$9.04\pm0.22\pm0.27$&$8.79\pm0.18\pm0.20$&$9.42\pm0.16\pm0.29$\\
2-4&$13.03\pm0.21\pm0.29$&$12.38\pm0.16\pm0.24$&$13.53\pm0.17\pm0.29$\\
4-6&$20.18\pm0.28\pm0.52$&$17.62\pm0.21\pm0.33$&$18.83\pm0.23\pm0.44$\\
6-8&$25.17\pm0.38\pm0.68$&$23.94\pm0.31\pm0.52$&$23.65\pm0.37\pm0.60$\\
8-20&$31.32\pm0.43\pm0.84$&$29.02\pm0.41\pm0.71$&$27.94\pm0.50\pm0.88$\\
\hline
\multicolumn{4}{|c|}{non-prompt}\\\hline
\pt(\gevc)& $2<y<2.8$& $2.8<y<3.5$& $3.5<y<4.5$ \\
\hline
0-2&$20.93\pm0.68\pm0.83$&$18.99\pm0.56\pm0.81$&$19.55\pm0.62\pm0.70$\\
2-4&$23.54\pm0.54\pm0.68$&$24.22\pm0.45\pm0.68$&$23.76\pm0.55\pm0.87$\\
4-6&$31.83\pm0.66\pm0.82$&$26.85\pm0.49\pm0.78$&$27.46\pm0.67\pm1.16$\\
6-8&$32.65\pm0.77\pm1.03$&$31.71\pm0.69\pm1.02$&$34.16\pm0.96\pm1.55$\\
8-20&$39.44\pm0.74\pm1.12$&$37.75\pm0.73\pm1.14$&$38.58\pm1.07\pm1.71$\\
\hline
\end{tabular}
}
\label{RatioPVN3}
\end{table}

\begin{table}[H]
\caption{Ratios (\%) of double differential production cross-section for \psitwos to \jpsi in bins of (\pt,$y$) for 70 $\leq$ $N_{\rm tracks}^{\rm PV}$ $<$ 95. The first uncertainties are statistical, the second are systematic.}
\centering
\resizebox{0.7\linewidth}{!}{
\begin{tabular}{|c|ccc|}
\hline
\multicolumn{4}{|c|}{prompt}\\\hline
\pt(\gevc)& $2<y<2.8$& $2.8<y<3.5$& $3.5<y<4.5$ \\
\hline
0-2&$7.71\pm0.44\pm0.28$&$7.89\pm0.34\pm0.23$&$7.55\pm0.28\pm0.51$\\
2-4&$11.29\pm0.37\pm0.34$&$11.58\pm0.31\pm0.28$&$12.18\pm0.30\pm0.45$\\
4-6&$17.92\pm0.43\pm0.56$&$15.68\pm0.31\pm0.33$&$16.52\pm0.36\pm0.47$\\
6-8&$23.07\pm0.54\pm0.78$&$20.16\pm0.43\pm0.57$&$21.77\pm0.52\pm0.75$\\
8-20&$29.65\pm0.59\pm1.14$&$27.43\pm0.55\pm0.95$&$25.27\pm0.66\pm1.00$\\
\hline
\multicolumn{4}{|c|}{non-prompt}\\\hline
\pt(\gevc)& $2<y<2.8$& $2.8<y<3.5$& $3.5<y<4.5$ \\
\hline
0-2&$17.74\pm1.03\pm1.07$&$17.59\pm0.80\pm0.75$&$19.41\pm0.99\pm0.83$\\
2-4&$26.99\pm0.92\pm0.95$&$24.29\pm0.69\pm0.73$&$25.62\pm0.89\pm1.23$\\
4-6&$32.36\pm0.99\pm1.04$&$26.84\pm0.71\pm0.89$&$26.85\pm0.95\pm1.18$\\
6-8&$31.96\pm1.12\pm1.24$&$31.70\pm1.01\pm1.27$&$28.69\pm1.32\pm1.59$\\
8-20&$41.22\pm1.06\pm1.22$&$35.64\pm1.01\pm1.35$&$34.33\pm1.47\pm1.58$\\
\hline
\end{tabular}
}
\label{RatioPVN4}
\end{table}

\begin{table}[H]
\caption{Ratios (\%) of double differential production cross-section for \psitwos to \jpsi in bins of (\pt,$y$) for 95 $\leq$ $N_{\rm tracks}^{\rm PV}$ $<$ 200. The first uncertainties are statistical, the second are systematic.}
\centering
\resizebox{0.7\linewidth}{!}{
\begin{tabular}{|c|ccc|}
\hline
\multicolumn{4}{|c|}{prompt}\\\hline
\pt(\gevc)& $2<y<2.8$& $2.8<y<3.5$& $3.5<y<4.5$ \\
\hline
0-2&$9.91\pm1.05\pm0.83$&$7.13\pm0.79\pm0.51$&$8.93\pm0.77\pm0.69$\\
2-4&$9.18\pm0.74\pm0.55$&$9.23\pm0.59\pm0.40$&$10.99\pm0.61\pm0.51$\\
4-6&$17.20\pm0.86\pm1.02$&$14.79\pm0.61\pm0.67$&$15.31\pm0.66\pm1.08$\\
6-8&$22.16\pm1.04\pm1.89$&$19.04\pm0.78\pm1.61$&$22.02\pm1.00\pm1.60$\\
8-20&$28.98\pm1.05\pm3.33$&$26.00\pm0.93\pm1.53$&$24.66\pm1.17\pm1.85$\\
\hline
\multicolumn{4}{|c|}{non-prompt}\\\hline
\pt(\gevc)& $2<y<2.8$& $2.8<y<3.5$& $3.5<y<4.5$ \\
\hline
0-2&$23.68\pm2.65\pm3.06$&$20.28\pm1.84\pm1.64$&$21.47\pm2.45\pm2.05$\\
2-4&$24.12\pm1.55\pm1.89$&$25.31\pm1.39\pm1.30$&$24.29\pm1.61\pm1.71$\\
4-6&$30.40\pm1.81\pm1.85$&$28.96\pm1.41\pm1.83$&$28.78\pm1.85\pm2.25$\\
6-8&$35.56\pm2.30\pm3.36$&$36.19\pm1.99\pm4.03$&$25.80\pm2.46\pm2.30$\\
8-20&$35.20\pm1.79\pm2.08$&$33.68\pm1.78\pm1.74$&$37.92\pm2.77\pm3.16$\\
\hline
\end{tabular}
}
\label{RatioPVN5}
\end{table}

\begin{table}[H]
\caption{Ratios (\%) of double differential production cross-section for \psitwos to \jpsi in bins of (\pt,$y$) for 0 $\leq$ $N_{\rm bwd}^{\rm PV}$ $<$ 8. The first uncertainties are statistical, the second are systematic. }
\centering
\resizebox{0.7\linewidth}{!}{
\begin{tabular}{|c|ccc|}
\hline
\multicolumn{4}{|c|}{prompt}\\\hline
\pt(\gevc)& $2<y<2.8$& $2.8<y<3.5$& $3.5<y<4.5$ \\
\hline
0-2&$12.78\pm0.18\pm0.89$&$12.48\pm0.14\pm0.66$&$12.88\pm0.13\pm0.75$\\
2-4&$17.26\pm0.23\pm0.81$&$16.40\pm0.18\pm0.68$&$16.36\pm0.18\pm0.66$\\
4-6&$23.45\pm0.36\pm1.07$&$21.17\pm0.26\pm0.87$&$20.84\pm0.29\pm0.88$\\
6-8&$30.78\pm0.62\pm1.63$&$26.87\pm0.47\pm1.26$&$26.21\pm0.53\pm1.31$\\
8-20&$32.25\pm0.71\pm1.68$&$28.28\pm0.60\pm1.42$&$27.93\pm0.76\pm1.46$\\
\hline
\multicolumn{4}{|c|}{non-prompt}\\\hline
\pt(\gevc)& $2<y<2.8$& $2.8<y<3.5$& $3.5<y<4.5$ \\
\hline
0-2&$21.71\pm0.88\pm2.39$&$20.83\pm0.72\pm1.38$&$21.28\pm0.82\pm1.67$\\
2-4&$26.41\pm0.79\pm1.31$&$26.46\pm0.63\pm1.29$&$26.41\pm0.75\pm1.35$\\
4-6&$34.65\pm1.00\pm1.70$&$29.13\pm0.71\pm1.39$&$30.13\pm0.95\pm1.69$\\
6-8&$33.39\pm1.23\pm2.17$&$31.37\pm1.02\pm1.85$&$33.99\pm1.40\pm2.61$\\
8-20&$41.81\pm1.26\pm2.03$&$36.94\pm1.11\pm2.00$&$35.45\pm1.45\pm2.12$\\
\hline
\end{tabular}
}
\label{RatioBACK1}
\end{table}

\begin{table}[H]
\caption{Ratios (\%) of double differential production cross-section for \psitwos to \jpsi in bins of (\pt,$y$) for 8 $\leq$ $N_{\rm bwd}^{\rm PV}$ $<$ 15. The first uncertainties are statistical, the second are systematic. }
\centering
\resizebox{0.7\linewidth}{!}{
\begin{tabular}{|c|ccc|}
\hline
\multicolumn{4}{|c|}{prompt}\\\hline
\pt(\gevc)& $2<y<2.8$& $2.8<y<3.5$& $3.5<y<4.5$ \\
\hline
0-2&$10.88\pm0.21\pm0.53$&$10.31\pm0.17\pm0.40$&$11.35\pm0.16\pm0.55$\\
2-4&$15.30\pm0.24\pm0.56$&$13.74\pm0.18\pm0.43$&$14.85\pm0.19\pm0.47$\\
4-6&$21.98\pm0.33\pm0.78$&$18.69\pm0.23\pm0.60$&$19.23\pm0.27\pm0.64$\\
6-8&$27.76\pm0.49\pm1.19$&$25.11\pm0.38\pm0.91$&$24.65\pm0.47\pm1.04$\\
8-20&$33.44\pm0.60\pm1.44$&$28.62\pm0.51\pm1.09$&$29.46\pm0.66\pm1.52$\\
\hline
\multicolumn{4}{|c|}{non-prompt}\\\hline
\pt(\gevc)& $2<y<2.8$& $2.8<y<3.5$& $3.5<y<4.5$ \\
\hline
0-2&$21.38\pm0.82\pm1.35$&$19.44\pm0.63\pm1.01$&$19.59\pm0.71\pm1.30$\\
2-4&$25.22\pm0.68\pm1.05$&$24.62\pm0.53\pm0.91$&$26.11\pm0.70\pm1.00$\\
4-6&$32.12\pm0.80\pm1.27$&$27.71\pm0.60\pm1.04$&$28.15\pm0.81\pm1.36$\\
6-8&$33.65\pm0.99\pm1.64$&$33.25\pm0.88\pm1.46$&$32.86\pm1.20\pm2.38$\\
8-20&$39.68\pm0.95\pm1.75$&$37.19\pm0.94\pm1.58$&$35.36\pm1.31\pm2.08$\\
\hline
\end{tabular}
}
\label{RatioBACK2}
\end{table}

\begin{table}[H]
\caption{Ratios (\%) of double differential production cross-section for \psitwos to \jpsi in bins of (\pt,$y$) for 15 $\leq$ $N_{\rm bwd}^{\rm PV}$ $<$ 22. The first uncertainties are statistical, the second are systematic. }
\centering
\resizebox{0.7\linewidth}{!}{
\begin{tabular}{|c|ccc|}
\hline
\multicolumn{4}{|c|}{prompt}\\\hline
\pt(\gevc)& $2<y<2.8$& $2.8<y<3.5$& $3.5<y<4.5$ \\
\hline
0-2&$8.97\pm0.27\pm0.32$&$9.42\pm0.23\pm0.32$&$9.58\pm0.21\pm0.39$\\
2-4&$13.41\pm0.28\pm0.34$&$12.68\pm0.22\pm0.27$&$13.57\pm0.24\pm0.36$\\
4-6&$20.29\pm0.38\pm0.55$&$17.52\pm0.27\pm0.39$&$18.62\pm0.33\pm0.48$\\
6-8&$25.76\pm0.52\pm0.90$&$23.21\pm0.41\pm0.69$&$22.37\pm0.49\pm0.78$\\
8-20&$32.96\pm0.62\pm1.24$&$28.41\pm0.55\pm0.91$&$28.58\pm0.73\pm1.04$\\
\hline
\multicolumn{4}{|c|}{non-prompt}\\\hline
\pt(\gevc)& $2<y<2.8$& $2.8<y<3.5$& $3.5<y<4.5$ \\
\hline
0-2&$18.28\pm0.84\pm0.91$&$19.54\pm0.74\pm0.91$&$21.70\pm0.89\pm1.02$\\
2-4&$26.99\pm0.78\pm0.89$&$24.53\pm0.61\pm0.84$&$24.61\pm0.76\pm1.15$\\
4-6&$31.88\pm0.91\pm1.02$&$25.95\pm0.64\pm1.11$&$27.80\pm0.93\pm1.13$\\
6-8&$35.26\pm1.08\pm1.48$&$31.24\pm0.93\pm1.30$&$32.19\pm1.34\pm1.96$\\
8-20&$40.96\pm1.06\pm1.57$&$36.60\pm1.00\pm1.41$&$37.63\pm1.47\pm1.93$\\
\hline
\end{tabular}
}
\label{RatioBACK3}
\end{table}

\begin{table}[H]
\caption{Ratios (\%) of double differential production cross-section for \psitwos to \jpsi in bins of (\pt,$y$) for 22 $\leq$ $N_{\rm bwd}^{\rm PV}$ $<$ 30. The first uncertainties are statistical, the second are systematic. }
\centering
\resizebox{0.7\linewidth}{!}{
\begin{tabular}{|c|ccc|}
\hline
\multicolumn{4}{|c|}{prompt}\\\hline
\pt(\gevc)& $2<y<2.8$& $2.8<y<3.5$& $3.5<y<4.5$ \\
\hline
0-2&$8.84\pm0.42\pm0.38$&$8.44\pm0.31\pm0.30$&$9.44\pm0.30\pm0.62$\\
2-4&$12.34\pm0.39\pm0.44$&$11.70\pm0.29\pm0.25$&$13.21\pm0.32\pm0.31$\\
4-6&$17.14\pm0.46\pm0.52$&$16.60\pm0.35\pm0.39$&$17.25\pm0.42\pm0.48$\\
6-8&$24.94\pm0.65\pm1.00$&$21.02\pm0.49\pm0.65$&$23.60\pm0.66\pm0.93$\\
8-20&$30.78\pm0.71\pm1.12$&$27.47\pm0.66\pm0.91$&$26.57\pm0.84\pm1.39$\\
\hline
\multicolumn{4}{|c|}{non-prompt}\\\hline
\pt(\gevc)& $2<y<2.8$& $2.8<y<3.5$& $3.5<y<4.5$ \\
\hline
0-2&$19.08\pm1.06\pm1.17$&$18.33\pm0.91\pm0.95$&$18.24\pm1.04\pm1.54$\\
2-4&$24.53\pm0.93\pm0.87$&$24.59\pm0.75\pm0.65$&$24.82\pm0.98\pm0.96$\\
4-6&$34.04\pm1.14\pm1.13$&$27.68\pm0.83\pm0.92$&$28.05\pm1.18\pm1.49$\\
6-8&$30.65\pm1.25\pm1.21$&$30.66\pm1.13\pm1.30$&$31.91\pm1.65\pm1.99$\\
8-20&$38.03\pm1.20\pm1.20$&$36.79\pm1.24\pm1.45$&$35.96\pm1.79\pm2.19$\\
\hline
\end{tabular}
}
\label{RatioBACK4}
\end{table}

\begin{table}[H]
\caption{Ratios (\%) of double differential production cross-section for \psitwos to \jpsi in bins of (\pt,$y$) for 30 $\leq$ $N_{\rm bwd}^{\rm PV}$ $<$ 80. The first uncertainties are statistical, the second are systematic. }
\centering
\resizebox{0.7\linewidth}{!}{
\begin{tabular}{|c|ccc|}
\hline
\multicolumn{4}{|c|}{prompt}\\\hline
\pt(\gevc)& $2<y<2.8$& $2.8<y<3.5$& $3.5<y<4.5$ \\
\hline
0-2&$10.05\pm0.65\pm0.70$&$8.74\pm0.49\pm0.34$&$8.60\pm0.43\pm0.40$\\
2-4&$11.00\pm0.52\pm0.40$&$12.93\pm0.46\pm0.58$&$13.28\pm0.44\pm0.39$\\
4-6&$19.54\pm0.67\pm0.72$&$15.82\pm0.44\pm0.45$&$17.01\pm0.52\pm0.70$\\
6-8&$21.82\pm0.78\pm1.10$&$21.98\pm0.68\pm1.17$&$21.93\pm0.79\pm0.90$\\
8-20&$28.63\pm0.87\pm1.25$&$25.73\pm0.78\pm0.98$&$22.99\pm0.96\pm1.05$\\
\hline
\multicolumn{4}{|c|}{non-prompt}\\\hline
\pt(\gevc)& $2<y<2.8$& $2.8<y<3.5$& $3.5<y<4.5$ \\
\hline
0-2&$20.03\pm1.60\pm1.77$&$19.62\pm1.20\pm1.13$&$20.23\pm1.49\pm1.20$\\
2-4&$26.02\pm1.26\pm1.00$&$23.62\pm1.01\pm0.83$&$24.16\pm1.25\pm0.91$\\
4-6&$33.55\pm1.45\pm1.31$&$28.41\pm1.07\pm1.14$&$27.64\pm1.44\pm1.67$\\
6-8&$35.20\pm1.81\pm2.00$&$34.75\pm1.62\pm2.78$&$27.33\pm2.10\pm2.32$\\
8-20&$40.75\pm1.57\pm1.72$&$33.67\pm1.46\pm1.26$&$34.87\pm2.28\pm1.82$\\
\hline
\end{tabular}
}
\label{RatioBACK5}
\end{table}

\begin{table}[H]
\caption{Ratios (\%) of double differential production cross-section for \psitwos to \jpsi in bins of (\pt,$y$) for 0 $\leq$ $N_{\rm fwd}^{\rm PV}$ $<$ 12. The first uncertainties are statistical, the second are systematic. }
\centering
\resizebox{0.7\linewidth}{!}{
\begin{tabular}{|c|ccc|}
\hline
\multicolumn{4}{|c|}{prompt}\\\hline
\pt(\gevc)& $2<y<2.8$& $2.8<y<3.5$& $3.5<y<4.5$ \\
\hline
0-2&$16.01\pm0.22\pm0.69$&$15.86\pm0.18\pm0.62$&$16.07\pm0.18\pm0.64$\\
2-4&$21.67\pm0.37\pm0.92$&$19.67\pm0.27\pm0.76$&$20.59\pm0.31\pm0.80$\\
4-6&$28.37\pm0.70\pm1.21$&$24.31\pm0.51\pm1.00$&$23.25\pm0.58\pm1.01$\\
6-8&$34.96\pm1.38\pm1.75$&$27.17\pm1.00\pm1.31$&$28.99\pm1.26\pm1.47$\\
8-20&$33.65\pm1.77\pm1.87$&$27.14\pm1.59\pm1.41$&$28.53\pm2.28\pm2.00$\\
\hline
\multicolumn{4}{|c|}{non-prompt}\\\hline
\pt(\gevc)& $2<y<2.8$& $2.8<y<3.5$& $3.5<y<4.5$ \\
\hline
0-2&$22.02\pm1.35\pm1.27$&$21.28\pm1.08\pm1.27$&$23.51\pm1.43\pm1.48$\\
2-4&$29.87\pm1.36\pm1.82$&$26.89\pm1.03\pm1.41$&$27.32\pm1.33\pm2.18$\\
4-6&$35.66\pm1.82\pm1.71$&$30.86\pm1.42\pm1.81$&$35.67\pm1.97\pm2.15$\\
6-8&$36.27\pm2.50\pm2.10$&$35.57\pm2.18\pm2.07$&$34.00\pm2.92\pm2.13$\\
8-20&$48.13\pm3.00\pm2.48$&$40.55\pm2.76\pm2.12$&$39.03\pm3.58\pm3.07$\\
\hline
\end{tabular}
}
\label{RatioFOR1}
\end{table}

\begin{table}[H]
\caption{Ratios (\%) of double differential production cross-section for \psitwos to \jpsi in bins of (\pt,$y$) for 12 $\leq$ $N_{\rm fwd}^{\rm PV}$ $<$ 24. The first uncertainties are statistical, the second are systematic. }
\centering
\resizebox{0.7\linewidth}{!}{
\begin{tabular}{|c|ccc|}
\hline
\multicolumn{4}{|c|}{prompt}\\\hline
\pt(\gevc)& $2<y<2.8$& $2.8<y<3.5$& $3.5<y<4.5$ \\
\hline
0-2&$12.52\pm0.16\pm0.51$&$12.00\pm0.12\pm0.45$&$12.79\pm0.12\pm0.51$\\
2-4&$17.29\pm0.20\pm0.69$&$16.64\pm0.15\pm0.61$&$16.81\pm0.16\pm0.62$\\
4-6&$24.11\pm0.32\pm0.95$&$21.80\pm0.23\pm0.81$&$22.20\pm0.27\pm0.84$\\
6-8&$30.99\pm0.54\pm1.38$&$27.27\pm0.42\pm1.07$&$25.82\pm0.50\pm1.08$\\
8-20&$34.20\pm0.69\pm1.46$&$28.33\pm0.59\pm1.21$&$28.99\pm0.77\pm1.33$\\
\hline
\multicolumn{4}{|c|}{non-prompt}\\\hline
\pt(\gevc)& $2<y<2.8$& $2.8<y<3.5$& $3.5<y<4.5$ \\
\hline
0-2&$19.02\pm0.69\pm1.00$&$20.54\pm0.60\pm0.90$&$20.53\pm0.69\pm1.05$\\
2-4&$27.70\pm0.70\pm1.17$&$26.41\pm0.55\pm1.15$&$27.35\pm0.69\pm1.20$\\
4-6&$33.87\pm0.87\pm1.40$&$28.73\pm0.63\pm1.24$&$28.92\pm0.85\pm1.66$\\
6-8&$34.24\pm1.10\pm1.97$&$31.66\pm0.95\pm1.53$&$36.00\pm1.40\pm2.05$\\
8-20&$42.40\pm1.14\pm1.83$&$36.58\pm1.06\pm1.62$&$37.08\pm1.47\pm1.96$\\
\hline
\end{tabular}
}
\label{RatioFOR2}
\end{table}

\begin{table}[H]
\caption{Ratios (\%) of double differential production cross-section for \psitwos to \jpsi in bins of (\pt,$y$) for 24 $\leq$ $N_{\rm fwd}^{\rm PV}$ $<$ 36. The first uncertainties are statistical, the second are systematic. }
\centering
\resizebox{0.7\linewidth}{!}{
\begin{tabular}{|c|ccc|}
\hline
\multicolumn{4}{|c|}{prompt}\\\hline
\pt(\gevc)& $2<y<2.8$& $2.8<y<3.5$& $3.5<y<4.5$ \\
\hline
0-2&$10.25\pm0.19\pm0.50$&$9.94\pm0.15\pm0.45$&$10.47\pm0.14\pm0.49$\\
2-4&$15.01\pm0.21\pm0.74$&$14.06\pm0.16\pm0.63$&$15.27\pm0.17\pm0.69$\\
4-6&$22.14\pm0.30\pm1.03$&$19.72\pm0.21\pm0.88$&$19.45\pm0.24\pm0.89$\\
6-8&$27.15\pm0.43\pm1.36$&$24.99\pm0.34\pm1.15$&$24.87\pm0.42\pm1.26$\\
8-20&$34.19\pm0.55\pm1.69$&$29.95\pm0.48\pm1.41$&$28.70\pm0.59\pm1.56$\\
\hline
\multicolumn{4}{|c|}{non-prompt}\\\hline
\pt(\gevc)& $2<y<2.8$& $2.8<y<3.5$& $3.5<y<4.5$ \\
\hline
0-2&$20.86\pm0.70\pm1.14$&$19.55\pm0.57\pm1.06$&$20.70\pm0.66\pm1.15$\\
2-4&$24.64\pm0.59\pm1.30$&$25.44\pm0.49\pm1.22$&$25.49\pm0.61\pm1.24$\\
4-6&$33.39\pm0.75\pm1.71$&$28.38\pm0.55\pm1.50$&$28.53\pm0.75\pm1.59$\\
6-8&$34.21\pm0.91\pm1.88$&$32.26\pm0.78\pm1.66$&$33.53\pm1.12\pm2.48$\\
8-20&$39.71\pm0.87\pm2.00$&$36.66\pm0.85\pm1.87$&$34.49\pm1.16\pm2.10$\\
\hline
\end{tabular}
}
\label{RatioFOR3}
\end{table}

\begin{table}[H]
\caption{Ratios (\%) of double differential production cross-section for \psitwos to \jpsi in bins of (\pt,$y$) for 36 $\leq$ $N_{\rm fwd}^{\rm PV}$ $<$ 48. The first uncertainties are statistical, the second are systematic. }
\centering
\resizebox{0.7\linewidth}{!}{
\begin{tabular}{|c|ccc|}
\hline
\multicolumn{4}{|c|}{prompt}\\\hline
\pt(\gevc)& $2<y<2.8$& $2.8<y<3.5$& $3.5<y<4.5$ \\
\hline
0-2&$8.97\pm0.27\pm0.34$&$8.91\pm0.21\pm0.28$&$9.18\pm0.19\pm0.35$\\
2-4&$12.98\pm0.25\pm0.40$&$12.24\pm0.20\pm0.35$&$13.34\pm0.21\pm0.40$\\
4-6&$19.67\pm0.33\pm0.66$&$17.34\pm0.24\pm0.48$&$18.71\pm0.27\pm0.62$\\
6-8&$25.22\pm0.45\pm0.91$&$24.19\pm0.37\pm0.77$&$22.53\pm0.43\pm0.93$\\
8-20&$30.83\pm0.52\pm1.17$&$29.39\pm0.48\pm1.01$&$26.80\pm0.58\pm1.10$\\
\hline
\multicolumn{4}{|c|}{non-prompt}\\\hline
\pt(\gevc)& $2<y<2.8$& $2.8<y<3.5$& $3.5<y<4.5$ \\
\hline
0-2&$19.55\pm0.82\pm0.94$&$18.95\pm0.68\pm0.89$&$18.66\pm0.73\pm0.97$\\
2-4&$24.04\pm0.66\pm0.87$&$24.03\pm0.55\pm0.86$&$24.57\pm0.67\pm0.97$\\
4-6&$32.21\pm0.79\pm1.18$&$26.33\pm0.56\pm0.83$&$27.16\pm0.79\pm1.37$\\
6-8&$32.09\pm0.91\pm1.37$&$33.16\pm0.85\pm1.31$&$31.59\pm1.13\pm2.03$\\
8-20&$38.94\pm0.89\pm1.47$&$37.14\pm0.87\pm1.45$&$39.10\pm1.30\pm1.88$\\
\hline
\end{tabular}
}
\label{RatioFOR4}
\end{table}

\begin{table}[H]
\caption{Ratios (\%) of double differential production cross-section for \psitwos to \jpsi in bins of (\pt,$y$) for 48 $\leq$ $N_{\rm fwd}^{\rm PV}$ $<$ 130. The first uncertainties are statistical, the second are systematic. }
\centering
\resizebox{0.7\linewidth}{!}{
\begin{tabular}{|c|ccc|}
\hline
\multicolumn{4}{|c|}{prompt}\\\hline
\pt(\gevc)& $2<y<2.8$& $2.8<y<3.5$& $3.5<y<4.5$ \\
\hline
0-2&$7.83\pm0.35\pm0.31$&$7.63\pm0.28\pm0.31$&$8.44\pm0.25\pm0.25$\\
2-4&$10.93\pm0.30\pm0.35$&$10.96\pm0.24\pm0.28$&$11.58\pm0.23\pm0.29$\\
4-6&$18.12\pm0.35\pm0.63$&$15.04\pm0.24\pm0.35$&$16.48\pm0.28\pm0.60$\\
6-8&$23.28\pm0.44\pm0.83$&$19.97\pm0.32\pm0.54$&$22.83\pm0.42\pm0.82$\\
8-20&$29.91\pm0.46\pm0.98$&$27.16\pm0.41\pm1.04$&$25.68\pm0.51\pm1.07$\\
\hline
\multicolumn{4}{|c|}{non-prompt}\\\hline
\pt(\gevc)& $2<y<2.8$& $2.8<y<3.5$& $3.5<y<4.5$ \\
\hline
0-2&$19.00\pm0.87\pm1.09$&$17.31\pm0.66\pm1.00$&$19.72\pm0.79\pm0.85$\\
2-4&$24.67\pm0.68\pm0.83$&$24.22\pm0.55\pm0.70$&$24.98\pm0.68\pm0.83$\\
4-6&$30.68\pm0.76\pm0.98$&$27.65\pm0.57\pm0.94$&$27.95\pm0.74\pm1.13$\\
6-8&$32.93\pm0.89\pm1.21$&$32.79\pm0.78\pm1.22$&$29.56\pm1.01\pm1.59$\\
8-20&$39.20\pm0.79\pm1.35$&$35.10\pm0.75\pm1.30$&$34.72\pm1.11\pm1.71$\\
\hline
\end{tabular}
}
\label{RatioFOR5}
\end{table}

\clearpage

\addcontentsline{toc}{section}{References}

\bibliographystyle{LHCb}
\bibliography{main,standard,LHCb-PAPER,LHCb-CONF,LHCb-DP,LHCb-TDR}

\ifx\mcitethebibliography\mciteundefinedmacro
\PackageError{LHCb.bst}{mciteplus.sty has not been loaded}
{This bibstyle requires the use of the mciteplus package.}\fi
\providecommand{\href}[2]{#2}
\begin{mcitethebibliography}{10}
\mciteSetBstSublistMode{n}
\mciteSetBstMaxWidthForm{subitem}{\alph{mcitesubitemcount})}
\mciteSetBstSublistLabelBeginEnd{\mcitemaxwidthsubitemform\space}
{\relax}{\relax}

\bibitem{MATSUI1986416}
T.~Matsui and H.~Satz, \ifthenelse{\boolean{articletitles}}{\emph{J/$\psi$ suppression by quark-gluon plasma formation}, }{}\href{https://doi.org/https://doi.org/10.1016/0370-2693(86)91404-8}{Phys.\ Lett.\  \textbf{B178} (1986) 416}\relax
\mciteBstWouldAddEndPuncttrue
\mciteSetBstMidEndSepPunct{\mcitedefaultmidpunct}
{\mcitedefaultendpunct}{\mcitedefaultseppunct}\relax
\EndOfBibitem
\bibitem{Zhao:2020jqu}
J.~Zhao, K.~Zhou, S.~Chen, and P.~Zhuang, \ifthenelse{\boolean{articletitles}}{\emph{{Heavy flavors under extreme conditions in high energy nuclear collisions}}, }{}\href{https://doi.org/10.1016/j.ppnp.2020.103801}{Prog.\ Part.\ Nucl.\ Phys.\  \textbf{114} (2020) 103801}, \href{http://arxiv.org/abs/2005.08277}{{\normalfont\ttfamily arXiv:2005.08277}}\relax
\mciteBstWouldAddEndPuncttrue
\mciteSetBstMidEndSepPunct{\mcitedefaultmidpunct}
{\mcitedefaultendpunct}{\mcitedefaultseppunct}\relax
\EndOfBibitem
\bibitem{Thews:2000rj}
R.~L. Thews, M.~Schroedter, and J.~Rafelski, \ifthenelse{\boolean{articletitles}}{\emph{{Enhanced $J/\psi$ production in deconfined quark matter}}, }{}\href{https://doi.org/10.1103/PhysRevC.63.054905}{Phys.\ Rev.\  \textbf{C63} (2001) 054905}, \href{http://arxiv.org/abs/hep-ph/0007323}{{\normalfont\ttfamily arXiv:hep-ph/0007323}}\relax
\mciteBstWouldAddEndPuncttrue
\mciteSetBstMidEndSepPunct{\mcitedefaultmidpunct}
{\mcitedefaultendpunct}{\mcitedefaultseppunct}\relax
\EndOfBibitem
\bibitem{AtashbarTehrani:2017mzi}
S.~Atashbar~Tehrani, \ifthenelse{\boolean{articletitles}}{\emph{{Nuclear parton distribution functions (nPDFs) and their uncertainties in the LHC Era}}, }{}\href{https://doi.org/10.18502/ken.v3i1.1758}{KnE Energ.\ Phys.\  \textbf{3} (2018) 297}, \href{http://arxiv.org/abs/1712.02153}{{\normalfont\ttfamily arXiv:1712.02153}}\relax
\mciteBstWouldAddEndPuncttrue
\mciteSetBstMidEndSepPunct{\mcitedefaultmidpunct}
{\mcitedefaultendpunct}{\mcitedefaultseppunct}\relax
\EndOfBibitem
\bibitem{Arleo:2014oha}
F.~Arleo and S.~Peign\'e, \ifthenelse{\boolean{articletitles}}{\emph{{Quarkonium suppression in heavy-ion collisions from coherent energy loss in cold nuclear matter}}, }{}\href{https://doi.org/10.1007/JHEP10(2014)073}{JHEP \textbf{10} (2014) 073}, \href{http://arxiv.org/abs/1407.5054}{{\normalfont\ttfamily arXiv:1407.5054}}\relax
\mciteBstWouldAddEndPuncttrue
\mciteSetBstMidEndSepPunct{\mcitedefaultmidpunct}
{\mcitedefaultendpunct}{\mcitedefaultseppunct}\relax
\EndOfBibitem
\bibitem{NA50:2006yzz}
NA50 collaboration, B.~Alessandro {\em et~al.}, \ifthenelse{\boolean{articletitles}}{\emph{{psi-prime production in Pb-Pb collisions at 158-GeV/nucleon}}, }{}\href{https://doi.org/10.1140/epjc/s10052-006-0153-y}{Eur.\ Phys.\ J.\ C \textbf{49} (2007) 559}, \href{http://arxiv.org/abs/nucl-ex/0612013}{{\normalfont\ttfamily arXiv:nucl-ex/0612013}}\relax
\mciteBstWouldAddEndPuncttrue
\mciteSetBstMidEndSepPunct{\mcitedefaultmidpunct}
{\mcitedefaultendpunct}{\mcitedefaultseppunct}\relax
\EndOfBibitem
\bibitem{PHENIX:2022nrm}
PHENIX collaboration, U.~A. Acharya {\em et~al.}, \ifthenelse{\boolean{articletitles}}{\emph{{Measurement of $\psi(2S)$ nuclear modification at backward and forward rapidity in $p$ $+$ $p$, $p$ $+$ Al, and $p$ $+$ Au collisions at $\sqrt{s_{_{NN}}}=200$ GeV}}, }{}\href{https://doi.org/10.1103/PhysRevC.105.064912}{Phys.\ Rev.\ C \textbf{105} (2022) 064912}, \href{http://arxiv.org/abs/2202.03863}{{\normalfont\ttfamily arXiv:2202.03863}}\relax
\mciteBstWouldAddEndPuncttrue
\mciteSetBstMidEndSepPunct{\mcitedefaultmidpunct}
{\mcitedefaultendpunct}{\mcitedefaultseppunct}\relax
\EndOfBibitem
\bibitem{LHCb:2018psc}
LHCb collaboration, R.~Aaij {\em et~al.}, \ifthenelse{\boolean{articletitles}}{\emph{{Study of $\OneS$ production in $p$Pb collisions at $\sqrt{s_{NN}}=8.16$ TeV}}, }{}\href{https://doi.org/10.1007/JHEP11(2018)194}{JHEP \textbf{11} (2018) 194}, \href{http://arxiv.org/abs/1810.07655}{{\normalfont\ttfamily arXiv:1810.07655}}, [Erratum: JHEP 02, 093 (2020)]\relax
\mciteBstWouldAddEndPuncttrue
\mciteSetBstMidEndSepPunct{\mcitedefaultmidpunct}
{\mcitedefaultendpunct}{\mcitedefaultseppunct}\relax
\EndOfBibitem
\bibitem{ALICE:2020vjy}
ALICE collaboration, S.~Acharya {\em et~al.}, \ifthenelse{\boolean{articletitles}}{\emph{{Measurement of nuclear effects on $\psi\rm{(2S)}$ production in p-Pb collisions at $\sqrt{\textit{s}_{\rm NN}} = 8.16$ TeV}}, }{}\href{https://doi.org/10.1007/JHEP07(2020)237}{JHEP \textbf{07} (2020) 237}, \href{http://arxiv.org/abs/2003.06053}{{\normalfont\ttfamily arXiv:2003.06053}}\relax
\mciteBstWouldAddEndPuncttrue
\mciteSetBstMidEndSepPunct{\mcitedefaultmidpunct}
{\mcitedefaultendpunct}{\mcitedefaultseppunct}\relax
\EndOfBibitem
\bibitem{LHCb:2016vqr}
LHCb collaboration, R.~Aaij {\em et~al.}, \ifthenelse{\boolean{articletitles}}{\emph{{Study of $\psi(2S)$ production and cold nuclear matter effects in pPb collisions at $\sqrt{s_{NN}}=5~\mathrm{TeV}$}}, }{}\href{https://doi.org/10.1007/JHEP03(2016)133}{JHEP \textbf{03} (2016) 133}, \href{http://arxiv.org/abs/1601.07878}{{\normalfont\ttfamily arXiv:1601.07878}}\relax
\mciteBstWouldAddEndPuncttrue
\mciteSetBstMidEndSepPunct{\mcitedefaultmidpunct}
{\mcitedefaultendpunct}{\mcitedefaultseppunct}\relax
\EndOfBibitem
\bibitem{CMS:2022wfi}
CMS collaboration, A.~Tumasyan {\em et~al.}, \ifthenelse{\boolean{articletitles}}{\emph{{Nuclear modification of $\Upsilon$ states in pPb collisions at $\sqrt{s_\mathrm{NN}}$ = 5.02 TeV}}, }{}\href{https://doi.org/10.1016/j.physletb.2022.137397}{Phys.\ Lett.\ B \textbf{835} (2022) 137397}, \href{http://arxiv.org/abs/2202.11807}{{\normalfont\ttfamily arXiv:2202.11807}}\relax
\mciteBstWouldAddEndPuncttrue
\mciteSetBstMidEndSepPunct{\mcitedefaultmidpunct}
{\mcitedefaultendpunct}{\mcitedefaultseppunct}\relax
\EndOfBibitem
\bibitem{ATLAS:2017prf}
ATLAS collaboration, M.~Aaboud {\em et~al.}, \ifthenelse{\boolean{articletitles}}{\emph{{Measurement of quarkonium production in proton\textendash{}lead and proton\textendash{}proton collisions at $5.02~\mathrm {TeV}$ with the ATLAS detector}}, }{}\href{https://doi.org/10.1140/epjc/s10052-018-5624-4}{Eur.\ Phys.\ J.\ C \textbf{78} (2018) 171}, \href{http://arxiv.org/abs/1709.03089}{{\normalfont\ttfamily arXiv:1709.03089}}\relax
\mciteBstWouldAddEndPuncttrue
\mciteSetBstMidEndSepPunct{\mcitedefaultmidpunct}
{\mcitedefaultendpunct}{\mcitedefaultseppunct}\relax
\EndOfBibitem
\bibitem{ALICE:2016fzo}
ALICE collaboration, J.~Adam {\em et~al.}, \ifthenelse{\boolean{articletitles}}{\emph{{Enhanced production of multi-strange hadrons in high-multiplicity proton-proton collisions}}, }{}\href{https://doi.org/10.1038/nphys4111}{Nature Phys.\  \textbf{13} (2017) 535}, \href{http://arxiv.org/abs/1606.07424}{{\normalfont\ttfamily arXiv:1606.07424}}\relax
\mciteBstWouldAddEndPuncttrue
\mciteSetBstMidEndSepPunct{\mcitedefaultmidpunct}
{\mcitedefaultendpunct}{\mcitedefaultseppunct}\relax
\EndOfBibitem
\bibitem{CMS:2016fnw}
CMS collaboration, V.~Khachatryan {\em et~al.}, \ifthenelse{\boolean{articletitles}}{\emph{{Evidence for collectivity in pp collisions at the LHC}}, }{}\href{https://doi.org/10.1016/j.physletb.2016.12.009}{Phys.\ Lett.\  \textbf{B765} (2017) 193}, \href{http://arxiv.org/abs/1606.06198}{{\normalfont\ttfamily arXiv:1606.06198}}\relax
\mciteBstWouldAddEndPuncttrue
\mciteSetBstMidEndSepPunct{\mcitedefaultmidpunct}
{\mcitedefaultendpunct}{\mcitedefaultseppunct}\relax
\EndOfBibitem
\bibitem{ALICE:2012eyl}
ALICE collaboration, B.~Abelev {\em et~al.}, \ifthenelse{\boolean{articletitles}}{\emph{{Long-range angular correlations on the near and away side in $p$-Pb collisions at $\sqrt{s_{NN}}=5.02$ TeV}}, }{}\href{https://doi.org/10.1016/j.physletb.2013.01.012}{Phys.\ Lett.\  \textbf{B719} (2013) 29}, \href{http://arxiv.org/abs/1212.2001}{{\normalfont\ttfamily arXiv:1212.2001}}\relax
\mciteBstWouldAddEndPuncttrue
\mciteSetBstMidEndSepPunct{\mcitedefaultmidpunct}
{\mcitedefaultendpunct}{\mcitedefaultseppunct}\relax
\EndOfBibitem
\bibitem{ATLAS:2012cix}
ATLAS collaboration, G.~Aad {\em et~al.}, \ifthenelse{\boolean{articletitles}}{\emph{{Observation of associated near-side and away-side long-range correlations in $\sqrt{s_{NN}}$=5.02 TeV proton-lead collisions with the ATLAS detector}}, }{}\href{https://doi.org/10.1103/PhysRevLett.110.182302}{Phys.\ Rev.\ Lett.\  \textbf{110} (2013) 182302}, \href{http://arxiv.org/abs/1212.5198}{{\normalfont\ttfamily arXiv:1212.5198}}\relax
\mciteBstWouldAddEndPuncttrue
\mciteSetBstMidEndSepPunct{\mcitedefaultmidpunct}
{\mcitedefaultendpunct}{\mcitedefaultseppunct}\relax
\EndOfBibitem
\bibitem{CMS:2012qk}
CMS collaboration, S.~Chatrchyan {\em et~al.}, \ifthenelse{\boolean{articletitles}}{\emph{{Observation of long-range near-side angular correlations in proton-lead collisions at the LHC}}, }{}\href{https://doi.org/10.1016/j.physletb.2012.11.025}{Phys.\ Lett.\  \textbf{B718} (2013) 795}, \href{http://arxiv.org/abs/1210.5482}{{\normalfont\ttfamily arXiv:1210.5482}}\relax
\mciteBstWouldAddEndPuncttrue
\mciteSetBstMidEndSepPunct{\mcitedefaultmidpunct}
{\mcitedefaultendpunct}{\mcitedefaultseppunct}\relax
\EndOfBibitem
\bibitem{CMS:2010ifv}
CMS collaboration, V.~Khachatryan {\em et~al.}, \ifthenelse{\boolean{articletitles}}{\emph{{Observation of long-range near-side angular correlations in proton-proton collisions at the LHC}}, }{}\href{https://doi.org/10.1007/JHEP09(2010)091}{JHEP \textbf{09} (2010) 091}, \href{http://arxiv.org/abs/1009.4122}{{\normalfont\ttfamily arXiv:1009.4122}}\relax
\mciteBstWouldAddEndPuncttrue
\mciteSetBstMidEndSepPunct{\mcitedefaultmidpunct}
{\mcitedefaultendpunct}{\mcitedefaultseppunct}\relax
\EndOfBibitem
\bibitem{Ferreiro:2012rq}
E.~G. Ferreiro, \ifthenelse{\boolean{articletitles}}{\emph{{Charmonium dissociation and recombination at LHC: Revisiting comovers}}, }{}\href{https://doi.org/10.1016/j.physletb.2014.02.011}{Phys.\ Lett.\  \textbf{B731} (2014) 57}, \href{http://arxiv.org/abs/1210.3209}{{\normalfont\ttfamily arXiv:1210.3209}}\relax
\mciteBstWouldAddEndPuncttrue
\mciteSetBstMidEndSepPunct{\mcitedefaultmidpunct}
{\mcitedefaultendpunct}{\mcitedefaultseppunct}\relax
\EndOfBibitem
\bibitem{Gavin:1996yd}
S.~Gavin and R.~Vogt, \ifthenelse{\boolean{articletitles}}{\emph{{Charmonium suppression by Comover scattering in Pb + Pb collisions}}, }{}\href{https://doi.org/10.1103/PhysRevLett.78.1006}{Phys.\ Rev.\ Lett.\  \textbf{78} (1997) 1006}, \href{http://arxiv.org/abs/hep-ph/9606460}{{\normalfont\ttfamily arXiv:hep-ph/9606460}}\relax
\mciteBstWouldAddEndPuncttrue
\mciteSetBstMidEndSepPunct{\mcitedefaultmidpunct}
{\mcitedefaultendpunct}{\mcitedefaultseppunct}\relax
\EndOfBibitem
\bibitem{Ferreiro:2018wbd}
E.~G. Ferreiro and J.-P. Lansberg, \ifthenelse{\boolean{articletitles}}{\emph{{Is bottomonium suppression in proton-nucleus and nucleus-nucleus collisions at LHC energies due to the same effects?}}, }{}\href{https://doi.org/10.1007/JHEP10(2018)094}{JHEP \textbf{10} (2018) 094}, \href{http://arxiv.org/abs/1804.04474}{{\normalfont\ttfamily arXiv:1804.04474}}, [Erratum: JHEP 03, 063 (2019)]\relax
\mciteBstWouldAddEndPuncttrue
\mciteSetBstMidEndSepPunct{\mcitedefaultmidpunct}
{\mcitedefaultendpunct}{\mcitedefaultseppunct}\relax
\EndOfBibitem
\bibitem{Esposito:2020ywk}
A.~Esposito {\em et~al.}, \ifthenelse{\boolean{articletitles}}{\emph{{The nature of X(3872) from high-multiplicity pp collisions}}, }{}\href{https://doi.org/10.1140/epjc/s10052-021-09425-w}{Eur.\ Phys.\ J.\  \textbf{C81} (2021) 669}, \href{http://arxiv.org/abs/2006.15044}{{\normalfont\ttfamily arXiv:2006.15044}}\relax
\mciteBstWouldAddEndPuncttrue
\mciteSetBstMidEndSepPunct{\mcitedefaultmidpunct}
{\mcitedefaultendpunct}{\mcitedefaultseppunct}\relax
\EndOfBibitem
\bibitem{Braaten:2020iqw}
E.~Braaten, L.-P. He, K.~Ingles, and J.~Jiang, \ifthenelse{\boolean{articletitles}}{\emph{{Production of $X(3872)$ at High Multiplicity}}, }{}\href{https://doi.org/10.1103/PhysRevD.103.L071901}{Phys.\ Rev.\  \textbf{D103} (2021) L071901}, \href{http://arxiv.org/abs/2012.13499}{{\normalfont\ttfamily arXiv:2012.13499}}\relax
\mciteBstWouldAddEndPuncttrue
\mciteSetBstMidEndSepPunct{\mcitedefaultmidpunct}
{\mcitedefaultendpunct}{\mcitedefaultseppunct}\relax
\EndOfBibitem
\bibitem{LHCb-DP-2008-001}
LHCb collaboration, A.~A. Alves~Jr.\ {\em et~al.}, \ifthenelse{\boolean{articletitles}}{\emph{{The \lhcb detector at the LHC}}, }{}\href{https://doi.org/10.1088/1748-0221/3/08/S08005}{JINST \textbf{3} (2008) S08005}\relax
\mciteBstWouldAddEndPuncttrue
\mciteSetBstMidEndSepPunct{\mcitedefaultmidpunct}
{\mcitedefaultendpunct}{\mcitedefaultseppunct}\relax
\EndOfBibitem
\bibitem{LHCb-DP-2014-002}
LHCb collaboration, R.~Aaij {\em et~al.}, \ifthenelse{\boolean{articletitles}}{\emph{{LHCb detector performance}}, }{}\href{https://doi.org/10.1142/S0217751X15300227}{Int.\ J.\ Mod.\ Phys.\  \textbf{A30} (2015) 1530022}, \href{http://arxiv.org/abs/1412.6352}{{\normalfont\ttfamily arXiv:1412.6352}}\relax
\mciteBstWouldAddEndPuncttrue
\mciteSetBstMidEndSepPunct{\mcitedefaultmidpunct}
{\mcitedefaultendpunct}{\mcitedefaultseppunct}\relax
\EndOfBibitem
\bibitem{LHCb-DP-2014-001}
R.~Aaij {\em et~al.}, \ifthenelse{\boolean{articletitles}}{\emph{{Performance of the LHCb Vertex Locator}}, }{}\href{https://doi.org/10.1088/1748-0221/9/09/P09007}{JINST \textbf{9} (2014) P09007}, \href{http://arxiv.org/abs/1405.7808}{{\normalfont\ttfamily arXiv:1405.7808}}\relax
\mciteBstWouldAddEndPuncttrue
\mciteSetBstMidEndSepPunct{\mcitedefaultmidpunct}
{\mcitedefaultendpunct}{\mcitedefaultseppunct}\relax
\EndOfBibitem
\bibitem{LHCb-DP-2013-003}
R.~Arink {\em et~al.}, \ifthenelse{\boolean{articletitles}}{\emph{{Performance of the LHCb Outer Tracker}}, }{}\href{https://doi.org/10.1088/1748-0221/9/01/P01002}{JINST \textbf{9} (2014) P01002}, \href{http://arxiv.org/abs/1311.3893}{{\normalfont\ttfamily arXiv:1311.3893}}\relax
\mciteBstWouldAddEndPuncttrue
\mciteSetBstMidEndSepPunct{\mcitedefaultmidpunct}
{\mcitedefaultendpunct}{\mcitedefaultseppunct}\relax
\EndOfBibitem
\bibitem{LHCb-DP-2012-003}
M.~Adinolfi {\em et~al.}, \ifthenelse{\boolean{articletitles}}{\emph{{Performance of the \lhcb RICH detector at the LHC}}, }{}\href{https://doi.org/10.1140/epjc/s10052-013-2431-9}{Eur.\ Phys.\ J.\  \textbf{C73} (2013) 2431}, \href{http://arxiv.org/abs/1211.6759}{{\normalfont\ttfamily arXiv:1211.6759}}\relax
\mciteBstWouldAddEndPuncttrue
\mciteSetBstMidEndSepPunct{\mcitedefaultmidpunct}
{\mcitedefaultendpunct}{\mcitedefaultseppunct}\relax
\EndOfBibitem
\bibitem{LHCb-DP-2012-002}
A.~A. Alves~Jr.\ {\em et~al.}, \ifthenelse{\boolean{articletitles}}{\emph{{Performance of the LHCb muon system}}, }{}\href{https://doi.org/10.1088/1748-0221/8/02/P02022}{JINST \textbf{8} (2013) P02022}, \href{http://arxiv.org/abs/1211.1346}{{\normalfont\ttfamily arXiv:1211.1346}}\relax
\mciteBstWouldAddEndPuncttrue
\mciteSetBstMidEndSepPunct{\mcitedefaultmidpunct}
{\mcitedefaultendpunct}{\mcitedefaultseppunct}\relax
\EndOfBibitem
\bibitem{Sjostrand:2006za}
T.~Sj\"{o}strand, S.~Mrenna, and P.~Skands, \ifthenelse{\boolean{articletitles}}{\emph{{PYTHIA 6.4 physics and manual}}, }{}\href{https://doi.org/10.1088/1126-6708/2006/05/026}{JHEP \textbf{05} (2006) 026}, \href{http://arxiv.org/abs/hep-ph/0603175}{{\normalfont\ttfamily arXiv:hep-ph/0603175}}\relax
\mciteBstWouldAddEndPuncttrue
\mciteSetBstMidEndSepPunct{\mcitedefaultmidpunct}
{\mcitedefaultendpunct}{\mcitedefaultseppunct}\relax
\EndOfBibitem
\bibitem{LHCb:2011dpk}
LHCb collaboration, I.~Belyaev {\em et~al.}, \ifthenelse{\boolean{articletitles}}{\emph{{Handling of the generation of primary events in Gauss, the LHCb simulation framework}}, }{}\href{https://doi.org/10.1088/1742-6596/331/3/032047}{J.\ Phys.\ Conf.\ Ser.\  \textbf{331} (2011) 032047}\relax
\mciteBstWouldAddEndPuncttrue
\mciteSetBstMidEndSepPunct{\mcitedefaultmidpunct}
{\mcitedefaultendpunct}{\mcitedefaultseppunct}\relax
\EndOfBibitem
\bibitem{Bargiotti:2007zz}
M.~Bargiotti and V.~Vagnoni,  \ifthenelse{\boolean{articletitles}}{\emph{{Heavy Quarkonia sector in PYTHIA 6.324: tuning, validation and perspectives at LHCb}}}{}, \href{https://cds.cern.ch/record/1042611}{LHCb-2007-042, CERN-LHCb-2007-042}, CERN, Geneva, 2007\relax
\mciteBstWouldAddEndPuncttrue
\mciteSetBstMidEndSepPunct{\mcitedefaultmidpunct}
{\mcitedefaultendpunct}{\mcitedefaultseppunct}\relax
\EndOfBibitem
\bibitem{Lange:2001uf}
D.~J. Lange, \ifthenelse{\boolean{articletitles}}{\emph{{The EvtGen particle decay simulation package}}, }{}\href{https://doi.org/10.1016/S0168-9002(01)00089-4}{Nucl.\ Instrum.\ Meth.\  \textbf{A462} (2001) 152}\relax
\mciteBstWouldAddEndPuncttrue
\mciteSetBstMidEndSepPunct{\mcitedefaultmidpunct}
{\mcitedefaultendpunct}{\mcitedefaultseppunct}\relax
\EndOfBibitem
\bibitem{Golonka:2005pn}
P.~Golonka and Z.~Was, \ifthenelse{\boolean{articletitles}}{\emph{{PHOTOS Monte Carlo: A precision tool for QED corrections in $Z$ and $W$ decays}}, }{}\href{https://doi.org/10.1140/epjc/s2005-02396-4}{Eur.\ Phys.\ J.\  \textbf{C45} (2006) 97}, \href{http://arxiv.org/abs/hep-ph/0506026}{{\normalfont\ttfamily arXiv:hep-ph/0506026}}\relax
\mciteBstWouldAddEndPuncttrue
\mciteSetBstMidEndSepPunct{\mcitedefaultmidpunct}
{\mcitedefaultendpunct}{\mcitedefaultseppunct}\relax
\EndOfBibitem
\bibitem{GEANT4:2002zbu}
GEANT4 collaboration, S.~Agostinelli {\em et~al.}, \ifthenelse{\boolean{articletitles}}{\emph{{GEANT4--a simulation toolkit}}, }{}\href{https://doi.org/10.1016/S0168-9002(03)01368-8}{Nucl.\ Instrum.\ Meth.\  \textbf{A506} (2003) 250}\relax
\mciteBstWouldAddEndPuncttrue
\mciteSetBstMidEndSepPunct{\mcitedefaultmidpunct}
{\mcitedefaultendpunct}{\mcitedefaultseppunct}\relax
\EndOfBibitem
\bibitem{Clemencic:2011zza}
LHCb collaboration, M.~Clemencic {\em et~al.}, \ifthenelse{\boolean{articletitles}}{\emph{{The LHCb simulation application, Gauss: Design, evolution and experience}}, }{}\href{https://doi.org/10.1088/1742-6596/331/3/032023}{J.\ Phys.\ Conf.\ Ser.\  \textbf{331} (2011) 032023}\relax
\mciteBstWouldAddEndPuncttrue
\mciteSetBstMidEndSepPunct{\mcitedefaultmidpunct}
{\mcitedefaultendpunct}{\mcitedefaultseppunct}\relax
\EndOfBibitem
\bibitem{LHCb-PAPER-2013-008}
LHCb collaboration, R.~Aaij {\em et~al.}, \ifthenelse{\boolean{articletitles}}{\emph{{Measurement of \jpsi polarization in \proton\proton collisions at \mbox{$\sqs=$7\tev}}}, }{}\href{https://doi.org/10.1140/epjc/s10052-013-2631-3}{Eur.\ Phys.\ J.\  \textbf{C73} (2013) 2631}, \href{http://arxiv.org/abs/1307.6379}{{\normalfont\ttfamily arXiv:1307.6379}}\relax
\mciteBstWouldAddEndPuncttrue
\mciteSetBstMidEndSepPunct{\mcitedefaultmidpunct}
{\mcitedefaultendpunct}{\mcitedefaultseppunct}\relax
\EndOfBibitem
\bibitem{LHCb-PAPER-2013-067}
LHCb collaboration, R.~Aaij {\em et~al.}, \ifthenelse{\boolean{articletitles}}{\emph{{Measurement of \psitwos polarisation in \proton\proton collisions at \mbox{$\sqs=$7\tev}}}, }{}\href{https://doi.org/10.1140/epjc/s10052-014-2872-9}{Eur.\ Phys.\ J.\  \textbf{C74} (2014) 2872}, \href{http://arxiv.org/abs/1403.1339}{{\normalfont\ttfamily arXiv:1403.1339}}\relax
\mciteBstWouldAddEndPuncttrue
\mciteSetBstMidEndSepPunct{\mcitedefaultmidpunct}
{\mcitedefaultendpunct}{\mcitedefaultseppunct}\relax
\EndOfBibitem
\bibitem{Workman:2022ynf}
Particle Data Group, R.~L. Workman and Others, \ifthenelse{\boolean{articletitles}}{\emph{{Review of Particle Physics}}, }{}\href{https://doi.org/10.1093/ptep/ptac097}{PTEP \textbf{2022} (2022) 083C01}\relax
\mciteBstWouldAddEndPuncttrue
\mciteSetBstMidEndSepPunct{\mcitedefaultmidpunct}
{\mcitedefaultendpunct}{\mcitedefaultseppunct}\relax
\EndOfBibitem
\bibitem{Aaij:2014zzy}
R.~Aaij {\em et~al.}, \ifthenelse{\boolean{articletitles}}{\emph{{Performance of the LHCb Vertex Locator}}, }{}\href{https://doi.org/10.1088/1748-0221/9/09/P09007}{JINST \textbf{9} (2014) P09007}, \href{http://arxiv.org/abs/1405.7808}{{\normalfont\ttfamily arXiv:1405.7808}}\relax
\mciteBstWouldAddEndPuncttrue
\mciteSetBstMidEndSepPunct{\mcitedefaultmidpunct}
{\mcitedefaultendpunct}{\mcitedefaultseppunct}\relax
\EndOfBibitem
\bibitem{Kucharczyk:1756296}
M.~Kucharczyk, P.~Morawski, and M.~Witek, \ifthenelse{\boolean{articletitles}}{\emph{{Primary Vertex Reconstruction at LHCb}}, }{} \href{http://cdsweb.cern.ch/search?p=LHCb-PUB-2014-044&f=reportnumber&action_search=Search&c=LHCb+Notes} {LHCb-PUB-2014-044}, 2014\relax
\mciteBstWouldAddEndPuncttrue
\mciteSetBstMidEndSepPunct{\mcitedefaultmidpunct}
{\mcitedefaultendpunct}{\mcitedefaultseppunct}\relax
\EndOfBibitem
\bibitem{Ma:2022rfl}
Z.-L. Ma, Z.~Lu, and L.~Zhang, \ifthenelse{\boolean{articletitles}}{\emph{{Inelastic heavy quarkonium photoproduction in ultrarelativistic heavy-ion collisions}}, }{}\href{http://arxiv.org/abs/2203.12153}{{\normalfont\ttfamily arXiv:2203.12153}}\relax
\mciteBstWouldAddEndPuncttrue
\mciteSetBstMidEndSepPunct{\mcitedefaultmidpunct}
{\mcitedefaultendpunct}{\mcitedefaultseppunct}\relax
\EndOfBibitem
\bibitem{Skwarnicki:1986xj}
T.~Skwarnicki, {\em {A study of the radiative cascade transitions between the Upsilon-prime and Upsilon resonances}}, PhD thesis, Institute of Nuclear Physics, Krakow, 1986, {\href{http://inspirehep.net/record/230779/}{DESY-F31-86-02}}\relax
\mciteBstWouldAddEndPuncttrue
\mciteSetBstMidEndSepPunct{\mcitedefaultmidpunct}
{\mcitedefaultendpunct}{\mcitedefaultseppunct}\relax
\EndOfBibitem
\bibitem{Anderlini:2202412}
L.~Anderlini {\em et~al.},  \ifthenelse{\boolean{articletitles}}{\emph{{The PIDCalib package}}}{}, \href{https://cds.cern.ch/record/2202412}{LHCb-PUB-2016-021, CERN-LHCb-PUB-2016-021}, CERN, Geneva, 2016\relax
\mciteBstWouldAddEndPuncttrue
\mciteSetBstMidEndSepPunct{\mcitedefaultmidpunct}
{\mcitedefaultendpunct}{\mcitedefaultseppunct}\relax
\EndOfBibitem
\bibitem{LHCb-DP-2013-002}
LHCb collaboration, R.~Aaij {\em et~al.}, \ifthenelse{\boolean{articletitles}}{\emph{{Measurement of the track reconstruction efficiency at LHCb}}, }{}\href{https://doi.org/10.1088/1748-0221/10/02/P02007}{JINST \textbf{10} (2015) P02007}, \href{http://arxiv.org/abs/1408.1251}{{\normalfont\ttfamily arXiv:1408.1251}}\relax
\mciteBstWouldAddEndPuncttrue
\mciteSetBstMidEndSepPunct{\mcitedefaultmidpunct}
{\mcitedefaultendpunct}{\mcitedefaultseppunct}\relax
\EndOfBibitem
\bibitem{Tolk:2014llp}
S.~Tolk, J.~Albrecht, F.~Dettori, and A.~Pellegrino,  \ifthenelse{\boolean{articletitles}}{\emph{{Data driven trigger efficiency determination at LHCb}}}{}, \href{https://cds.cern.ch/record/1701134}{LHCb-PUB-2014-039, CERN-LHCb-PUB-2014-039}, CERN, Geneva, 2014\relax
\mciteBstWouldAddEndPuncttrue
\mciteSetBstMidEndSepPunct{\mcitedefaultmidpunct}
{\mcitedefaultendpunct}{\mcitedefaultseppunct}\relax
\EndOfBibitem
\bibitem{Cranmer:2000du}
K.~S. Cranmer, \ifthenelse{\boolean{articletitles}}{\emph{{Kernel estimation in high-energy physics}}, }{}\href{https://doi.org/10.1016/S0010-4655(00)00243-5}{Comput.\ Phys.\ Commun.\  \textbf{136} (2001) 198}, \href{http://arxiv.org/abs/hep-ex/0011057}{{\normalfont\ttfamily arXiv:hep-ex/0011057}}\relax
\mciteBstWouldAddEndPuncttrue
\mciteSetBstMidEndSepPunct{\mcitedefaultmidpunct}
{\mcitedefaultendpunct}{\mcitedefaultseppunct}\relax
\EndOfBibitem
\bibitem{Pivk:2004ty}
M.~Pivk and F.~R. Le~Diberder, \ifthenelse{\boolean{articletitles}}{\emph{{sPlot: A statistical tool to unfold data distributions}}, }{}\href{https://doi.org/10.1016/j.nima.2005.08.106}{Nucl.\ Instrum.\ Meth.\  \textbf{A555} (2005) 356}, \href{http://arxiv.org/abs/physics/0402083}{{\normalfont\ttfamily arXiv:physics/0402083}}\relax
\mciteBstWouldAddEndPuncttrue
\mciteSetBstMidEndSepPunct{\mcitedefaultmidpunct}
{\mcitedefaultendpunct}{\mcitedefaultseppunct}\relax
\EndOfBibitem
\bibitem{DeCian:1402577}
M.~De~Cian {\em et~al.},  \ifthenelse{\boolean{articletitles}}{\emph{{Measurement of the track finding efficiency}}}{}, \href{https://cds.cern.ch/record/1402577}{LHCb-PUB-2011-025, CERN-LHCb-PUB-2011-025}, CERN, Geneva, 2012\relax
\mciteBstWouldAddEndPuncttrue
\mciteSetBstMidEndSepPunct{\mcitedefaultmidpunct}
{\mcitedefaultendpunct}{\mcitedefaultseppunct}\relax
\EndOfBibitem
\bibitem{Archilli:2013npa}
F.~Archilli {\em et~al.}, \ifthenelse{\boolean{articletitles}}{\emph{{Performance of the Muon Identification at LHCb}}, }{}\href{https://doi.org/10.1088/1748-0221/8/10/P10020}{JINST \textbf{8} (2013) P10020}, \href{http://arxiv.org/abs/1306.0249}{{\normalfont\ttfamily arXiv:1306.0249}}\relax
\mciteBstWouldAddEndPuncttrue
\mciteSetBstMidEndSepPunct{\mcitedefaultmidpunct}
{\mcitedefaultendpunct}{\mcitedefaultseppunct}\relax
\EndOfBibitem
\bibitem{LHCb-DP-2012-004}
R.~Aaij {\em et~al.}, \ifthenelse{\boolean{articletitles}}{\emph{{The \lhcb trigger and its performance in 2011}}, }{}\href{https://doi.org/10.1088/1748-0221/8/04/P04022}{JINST \textbf{8} (2013) P04022}, \href{http://arxiv.org/abs/1211.3055}{{\normalfont\ttfamily arXiv:1211.3055}}\relax
\mciteBstWouldAddEndPuncttrue
\mciteSetBstMidEndSepPunct{\mcitedefaultmidpunct}
{\mcitedefaultendpunct}{\mcitedefaultseppunct}\relax
\EndOfBibitem
\bibitem{CMS:2020fae}
CMS collaboration, A.~M. Sirunyan {\em et~al.}, \ifthenelse{\boolean{articletitles}}{\emph{{Investigation into the event-activity dependence of $\Upsilon$(nS) relative production in proton-proton collisions at $ \sqrt{s} $ = 7 TeV}}, }{}\href{https://doi.org/10.1007/JHEP11(2020)001}{JHEP \textbf{11} (2020) 001}, \href{http://arxiv.org/abs/2007.04277}{{\normalfont\ttfamily arXiv:2007.04277}}\relax
\mciteBstWouldAddEndPuncttrue
\mciteSetBstMidEndSepPunct{\mcitedefaultmidpunct}
{\mcitedefaultendpunct}{\mcitedefaultseppunct}\relax
\EndOfBibitem
\bibitem{ATLAS-CONF-2022-023}
ATLAS collaboration,  \ifthenelse{\boolean{articletitles}}{\emph{{Correlation of $\Upsilon$ meson production with the underlying event in $pp$ collisions measured by the ATLAS experiment}}}{}, \href{https://cds.cern.ch/record/2806464}{ATLAS-CONF-2022-023}, CERN, Geneva, 2022\relax
\mciteBstWouldAddEndPuncttrue
\mciteSetBstMidEndSepPunct{\mcitedefaultmidpunct}
{\mcitedefaultendpunct}{\mcitedefaultseppunct}\relax
\EndOfBibitem
\bibitem{PHENIX:2016vmz}
PHENIX collaboration, A.~Adare {\em et~al.}, \ifthenelse{\boolean{articletitles}}{\emph{{Measurement of the relative yields of $\psitwos$ to $\jpsi$ mesons produced at forward and backward rapidity in $p+p$, $p+$Al, $p+$Au, and $^{3}$He$+$Au collisions at $\sqrt{s_{_{NN}}}=200$ GeV}}, }{}\href{https://doi.org/10.1103/PhysRevC.95.034904}{Phys.\ Rev.\  \textbf{C95} (2017) 034904}, \href{http://arxiv.org/abs/1609.06550}{{\normalfont\ttfamily arXiv:1609.06550}}\relax
\mciteBstWouldAddEndPuncttrue
\mciteSetBstMidEndSepPunct{\mcitedefaultmidpunct}
{\mcitedefaultendpunct}{\mcitedefaultseppunct}\relax
\EndOfBibitem
\bibitem{NA50:2006rdp}
NA50 collaboration, B.~Alessandro {\em et~al.}, \ifthenelse{\boolean{articletitles}}{\emph{{$J/\psi$ and $\psi\prime$ production and their normal nuclear absorption in proton-nucleus collisions at 400-GeV}}, }{}\href{https://doi.org/10.1140/epjc/s10052-006-0079-4}{Eur.\ Phys.\ J.\  \textbf{C48} (2006) 329}, \href{http://arxiv.org/abs/nucl-ex/0612012}{{\normalfont\ttfamily arXiv:nucl-ex/0612012}}\relax
\mciteBstWouldAddEndPuncttrue
\mciteSetBstMidEndSepPunct{\mcitedefaultmidpunct}
{\mcitedefaultendpunct}{\mcitedefaultseppunct}\relax
\EndOfBibitem
\bibitem{PHENIX:2011gyb}
PHENIX collaboration, A.~Adare {\em et~al.}, \ifthenelse{\boolean{articletitles}}{\emph{{Ground and excited charmonium state production in $p+p$ collisions at $\sqrt{s}=200$ GeV}}, }{}\href{https://doi.org/10.1103/PhysRevD.85.092004}{Phys.\ Rev.\  \textbf{D85} (2012) 092004}, \href{http://arxiv.org/abs/1105.1966}{{\normalfont\ttfamily arXiv:1105.1966}}\relax
\mciteBstWouldAddEndPuncttrue
\mciteSetBstMidEndSepPunct{\mcitedefaultmidpunct}
{\mcitedefaultendpunct}{\mcitedefaultseppunct}\relax
\EndOfBibitem
\bibitem{E705:1992vec}
E705 collaboration, L.~Antoniazzi {\em et~al.}, \ifthenelse{\boolean{articletitles}}{\emph{{Production of $\jpsi$ via $\psitwos$ and $\Xi$ decay in 300-GeV/c proton and $\pi^{\pm}$ nucleon interactions}}, }{}\href{https://doi.org/10.1103/PhysRevLett.70.383}{Phys.\ Rev.\ Lett.\  \textbf{70} (1993) 383}\relax
\mciteBstWouldAddEndPuncttrue
\mciteSetBstMidEndSepPunct{\mcitedefaultmidpunct}
{\mcitedefaultendpunct}{\mcitedefaultseppunct}\relax
\EndOfBibitem
\bibitem{NA51:1998uun}
NA51 collaboration, M.~C. Abreu {\em et~al.}, \ifthenelse{\boolean{articletitles}}{\emph{{$J/\psi$, $\psi\prime$ and Drell-Yan production in pp and pd interactions at 450-GeV/c}}, }{}\href{https://doi.org/10.1016/S0370-2693(98)01014-4}{Phys.\ Lett.\  \textbf{B438} (1998) 35}\relax
\mciteBstWouldAddEndPuncttrue
\mciteSetBstMidEndSepPunct{\mcitedefaultmidpunct}
{\mcitedefaultendpunct}{\mcitedefaultseppunct}\relax
\EndOfBibitem
\bibitem{Clark:1978mg}
A.~G. Clark {\em et~al.}, \ifthenelse{\boolean{articletitles}}{\emph{{Electron pair production at the {CERN} {ISR}}}, }{}\href{https://doi.org/10.1016/0550-3213(78)90400-5}{Nucl.\ Phys.\  \textbf{B142} (1978) 29}\relax
\mciteBstWouldAddEndPuncttrue
\mciteSetBstMidEndSepPunct{\mcitedefaultmidpunct}
{\mcitedefaultendpunct}{\mcitedefaultseppunct}\relax
\EndOfBibitem
\bibitem{UA1:1990eni}
UA1 collaboration, C.~Albajar {\em et~al.}, \ifthenelse{\boolean{articletitles}}{\emph{{$J/\psi$ and $\psi\prime$ production at the CERN $p\bar{p}$ collider}}, }{}\href{https://doi.org/10.1016/0370-2693(91)90227-H}{Phys.\ Lett.\  \textbf{B256} (1991) 112}\relax
\mciteBstWouldAddEndPuncttrue
\mciteSetBstMidEndSepPunct{\mcitedefaultmidpunct}
{\mcitedefaultendpunct}{\mcitedefaultseppunct}\relax
\EndOfBibitem
\bibitem{CDF:1997ykw}
CDF collaboration, F.~Abe {\em et~al.}, \ifthenelse{\boolean{articletitles}}{\emph{{$J/\psi$ and $\psi(2S)$ production in $p\bar{p}$ collisions at $\sqrt{s} = 1.8$ TeV}}, }{}\href{https://doi.org/10.1103/PhysRevLett.79.572}{Phys.\ Rev.\ Lett.\  \textbf{79} (1997) 572}\relax
\mciteBstWouldAddEndPuncttrue
\mciteSetBstMidEndSepPunct{\mcitedefaultmidpunct}
{\mcitedefaultendpunct}{\mcitedefaultseppunct}\relax
\EndOfBibitem
\bibitem{LHCb:2013nqs}
LHCb collaboration, R.~Aaij {\em et~al.}, \ifthenelse{\boolean{articletitles}}{\emph{{Exclusive $J/\psi$ and $\psi$(2S) production in pp collisions at $ \sqrt{s} = 7$ TeV}}, }{}\href{https://doi.org/10.1088/0954-3899/40/4/045001}{J.\ Phys.\ G \textbf{G40} (2013) 045001}, \href{http://arxiv.org/abs/1301.7084}{{\normalfont\ttfamily arXiv:1301.7084}}\relax
\mciteBstWouldAddEndPuncttrue
\mciteSetBstMidEndSepPunct{\mcitedefaultmidpunct}
{\mcitedefaultendpunct}{\mcitedefaultseppunct}\relax
\EndOfBibitem
\bibitem{CMS:2011rxs}
CMS collaboration, S.~Chatrchyan {\em et~al.}, \ifthenelse{\boolean{articletitles}}{\emph{{$\jpsi$ and $\psitwos$ production in $pp$ collisions at $\sqrt{s}=7$ TeV}}, }{}\href{https://doi.org/10.1007/JHEP02(2012)011}{JHEP \textbf{02} (2012) 011}, \href{http://arxiv.org/abs/1111.1557}{{\normalfont\ttfamily arXiv:1111.1557}}\relax
\mciteBstWouldAddEndPuncttrue
\mciteSetBstMidEndSepPunct{\mcitedefaultmidpunct}
{\mcitedefaultendpunct}{\mcitedefaultseppunct}\relax
\EndOfBibitem
\bibitem{ALICE:2017leg}
ALICE collaboration, S.~Acharya {\em et~al.}, \ifthenelse{\boolean{articletitles}}{\emph{{Energy dependence of forward-rapidity $\jpsi $ and $\psi \mathrm {(2S)}$ production in pp collisions at the LHC}}, }{}\href{https://doi.org/10.1140/epjc/s10052-017-4940-4}{Eur.\ Phys.\ J.\  \textbf{C77} (2017) 392}, \href{http://arxiv.org/abs/1702.00557}{{\normalfont\ttfamily arXiv:1702.00557}}\relax
\mciteBstWouldAddEndPuncttrue
\mciteSetBstMidEndSepPunct{\mcitedefaultmidpunct}
{\mcitedefaultendpunct}{\mcitedefaultseppunct}\relax
\EndOfBibitem
\bibitem{Ma:2014mri}
Y.-Q. Ma and R.~Venugopalan, \ifthenelse{\boolean{articletitles}}{\emph{{Comprehensive Description of J/\ensuremath{\psi} Production in Proton-Proton Collisions at Collider Energies}}, }{}\href{https://doi.org/10.1103/PhysRevLett.113.192301}{Phys.\ Rev.\ Lett.\  \textbf{113} (2014) 192301}, \href{http://arxiv.org/abs/1408.4075}{{\normalfont\ttfamily arXiv:1408.4075}}\relax
\mciteBstWouldAddEndPuncttrue
\mciteSetBstMidEndSepPunct{\mcitedefaultmidpunct}
{\mcitedefaultendpunct}{\mcitedefaultseppunct}\relax
\EndOfBibitem
\end{mcitethebibliography}
\centerline
{\large\bf LHCb collaboration}
\begin
{flushleft}
\small
R.~Aaij$^{35}$\lhcborcid{0000-0003-0533-1952},
A.S.W.~Abdelmotteleb$^{54}$\lhcborcid{0000-0001-7905-0542},
C.~Abellan~Beteta$^{48}$,
F.~Abudin{\'e}n$^{54}$\lhcborcid{0000-0002-6737-3528},
T.~Ackernley$^{58}$\lhcborcid{0000-0002-5951-3498},
B.~Adeva$^{44}$\lhcborcid{0000-0001-9756-3712},
M.~Adinolfi$^{52}$\lhcborcid{0000-0002-1326-1264},
P.~Adlarson$^{78}$\lhcborcid{0000-0001-6280-3851},
C.~Agapopoulou$^{46}$\lhcborcid{0000-0002-2368-0147},
C.A.~Aidala$^{79}$\lhcborcid{0000-0001-9540-4988}
% ...

{\footnotesize \it

% ...
$^{35}$Nikhef National Institute for Subatomic Physics, Amsterdam, Netherlands\\
% ...
$^{44}$Instituto Galego de F{\'\i}sica de Altas Enerx{\'\i}as (IGFAE), Universidade de Santiago de Compostela, Santiago de Compostela, Spain\\
% ...
$^{46}$European Organization for Nuclear Research (CERN), Geneva, Switzerland\\
% ...
$^{48}$Physik-Institut, Universit{\"a}t Z{\"u}rich, Z{\"u}rich, Switzerland\\
% ...
$^{52}$H.H. Wills Physics Laboratory, University of Bristol, Bristol, United Kingdom\\
% ...
$^{54}$Department of Physics, University of Warwick, Coventry, United Kingdom\\
% ...
$^{57}$School of Physics and Astronomy, University of Glasgow, Glasgow, United Kingdom\\
$^{58}$Oliver Lodge Laboratory, University of Liverpool, Liverpool, United Kingdom\\
% ...
$^{66}$Syracuse University, Syracuse, NY, United States\\
% ...
$^{78}$Department of Physics and Astronomy, Uppsala University, Uppsala, Sweden, associated to $^{57}$\\
$^{79}$University of Michigan, Ann Arbor, MI, United States, associated to $^{66}$\\
% ...
}
\end{flushleft}

\end{document}